\RequirePackage{fix-cm} 
\documentclass[a4paper, twoside, reqno, 12pt, dvips]{amsart}
\usepackage{fixltx2e}     

\usepackage[english]{babel}
\usepackage[latin1]{inputenc}

\usepackage{eucal}
\usepackage{esint}
\usepackage{amsgen}
\usepackage{amsthm}
\usepackage{amssymb}
\usepackage{amsmath}
\usepackage{amsfonts}
\usepackage{verbatim}
\usepackage[nice]{nicefrac}
\usepackage{mathrsfs, dsfont}

\usepackage{a4wide}
\usepackage{xspace}

\usepackage{indentfirst}
\usepackage{graphicx}
\usepackage{subfigure}
\usepackage[section]{placeins}
\usepackage{psfrag}

\usepackage{microtype}
\usepackage[bf, up, hang]{caption}

\usepackage[usenames, dvipsnames]{pstricks}
\usepackage{epsfig}
\usepackage{pst-grad} 
\usepackage{pst-plot} 

\usepackage{color}
\definecolor{oneblue}{rgb}{0.0, 0.0, 0.85}
\definecolor{darkgrey}{rgb}{0.273, 0.281, 0.30}

\usepackage[colorlinks,
            urlcolor=oneblue,
            linkcolor=oneblue,
            citecolor=oneblue,
            bookmarksopen=false,
            pagebackref]{hyperref}

\usepackage[compact]{titlesec}

\titleformat{\section}{\normalfont\Large\bfseries\sffamily\center\color{darkgrey}}{\thesection.}{0.5em}{}{}
\titleformat{\subsection}{\normalfont\large\bfseries\sffamily\color{darkgrey}}{\thesubsection.}{0.4em}{}{}
\titleformat{\subsubsection}{\normalfont\normalsize\bfseries\sffamily\color{darkgrey}}{\thesubsubsection.}{0.3em}{}{}

\titlespacing*{\section}{1.0em}{1.0em}{0.8em}[0em]
\titlespacing*{\subsection}{1.0em}{1.0em}{0.8em}[0em]
\titlespacing*{\subsubsection}{1.0em}{0.7em}{0.6em}[0em]

\usepackage{titletoc}

\contentsmargin{0.0em}
\dottedcontents{section}[2.5em]{\addvspace{0.45em}\bfseries}{1.0em}{0pc}
\dottedcontents{subsection}[4.5em]{}{2.6em}{0pc}
\dottedcontents{subsubsection}[5.5em]{}{3.0em}{0pc}

\usepackage{fancyhdr}
\usepackage{lastpage}

\newcommand*\Title{On the Galilean invariance of dispersive equations}
\newcommand*\Authors{A.~Duran, D.~Dutykh \& D.~Mitsotakis}

\usepackage{acronym}

\acrodef{SW}{Solitary Wave}
\acrodef{KdV}{Korteweg--de Vries}
\acrodef{BBM}{Benjamin--Bona--Mahony}
\acrodef{NSWE}{Nonlinear Shallow Water Equations}
\acrodef{iBBM}{invariant Benjamin--Bona--Mahony}
\acrodef{cPer}{classical Peregrine}
\acrodef{iPer}{invariant Peregrine}

\numberwithin{equation}{section}

\newtheorem{remark}{Remark}

\newcommand{\J}{\mathbb{J}}
\newcommand{\R}{\mathds{R}}

\newcommand{\St}{\mathrm{S}}
\newcommand{\N}{\mathcal{N}}

\newcommand{\eps}{\varepsilon}
\renewcommand{\O}{\mathcal{O}}
\renewcommand{\L}{\mathcal{L}}
\renewcommand{\H}{\mathcal{H}}

\newcommand{\ud}{\mathrm{d}}

\newcommand{\sech}{\mathrm{sech}}

\newcommand{\scal}{\boldsymbol{\cdot}}

\newcommand{\half}{{\textstyle{1\over2}}}

\newcommand{\third}{{\textstyle{1\over3}}}

\newcommand{\tthird}{{\textstyle{3\over2}}}

\begin{document}

\title[\Title]{On the Galilean invariance of some dispersive wave equations}

\author[A.~Duran]{Angel Duran}
\address{Departamento de Matem\'atica Aplicada, E.T.S.I. Telecomunicaci\'on, Campus Miguel Delibes, Universidad de Valladolid, Paseo de Belen 15, 47011 Valladolid, Spain}
\email{angel@mac.uva.es}

\author[D.~Dutykh]{Denys Dutykh$^*$}
\address{University College Dublin, School of Mathematical Sciences, Belfield, Dublin 4, Ireland \and LAMA, UMR 5127 CNRS, Universit\'e de Savoie, Campus Scientifique, 73376 Le Bourget-du-Lac Cedex, France}
\email{Denys.Dutykh@ucd.ie}
\urladdr{http://www.denys-dutykh.com/}
\thanks{$^*$ Corresponding author}

\author[D.~Mitsotakis]{Dimitrios Mitsotakis}
\address{University of California, Merced, 5200 North Lake Road, Merced, CA 94353, USA}
\email{dmitsot@gmail.com}
\urladdr{http://dmitsot.googlepages.com/}

\begin{abstract}
Surface water waves in ideal fluids have been typically modeled by asymptotic approximations of the full Euler equations. Some of these simplified models lose relevant properties of the full water wave problem. One of these properties is the Galilean symmetry, i.e. the invariance under Galilean transformations. In this paper, a mechanism to incorporate Galilean invariance in classical water wave models is proposed. The technique is applied to the Benajmin-Bona-Mahony (BBM) equation and the Peregrine (classical Boussinesq) system, leading to the corresponding Galilean invariant versions of these models. Some properties of the new equations are presented, with special emphasis on the computation and interaction of solitary wave solutions. A comparison with the Euler equations demonstrates the relevance of the Galilean invariance in the description of water waves.

\bigskip
\noindent \textbf{\keywordsname:} water waves; Galilean invariance; Boussinesq equations; Peregrine system; BBM equation; dispersive waves; solitary waves
\end{abstract}
 

\maketitle
\tableofcontents
\thispagestyle{empty}

\section{Introduction}

The purpose of this paper is to investigate the relevance of the Galilean invariance in the description of water waves. A mechanism to incorporate invariance under Galilean transformations in some classical approximate models without this property is introduced. The corresponding Galilean invariant versions of two models, the BBM equation and the Peregrine system (also known to as the `classical' Boussinesq system), are formulated. These new versions are compared numerically with their non-invariant counterparts, with some other classical models and with the Euler equations. The comparison is focused on the existence and the dynamics of solitary waves.

A natural argument in mathematical modeling  is the inheritance of the physical properties of the phenomenon under consideration through the introduction of mathematical devices. In the case of the water wave theory, approximations to the Euler equations lead to some mathematical models where some fundamental properties of the original problem can be lost. This is relevant in the case of symmetries. For example, when asymptotic expansions around the still water level are performed, the invariance under vertical translations can be broken and the derived model is valid only in this particular frame of reference. Dispersive wave models possessing the property of invariance under vertical translations have been shown to be particularly robust for the simulation of the long wave runup, cf. \cite{Dutykh2011, Dutykh2011e}. A second symmetry, on which this paper is focused on, is related to the universality of mechanical laws in all inertial frames of reference. The Galilean invariance (or Galilean relativity) is one of the fundamental properties of any mathematical model arising in classical mechanics. This principle was empirically established by Galileo \textsc{Galilei} 55 years before the formulation of Newton's laws of mechanics in 1687, \cite{Newton1687}. Nowadays, it is common to speak about this principle in terms of a symmetry of the governing equations. Hereinbelow, the Galilean invariance property is referred to the fact that the governing equations are invariant under a Galilean boost transformation, in the sense that this kind of transformations preserves the space of the solutions of the respective problem, \cite{Olver1993}. For example, the particular form of the Galilean boost will depend on the fact that if the KdV (or BBM) equation is written in terms of the free surface elevation or the horizontal velocity variable. Nevertheless, in both cases one can easily check whether a model admits Galilean invariance as a symmetry or not.

The complete water wave problem possesses naturally this property. (For a systematic study of the symmetries and the conservation laws of the full water wave formulation we refer to the work of \textsc{Benjamin} \& \textsc{Olver} (1982), \cite{Benjamin1982}, and in particular \S4.1). Nevertheless, numerous dispersive wave equations are not invariant under the Galilean transformation. This issue was already addressed in a previous study by \textsc{Christov} (2001), \cite{Christov2001}. We note that some fully nonlinear approximations such as the \acf{NSWE}, \cite{SV1871}, the improved Shallow Water Equations, \cite{Dutykh2011b}, and the Serre--Green--Naghdi equations (sometimes also referred to as the Su--Gardner equations), \cite{Serre1953, Su1969, Green1974, Green1976, Lannes2009}, are invariant under the vertical translation and the Galilean boost. On the other hand most of the Boussinesq models are not Galilean invariant, cf. \cite{BS, Nwogu1993, BCS}. 

Some consequences of the presence of symmetries in the models are well known. They include, for example, the Hamiltonian formulation and the generation of conserved quantities. In the case of Galilean invariance, it is worth mentioning its influence in the existence and stability of periodic traveling wave solutions (see e.~g., \cite{Pava2006, Carter2011}). It is also noted that the idea of exploiting symmetries of continuous equations has already been shown very beneficial in improving the behavior of underlying numerical discretizations, \cite{Kim2007a, Kim2008, Chhay2011a, Chhay2011}. However, to our knowledge, the practical implications of the loss of the Galilean symmetry are not sufficiently known. In the present study we try to shed some light on this issue and its influence on the approximate dispersive wave models. Some other properties, such as mass, momentum and energy conservation, are at least equally important. These questions have been already addressed in the literature and, consequently, are not central to our study.

The paper is organized as follows. In Section~\ref{sec:invar} some classical models of water wave theory are reviewed and the invariant counterparts of the \acs{BBM} equation and the classical Peregrine system are derived. The methodology of this derivation can be extended to other non invariant systems. These new equations are not attempted to provide new asymptotic models, but to investigate the implications of the introduction of the Galilean symmetry by comparison with the original models, other classical equations and the full Euler system. To this end, in Section ~\ref{sec:sols}, solitary wave profiles of the new equations are numerically generated. The comparison with the wave profiles of other approximations (some with exact formulas) and of the Euler equations (by using Tanaka's algorithm and Fenton's asymptotic solution) is established in terms of the amplitude-speed and amplitude-shape relations. The interactions of solitary waves for the new models, with head-on and overtaking collisions, are studied in Section~\ref{sec:interact}. Finally, the main conclusions of this study are outlined in Section~\ref{sec:concl}.

\section{Mathematical models}\label{sec:invar}

Consider an ideal fluid of constant density along with a cartesian coordinate system in two space dimensions $(x,y)$. The $y$-axis is taken vertically upwards and the $x$-axis is horizontal and coincides traditionally with the still water level. The fluid is bounded below by an impermeable horizontal bottom at $y = -d$ and above by an impermeable free surface at $y = \eta(x,t)$. We assume that the total depth $h(x, t) \equiv d + \eta (x,t)$ remains positive $h(x,t) \geqslant h_0>0$ at all times $t$. 

Assuming that the flow is incompressible and irrotational, the governing equations of the classical water wave problem, \cite{Stoker1958, Whitham1999}, are the following:
\begin{align}
  \phi^{\,2}_{xx}\ +\ \phi^{\,2}_{yy}\ &=\ 0 \qquad
  -d\,\leqslant y\leqslant\eta(x,t), \label{eq:laplace} \\
  \eta_t\ +\ \phi_x\,\eta_x\ -\ \phi_y\ &=\ 0
  \qquad y = \eta(x, t), \label{eq:kinematic} \\
  \phi_t\ +\ \half\,(\phi_x)^2\ +\ \half\,(\phi_y)^2\ +\ g\,\eta\ &=\ 0
  \qquad y = \eta(x,t), \label{eq:bernoulli} \\
  \phi_y\ &=\ 0 \qquad y = -d, \label{eq:bottomkin}
\end{align}
with $\phi$ being the velocity potential (by definition, the irrotational velocity field $u = \phi_x$ and $g$ the acceleration due to the gravity force. The water wave problem possesses Hamiltonian, \cite{Petrov1964, Zakharov1968, Broer1974, Salmon1988}, and Lagrangian, \cite{Luke1967, Clamond2009}, variational structures.

The symmetry group of the complete water wave problem \eqref{eq:laplace} -- \eqref{eq:bottomkin} was described by Benjamin \& Olver (1982) in \cite{Benjamin1982}. In particular, the full formulation of the water wave equations admits the Galilean boost symmetry and the invariance under the vertical translations (the latter issue will be addressed by the authors in a future work). However, the water wave theory has been developed from the beginning by constructing various approximate models which may conserve or break some of the symmetries, \cite{Craik2004}. Below we consider several classical models and discuss their Galilean invariance property.

\subsection{The \acs{KdV} equation}

The unidirectional propagation of long waves in the so-called Boussinesq regime (where the nonlinearity and the dispersion are of the same order of magnitude), \cite{Boussinesq1872, BCS}, can be described by the celebrated \acf{KdV} equation \cite{KdV, Johnson2004}, which in dimensional variables can be written in the form:
\begin{equation}\label{eq:KdV}
  u_t + \sqrt{gd}\,u_x + \frac{3}{2}uu_x + \frac{d^2}{6}\sqrt{gd}\,u_{xxx} = 0,
\end{equation}
where $u(x,t)$ is the horizontal velocity variable which is usually defined as the depth-averaged velocity of the fluid, \cite{Peregrine1967}, or the fluid velocity measured at some specific water depth, \cite{BS}. Some well-known properties of (\ref{eq:KdV}) are reviewed (see e.g. \cite{Gardner1974, Lax1968, Miura1976}). First, the \acs{KdV} equation is an integrable model with the following two-parameter family of solitary wave solutions:
\begin{equation*}
  u(x,t) = u_0\,\sech^2\bigl(\half\kappa(x-c_s t-x_{0})\bigr), \quad c_s = \sqrt{gd}+\frac{u_0}{2}, \quad (\kappa d)^2 = \frac{3u_0}{\sqrt{gd}},\quad u_{0}>0, x_{0}\in \mathbb{R}.
\end{equation*}

The initial value problem of the \acf{KdV} equation possesses a Hamiltonian structure
\begin{equation*}
  u_t = \J\frac{\delta\H}{\delta\/ u},
\end{equation*}
(where $\delta$ denotes the variational derivative) in a suitable phase space of functions vanishing, along with some of their derivatives, at infinity. The
skew-symmetric operator $\J$ and the Hamiltonian functional $\H$ are
\begin{equation*}
  \J = -\partial_x, \qquad
  \H = \frac12\int_\R\Bigl[\sqrt{gd}u^2 + \half u^3
       -\sqrt{gd}\,\frac{d^2}{6}u_x^2\Bigr]\ud\/x.
\end{equation*}
The Hamiltonian $\H$ is the third conserved quantity of the well-known hierarchy of invariants for \eqref{eq:KdV}, \cite{Lax1968}.

The central question in our study is the Galilean invariance of model equations. We recall that the \acs{KdV} equation \eqref{eq:KdV} possesses this property. For a systematic derivation of symmetries of (\ref{eq:KdV}) we refer to \cite{Miura1976} and \cite{Olver1993}. To verify the invariant under Galilean transformations we choose another frame of reference which moves uniformly, for example, rightwards with constant celerity $c$. This symmetry is expressed by the following transformation of variables:
\begin{equation}\label{eq:gal32}
  x \to x-\tthird ct, \quad t \to t, \quad u \to u + c.
\end{equation}
In this moving frame of reference \eqref{eq:KdV} becomes:
\begin{equation*}
  u_t -\frac{3}{2} cu_x + \sqrt{gd}\,u_x + \frac{3}{2}(u+c)u_x + \frac{d^2}{6}\sqrt{gd}\,u_{xxx} = 0.
\end{equation*}
After some simplifications one can recover the original \acs{KdV} equation, which completes the proof of the invariance.

In order to assess the relative magnitude of various terms in equation \eqref{eq:KdV}  scaled variables are introduced. The classical long wave scaling is the following:
\begin{equation}\label{eq:scaling1}
  x' := \frac{x}{\ell}, \quad y' := \frac{y}{d}, \quad t' := \frac{g}{d}t, \quad
  \eta' := \frac{\eta}{a}, \quad u' := \frac{u}{\sqrt{gd}},
\end{equation}
where $h_0$, $a$, $\ell$ are the characteristic water depth, wave amplitude and wave length respectively. Using these three characteristic lengths we can form three following important dimensionless numbers:
\begin{equation}\label{eq:scaling2}
  \varepsilon := \frac{a}{d}, \quad \mu^2 := \Bigl(\frac{d}{\ell}\Bigr)^2, \quad
  \St := \frac{\varepsilon}{\mu^2}.
\end{equation}
Parameters $\eps \ll 1$ and $\mu^2 \ll 1$ characterize the wave nonlinearity and dispersion, while the so-called Stokes number $\St$ measures the analogy between these two effects. In the Boussinesq regime the Stokes number is of order of one, $\St=\O(1)$, which establishes that the dispersion and the nonlinear effects are comparable. We note that in \cite{Bona1980a} the range of validity of the KdV had been studied and appeared that for the values of the Stokes number $0.5\leq S\leq 10$ appears to have excellent performance, while even for larger values the results are acceptable too.

Using these dimensionless and scaled variables the \acs{KdV} equation \eqref{eq:KdV} can be written in the form:
\begin{equation}\label{KdVE}
  u_t + u_x + \frac{3}{2}\eps uu_x + \frac{\mu^2}{6} u_{xxx} = 0.
\end{equation}
where the primes have been dropped. Formulas \eqref{eq:scaling1} and \eqref{eq:scaling2} will also be used in some of the developments below.

\subsection{The \acs{BBM} equation}

Benjamin, Bona \& Mahony (1970), \cite{bona}, (see also \cite{peregr}) proposed the following modification of the KdV equation, known as the \acs{BBM} equation:
\begin{equation}\label{eq:BBM}
  u_t + \sqrt{gd}\,u_x + \frac{3}{2}uu_x - \frac{d^2}{6} u_{xxt} = 0.
\end{equation}
The main idea for the derivation of this model is to use the lower order relation between time and space derivatives from (\ref{KdVE}) $u_t = -u_x + \O(\eps,\mu^2)$ to modify the higher-order dispersive term $u_{xxx} = -u_{xxt} + \O(\eps,\mu^2)$, cf. \cite{bona}.

One of the main practical motivations for this modification is to improve the dispersion relation properties of the \acs{KdV} equation.  Specifically, unlike the \acs{KdV} equation, the phase and group velocities of the \acs{BBM} equation have a lower bound. It is also noted that the \acs{BBM} equation has the following solitary wave solutions:
\begin{equation*}
  u(x,t) = u_0\,\sech^2\bigl(\half\kappa(x-c_s t-x_{0})\bigr), \quad c_s = \sqrt{gd}+\frac{u_0}{2}, \quad (\kappa d)^2 = \frac{3u_0}{\sqrt{gd} + \half u_0},\quad x_{0}\in \mathbb{R}.
\end{equation*}
The \acs{BBM} equation is not integrable, but it can also be written as an infinite-dimensional Hamiltonian system
\begin{equation*}
  u_t = \J\frac{\delta\H}{\delta\/ u},
\end{equation*}
where the operator $\J$ and the Hamiltonian functional $\H$ are defined as:
\begin{equation}\label{eq:BBM_ham}
  \J = (1-\frac{d^2}{6}\partial_{xx})^{-1}\cdot(-\partial_x), \qquad
  \H = \frac12\int_\R\bigl[\sqrt{gd}u^2 + \half u^3\bigr]\,\ud\/x.
\end{equation}
and the structure is defined on a phase space similar to that of the \acs{KdV} equation.

As far as the Galilean invariance is concerned, the change of variables \eqref{eq:gal32} applied to \eqref{eq:BBM} leads to
\begin{equation*}
  u_t -\frac{3}{2} cu_x + \sqrt{gd}\,u_x + \frac{3}{2}(u+c)u_x - \frac{d^2}{6}u_{xxt} + \frac{d^2}{4}cu_{xxx} = 0,
\end{equation*}
and after some algebraic simplifications we obtain:
\begin{equation*}
 u_t + \sqrt{gd}\,u_x + \frac{3}{2}uu_x - \frac{d^2}{6}u_{xxt} + \frac{d^2}{4}cu_{xxx} = 0.
\end{equation*}
Since there is at least one new term ($\frac{d^2}{4}cu_{xxx}$) appeared in the previous moving frame of reference, the \acs{BBM} equation is not Galilean invariant. (The relevance of this drawback always puzzled the researchers, cf. \cite{Christov2001}.).

\subsection{The \acs{iBBM} equation}

We now present a modification of the classical \acs{BBM} equation which allows us to recover the Galilean invariance property. Furthermore, the idea behind the arguments below can be applied to other models. The strategy is to add a new \emph{ad-hoc} term which will vanish the non-invariant contribution of the \acs{BBM} dispersion $u_{xxt}$ under the transformation \eqref{eq:gal32}. The resulting equation, which will be called \acf{iBBM} equation, takes the form:
\begin{equation}\label{eq:iBBM}
  u_t + \sqrt{gd}\,u_x + \frac{3}{2}uu_x - \frac{d^2}{6} u_{xxt} - \frac{d^2}{4}uu_{xxx} = 0.
\end{equation}
It is straightforward to see that \eqref{eq:iBBM} is invariant under the Galilean transformation \eqref{eq:gal32}.

The modification proposed above becomes more transparent in scaled variables. The application of the long wave limit (\ref{eq:scaling1}) to \eqref{eq:iBBM} leads to
\begin{equation*}
  u_t + u_x + \frac{3}{2}\eps uu_x - \frac{\mu^2}{6} u_{xxt} - \frac{\eps\mu^2}{4}uu_{xxx} = 0.
\end{equation*}
One can observe that the last term on the left hand side, responsible for the Galilean invariance of \eqref{eq:iBBM}, is a nonlinear term of order $\O(\eps\mu^2)$ and consequently, it is asymptotically negligible in the \acs{BBM} formulation. Since this additional term is nonlinear, the linear dispersion relation of \eqref{eq:BBM} is not modified. Its effect will be studied thoroughly in the following sections. We stress out that the original \acs{BBM} and \acs{iBBM} equations are coincide up to order $\O(\varepsilon,\mu^2)$, cf. \cite{Dullin2004}.

\begin{remark}\label{rem:iiBBM}
Unlike the \acs{BBM} equation \eqref{eq:BBM}, this invariant version \eqref{eq:iBBM} does not possess, to our knowledge, a Hamiltonian structure. However, it is possible to propose an invariantization which preserves this variational formulation along with the Galilean invariance. The alternative given by the equation
\begin{equation}\label{eq:iiBBM}
  u_t + \sqrt{gd}\,u_x + \frac{3}{2}uu_x - \frac{d^2}{6} u_{xxt}
  -\frac{d^2}{4}\bigl(2u_xu_{xx} + uu_{xxx}\bigr) = 0.
\end{equation}
has an additional higher-order nonlinear term which allows for a non-canonical Hamiltonian structure. In this case, the operator $\J = (1-\frac{d^2}{6}\partial_{xx})^{-1} \cdot (-\partial_x)$ is the same as for the \acs{BBM} equation and the Hamiltonian $\H$ is
\begin{equation*}
  \H = \frac{1}{2}\int_\R\Bigl[\sqrt{gd}u^2 + \frac{1}{2}u^3 + \frac{d^2}{4}uu_x^2\Bigr] \, \ud\/x.
\end{equation*}
We underline that the new terms in \eqref{eq:iiBBM} can be also found in several models such as the Camassa-Holm \cite{Camassa1993}, Burgers-Poisson \cite{FL} and Degasperis-Procesi \cite{Degasperis1999} equations.

Both equations \eqref{eq:iBBM}, \eqref{eq:iiBBM}  are Galilean invariant versions of the 
\acs{BBM} equation. For the comparisons performed in Section ~\ref{sec:sols}, \eqref{eq:iBBM} has been considered, with the purpose of focusing exclusively on the effects of the Galilean invariance property, which is the main goal of the paper. A similar study could be done with the alternative  \eqref{eq:iiBBM}.
\end{remark}

\begin{figure}%
\centering
\subfigure[]{\includegraphics[width=.49\textwidth]{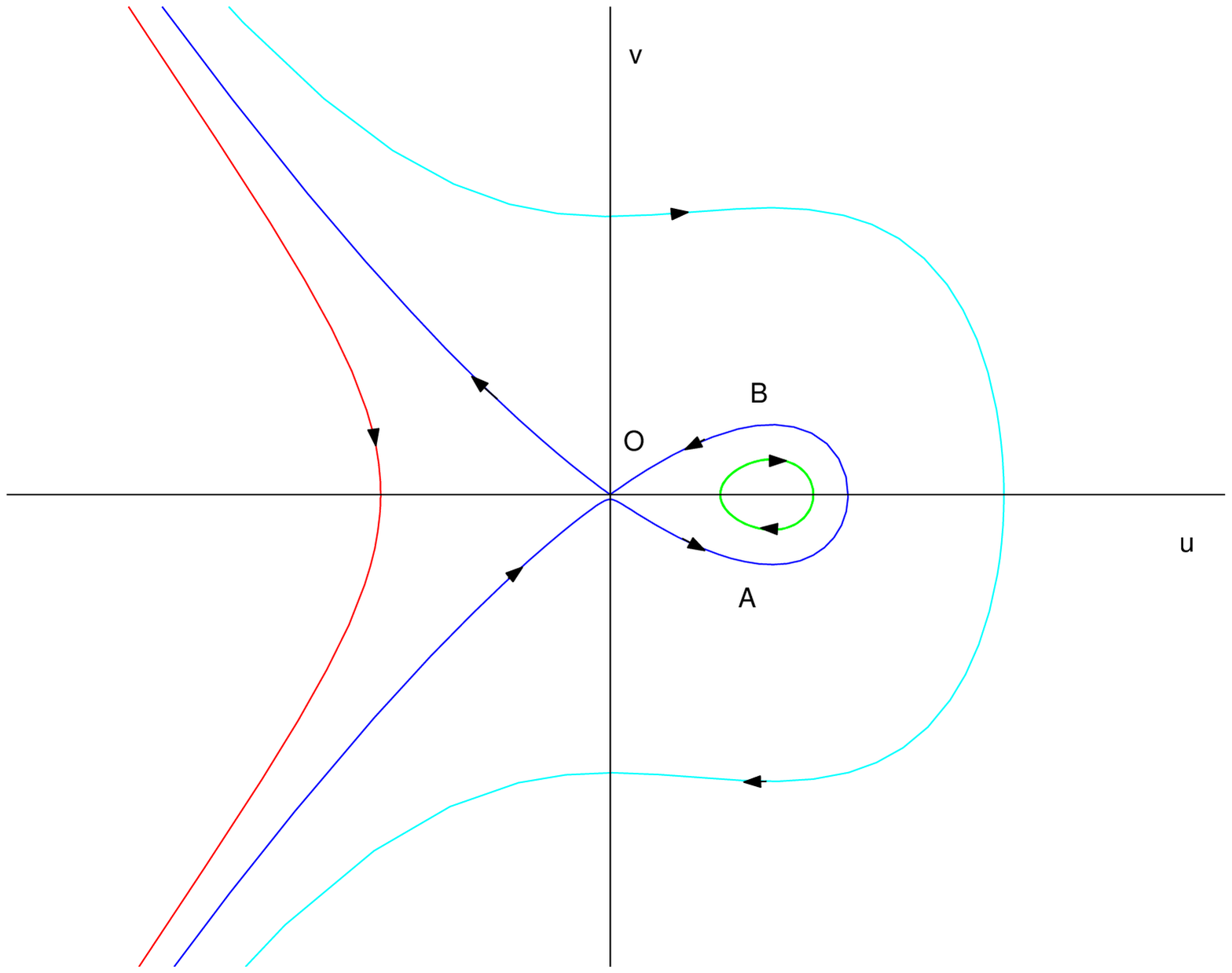}}
\subfigure[]{\includegraphics[width=.49\textwidth]{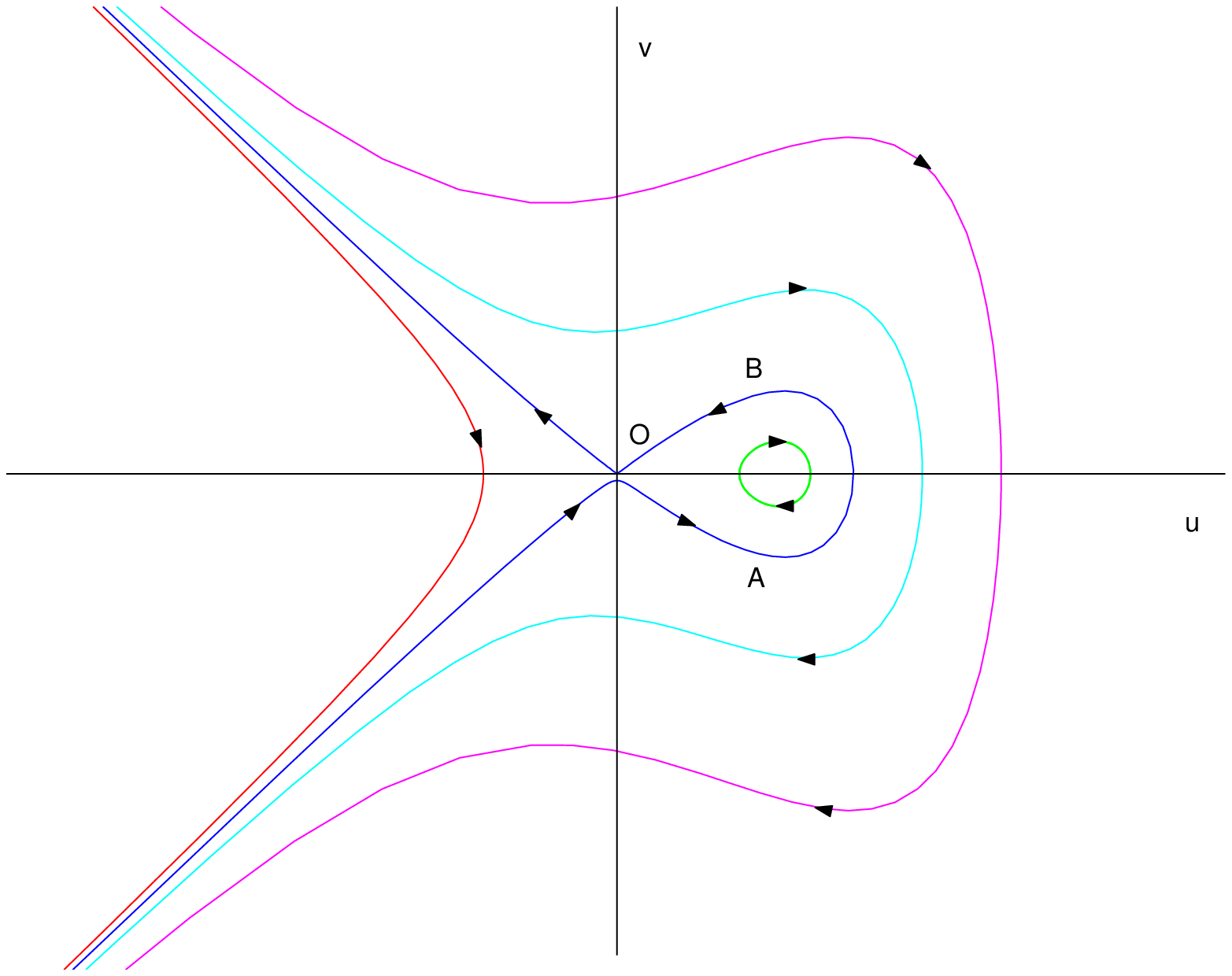}}
\caption{\em Phase plane of the invariant models with $c_{s}>\sqrt{gd}$: (a) iBBM. (b) iPer. In both cases, the trajectory $O\rightarrow A\rightarrow B\rightarrow O$ represents a solitary wave.}
\label{Fig:phasp}
\end{figure}

We now look for travelling wave solutions of \eqref{eq:iBBM} of the form:
\begin{equation}\label{sw1}
  u(x,t) = u(\xi), \qquad \xi := x - c_s t,
\end{equation}
where $c_s$ is the solitary wave speed. We also assume that $u(\xi)$ decays to zero along with all derivatives when $|\xi|\to \infty$. Substituting \eqref{sw1} into the \acs{iBBM} equation \eqref{eq:iBBM} leads to the ordinary differential equation:
\begin{equation}\label{eq:ib1}
  (\sqrt{gd}-c_s)u' + \frac{3}{4}(u^2)' + c_s\frac{d^2}{6}u''' - \frac{d^2}{4}uu''' = 0,
\end{equation}
where the prime denotes differentiation with respect to $\xi$. Using the boundary conditions at infinity, the identity $uu''' = (uu'' - \half (u')^2)'$, and after an integration, equation \eqref{eq:ib1} becomes:
\begin{equation}\label{eq:ib2}
  (\sqrt{gd}-c_s)u + \frac{3}{4} u^2 + c_s\frac{d^2}{6}u'' - \frac{d^2}{6}\Bigl(\frac{1}{2}(u')^2 - uu''\Bigr) = 0,
\end{equation}
that can be written as a system
\begin{eqnarray}
u'&=&v,\label{eq:ib3}\\
v'&=&\frac{2}{d^{2}\left(\frac{c_{s}}{3}-\frac{u}{2}\right)}\left((c_{s}-\sqrt{gd})u-\frac{3}{4}u^{2}-\frac{d^{2}}{8}v^{2}\right)
.\label{eq:ib4}
\end{eqnarray}
Now it can be verified that when $c_{s}>\sqrt{gd}$, the origin $u=v=0$ is a saddle point, as depicted in Figure~\ref{Fig:phasp}(a), which shows the corresponding phase plane. The homoclinic trajectory $O\rightarrow A\rightarrow B\rightarrow O$ represents a solitary wave. (The MATLAB code for this figure can be found in \cite{Olver2012}.). In Section \ref{sec:sols} we compute solitary wave solutions of (\ref{eq:ib2}) by numerical means.

\subsection{The Peregrine system}

Under the assumptions described above, D.H.~\textsc{Peregrine} in 1967 derived the following system of equations governing the two-way propagation of long waves of small amplitude in the Boussinesq regime \cite{Peregrine1967}:
\begin{eqnarray}\label{eq:p1}
  \eta_t + \bigl((d+\eta)u\bigr)_x &=& 0, \\
  u_t + uu_x + g\eta_x - \frac{d^2}{3}u_{xxt} &=& 0, \label{eq:p2}
\end{eqnarray}
where $u(x,t)$ is now defined as the depth averaged fluid velocity and $\eta(x,t)$ is the deviation of the free surface of the water from its rest position. This system is also known as the classical Boussinesq system, cf. \cite{BCS} and will be denoted by cPer. In \cite{Pego1997, Chen2000}, the existence and some properties of solitary wave solutions of \eqref{eq:p1}--\eqref{eq:p2} are obtained, without explicit formulas. On the other hand, to our knowledge, a Hamiltonian structure has not been found, \cite{BCS}.

We now study the Galilean invariance of \eqref{eq:p1}--\eqref{eq:p2}. In this case, the Galilean transformation takes the following form:
\begin{equation}\label{eq:gal}
  x \to x-ct, \quad t \to t, \quad \eta \to \eta, \quad u \to u + c.
\end{equation}
The mass conservation equation \eqref{eq:p1} in new variables reads:
\begin{equation*}
  \eta_t - c\eta_x + \bigl((d+\eta)(u+c)\bigr)_x = 0.
\end{equation*}
After simplifications one can see that this equation remains invariant under the transformation \eqref{eq:gal}.

Now let us consider the momentum balance equation \eqref{eq:p2}. In the moving frame of reference this equation becomes:
\begin{equation*}
  u_t - cu_x + (u+c)u_x + g\eta_x - \frac{d^2}{3}u_{xxt} + c\frac{d^2}{3}u_{xxx} = 0,
\end{equation*}
and after some simplifications
\begin{equation*}
  u_t + uu_x + g\eta_x - \frac{d^2}{3}u_{xxt} + c\frac{d^2}{3}u_{xxx} = 0.
\end{equation*}
As in the BBM case, a new dispersive term $c\frac{d^2}{3}u_{xxx}$ appears, showing that the system \eqref{eq:p1}--\eqref{eq:p2} is not Galilean invariant.

\subsection{The iPeregrine system}\label{Peregrine}

Following the same technique as in the case of the \acs{BBM} equation, it is possible to derive a modification of the classical Peregrine system \eqref{eq:p1}--\eqref{eq:p2} which will allow us to recover the Galilean invariance property. It can be done in a similar way leading to the \acs{iBBM} equation \eqref{eq:iBBM}. The corresponding system reads:
\begin{eqnarray}\label{eq:ip1}
  \eta_t + \bigl((d+\eta)u\bigr)_x &=& 0, \\
  u_t + uu_x + g\eta_x - \frac{d^2}{3}u_{xxt} - \frac{d^2}{3}uu_{xxx} &=& 0. \label{eq:ip2}
\end{eqnarray}
Note that since the mass conservation equation is invariant, it is not modified in the new version. Now it is straightforward to check the invariance of equation \eqref{eq:ip2}. Therefore, system \eqref{eq:ip1}--\eqref{eq:ip2}, which will be called invariant Peregrine system or iPer for the sake of conciseness, is Galilean invariant. In dimensionless and scaled variables, the system reads:
\begin{eqnarray*}
  \eta_t + \bigl((1+\eps\eta)u\bigr)_x &=& 0, \\
  u_t + \eps uu_x + \eta_x - \frac{\mu^2}{3}u_{xxt} - \frac{\eps\mu^2}{3}uu_{xxx} &=& 0,
\end{eqnarray*}
and one can see that the new \emph{ad-hoc} term is of higher-order and, asymptotically speaking, negligible. As in the case of the Peregrine system, equations \eqref{eq:ip1}--\eqref{eq:ip2} do not have, to our knowledge, a Hamiltonian structure.

Finally, we can look for travelling wave solutions of system \eqref{eq:ip1}--\eqref{eq:ip2} of the form
\begin{equation*}
  \eta(x,t) = \eta(\xi), \quad u(x,t) = u(\xi), \quad \xi := x - c_s t,
\end{equation*}
 where $\eta$ and $u$ decay to zero, along with their derivatives, as $|\xi|\rightarrow \infty$. After substituting this representation into the governing equations \eqref{eq:ip1}--\eqref{eq:ip2} they become:
\begin{eqnarray*}
  -c_s\eta' + \bigl((d+\eta)u\bigr)' &=& 0, \\
  -c_s u' + \frac{1}{2}(u^2)' + g\eta' + c_s\frac{d^2}{3}u''' - \frac{d^2}{3}uu''' &=& 0.
\end{eqnarray*}
An integration of the mass conservation equation and the decay at infinity lead to
\begin{equation}\label{eq:etau}
  u = \frac{c_s\eta}{d+\eta}.
\end{equation}
Then the momentum balance equation can be integrated once and we have
\begin{equation}\label{eq:etau2}
  -c_s\bigl(u - \frac{d^2}{3}u''\bigr) + \frac{1}{2} u^2 + \frac{gd\cdot u}{c_s-u} -\frac{d^2}{3}\Bigl(\frac{1}{2}(u')^2 - uu''\Bigr) = 0.
\end{equation}
Similarly to the case of the \acs{iBBM}, one can see that, \eqref{eq:etau2} written as a first order system and when $c_{s}^{2}-gd > 0$, the origin is a saddle point; the phase plane sketched in Figure~\ref{Fig:phasp}(b) also shows a solitary wave, in the form of a trajectory $O\rightarrow A\rightarrow B\rightarrow O$.

\subsection{The Serre equations}

In order to complete the presentation of our model equations, we consider the fully-nonlinear system of equations referred to as the Serre equations, \cite{Serre1953, Barthelemy2004, Dutykh2011a}:
\begin{eqnarray}\label{eq:ser1}
  h_t + [hu]_x &=& 0, \\
  u_t + uu_x + gh_x &=& \third\,h^{-1}\bigl[h^3(u_{xt} + uu_{xx} - u_x^2)\bigr]_x, \label{eq:ser2}
\end{eqnarray}
where $h(x,t) := d + \eta(x,t)$ is the total water depth variable. Solitary wave solutions of \eqref{eq:ser1}--\eqref{eq:ser2} are explicitly known. They are given by the formulas:
\begin{equation}\label{eq:ser_sol}
  \eta(x,t) = a_0\sech^2\bigl(\half\kappa(x-c_s t-x_{0})\bigr), \quad u = \frac{c_s\,\eta}{d+\eta},
\end{equation}
where $c_s = \sqrt{g(d+a_0)}$ and $(\kappa d)^2 = \frac{3a_0}{d+a_0}$.

As pointed out in \cite{Johnson2002}, \cite{Li2001} and \cite{Li2002}, the Serre equations possess a  Hamiltonian structure:
\begin{equation*}
  \begin{pmatrix}
      \tilde{q}_t \\
    h_t 
  \end{pmatrix}\ =\ -\begin{pmatrix}
  \partial_x [ \tilde{q}+\tilde{q}\partial_x[ & h\partial_x[ \\
  \partial_x[h & 0
  \end{pmatrix}
  \scal \begin{pmatrix}
                     \displaystyle{\delta\,\H\,/\,\delta\/ \tilde{q}}] \\
                     \displaystyle{\delta\,\H\,/\,\delta\/ h]}
                   \end{pmatrix},
\end{equation*}
where the Hamiltonian functional $\H$ is given by
\begin{equation*}
  \H\ =\ \half\int_\R\Bigl[\,h\,u^2\, +\, \third\, h^3\,u_x^2\, +\, g\,\eta^2\,\Bigr]\,\ud\/x,
 \end{equation*}
The variable $\tilde{q}$ is sometimes referred to as the \textit{potential vorticity flux} and is defined by
\begin{equation*}
  \tilde{q}\ :=\ h\,u\ -\ \third\,[\,h^3\,u_x\,]_x.
\end{equation*}

The Serre equations \eqref{eq:ser1}--\eqref{eq:ser2} can be shown to have the Galilean invariance property. For the mass conservation equation \eqref{eq:ser1} we refer to Section \ref{Peregrine}. Thus it remains to check this property for the momentum conservation equation \eqref{eq:ser2}. If we make the change of variables $t \to t$, $x \to x - ct$, $h \to h$ and $u \to u + c$ as before, equation \eqref{eq:ser2} becomes:
\begin{equation*}
  u_t - cu_x + (u+c)u_x + gh_x = \third\,h^{-1}\bigl[h^3(u_{xt} - cu_{xx} + (u+c)u_{xx} - u_x^2)\bigr]_x,
\end{equation*}
and after simple algebraic simplifications one can recover the original equation \eqref{eq:ser2}.

\section{Numerical computation of travelling waves}\label{sec:sols}

In the previous section we presented several models arising in water wave theory. Moreover, we proposed two novel equations, the \acs{iBBM} equation and the iPer system, with the aim of incorporating the property of invariance under the Galilean transformation, lost by the original \acs{BBM} equation and the cPer model. The purpose of this and the next sections is to compare these models through the computation of their respective solitary wave solutions and, whenever possible, with the solitary waves of full Euler equations. 
In some cases, approximate solitary waves must be generated numerically. In Section~\ref{sec:sols1} the great lines of the numerical procedure to this end is given, while the methods used to construct approximate solitary waves of the Euler equations are mentioned in Section~\ref{sec:sols2}, where a comparative study is carried out. In the sequel we consider all the models in nondimensional but unscaled variables (i.e. when $\eps = \mu^2 = 1$).


\subsection{Computation of travelling-wave profiles. The Petviashvili method}\label{sec:sols1}

The investigation for travelling wave solutions in one-dimensional systems typically leads to a set of differential equations of the form
\begin{equation}\label{eq:petv}
\L U = \N(U),
\end{equation}
for some differential operators $\L$ (linear) and $\N$ (nonlinear). The numerical resolution of the preceding system can be done in many different ways (see \cite{Yang2010} and the references therein as a modest representation of the related literature). Among all the possibilities, the so-called Petviashvili method will be used in our computations. This method stems from the pioneering work of V.I.~\textsc{Petviashvili} (1976), \cite{Petviashvili1976}. It is based on a modification of the classical fixed point iteration (which in these cases is usually divergent) and it is formulated as follows. Given an initial profile $U_{0}$, the Petviashvili iteration generates approximations $U_{n}$ of the original solution of (\ref{eq:petv}) following the formulas
\begin{eqnarray}
M_{n}&=&\frac{\langle \L U_{n},U_{n}\rangle}{\langle \N(U_{n}),U_{n}\rangle},\label{eq:fpetv1}\\
\L U_{n+1}&=&M_{n}^{\gamma}N(U_{n}),\label{eq:fpetv2}
\end{eqnarray}
where $\langle \cdot,\cdot\rangle$ denotes the usual $L^{2}$ inner product and $\gamma$ is a free parameter that controls the convergence of the method. The quantity $M_n$ in  \eqref{eq:fpetv1} is called the stabilizing factor. See \cite{Pelinovsky2004, Lakoba2007} for details, generalizations and local convergence results for some model equations.

In this study, the iteration \eqref{eq:fpetv1}--\eqref{eq:fpetv2} is applied to compute solitary wave profiles in the following cases with the corresponding operators, namely:
\begin{itemize}
\item (iBBM):
$\L u= (\sqrt{gd}-c_{s})u+ c_{s}\frac{d^{2}}{6}u'',\quad \N(u)=\frac{d^{2}}{4}\left(\frac{(u')^{2}}{2}-uu''\right)-\frac{3}{4}u^{2}.$
\item (cPer):
$\L u= c_{s}\left(u-\frac{d^{2}}{3}u''\right),\quad \N(u)=\frac{u^{2}}{2}+\frac{gdu}{c_{s}-u}.$
\item (iPer):
$\L u= c_{s}\left(u-\frac{d^{2}}{3}u''\right),\quad \N(u)=\frac{u^{2}}{2}+\frac{gdu}{c_{s}-u}-
\frac{d^{2}}{3}\left(\frac{(u')^{2}}{2}-uu''\right),$
\end{itemize}
where $c_{s}$ is the solitary wave speed. In order to reconstruct the free surface elevation profile from the horizontal velocity distribution $u(\xi)$ we use the following exact formula which can be derived for the travelling wave solutions to the full Euler equations provided that $u(\xi)$ is defined as the depth-averaged horizontal velocity:
\begin{equation*}
  u(\xi) = \frac{c_s\eta(\xi)}{d + \eta(\xi)}.
\end{equation*}

\subsubsection{Generation and numerical evolution of the profiles}\label{sec:sols2}

\begin{figure}%
\centering
\subfigure[\acf{iBBM} equation]%
{\includegraphics[width=0.49\textwidth]{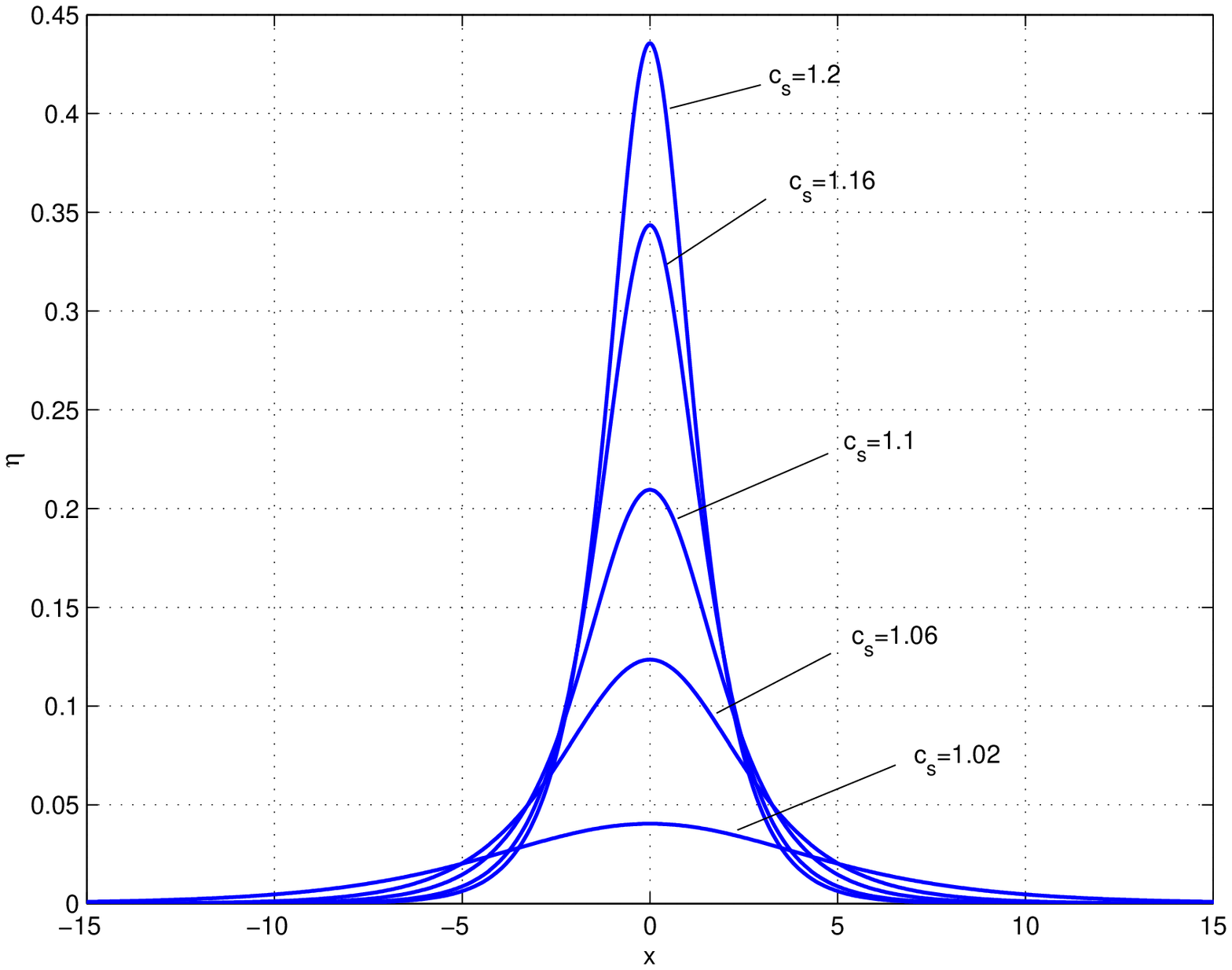}}
\subfigure[\acf{iPer} system]%
{\includegraphics[width=0.49\textwidth]{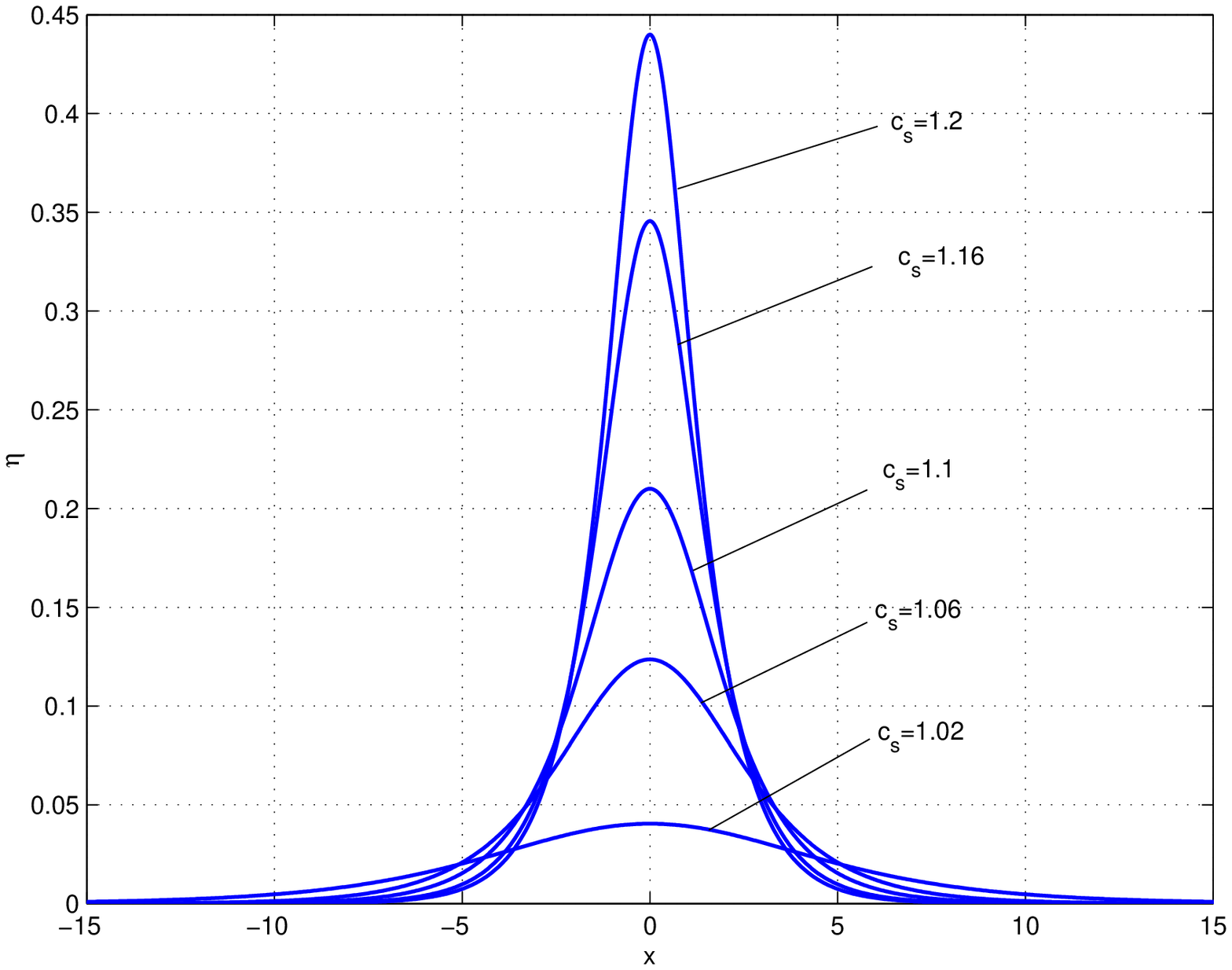}}
\caption{\em Approximate solitary wave profiles for different speeds $c_s$, generated by the Peviashvili method \eqref{eq:fpetv1}--\eqref{eq:fpetv2}.}
\label{Fig:isw}
\end{figure}

In many cases, the method \eqref{eq:fpetv1}--\eqref{eq:fpetv2} can be efficiently implemented by using Fourier techniques, \cite{Pelinovsky2004, Lakoba2007}. Specifically, our implementation for the three systems has been performed by considering the corresponding periodic problem and using a pseudospectral representation for the approximations to the profiles. As an initial approximation, a solitary wave solution \eqref{eq:ser_sol} of the Serre equations or the third-order asymptotic solution of Grimshaw, \cite{Grimshaw1971}, can be considered. The iterative procedure is continued until the difference between two consecutive iterations in the $L_\infty$-norm, or the $L_\infty$-norm of the residual is less than a prescribed small tolerance which, in our case, is of order $\O(10^{-15})$. The convergence is reached within 10--20 iterations. In order to illustrate better the transformations that solitary waves undergo while we gradually increase the propagation speed parameter $c_s$, we superpose several profiles on the same Figure (Figure~\ref{Fig:isw}(a) corresponds to the \acf{iBBM} system and Figure~\ref{Fig:isw}(b) to the \acf{iPer} system).

In order to assess the accuracy of the computations, the three models considered in this paper have been numerically integrated in time using the computed solitary wave profiles as initial conditions. Some error indicators measuring the accuracy of the numerical approximation of the solitary waves have been computed. The numerical method for the corresponding initial-periodic boundary value problem consists of a pseudospectral method for the semi-discretization in space and the classical, explicit fourth-order Runge-Kutta scheme for the time integration, \cite{DMII}.

In the experiments below, we study the propagation of a solitary-wave profile, generated by \eqref{eq:fpetv1}--\eqref{eq:fpetv2} with speed $c_{s} = 1.1$ in the interval $[-128,128]$, with $N=2048$ nodes for the pseudospectral approximation and spatial and time step sizes $\Delta x = 1.25\times 10^{-2}$, and $\Delta t = 1.25\times 10^{-3}$ respectively. The results correspond to \acs{cPer} and \acs{iPer} systems. The \acs{iBBM} equation has also been implemented, with similar results.

\begin{figure}%
\centering
\subfigure[Normalized amplitude error]%
{\includegraphics[width=0.49\textwidth]{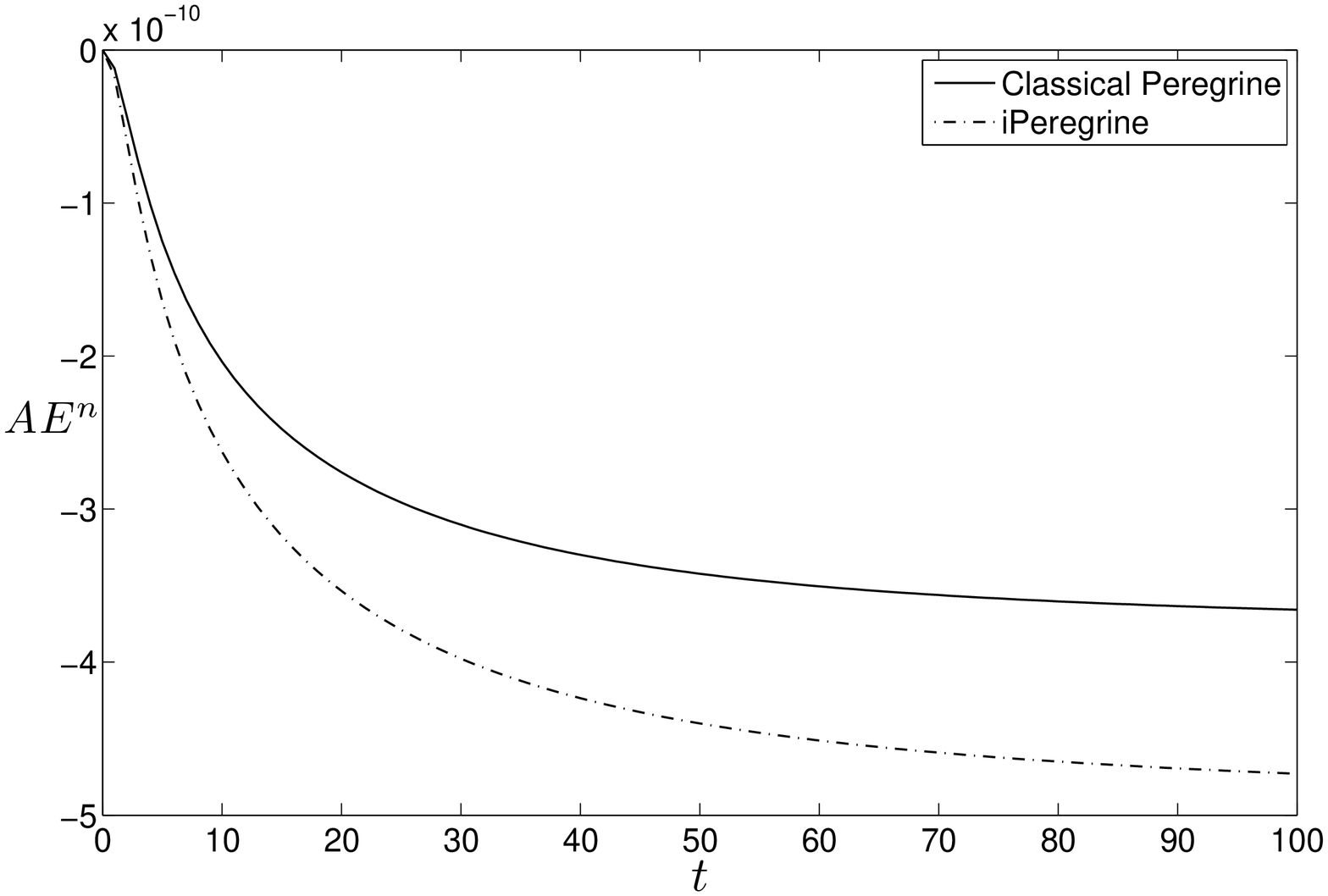}}
\subfigure[Shape error]%
{\includegraphics[width=0.49\textwidth]{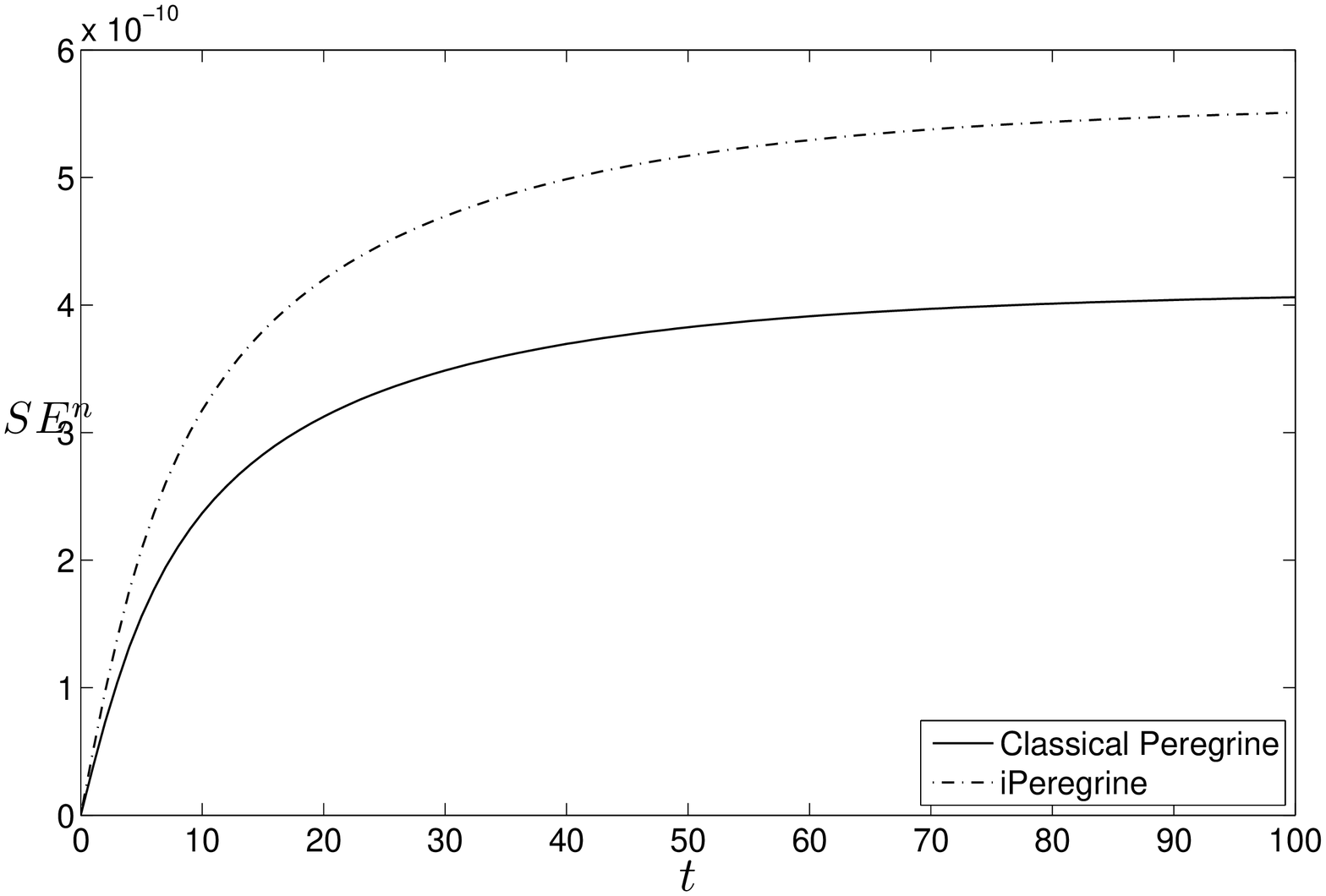}}
\caption{\em Some error indicators for the propagation of a solitary for the Peregrine and the invariant Peregrine systems.}
\label{Fig:errs1}
\end{figure}

We study the evolution of two parameters: the normalized amplitude error and the shape error. The first one is computed by comparing, at each timestep, the initial amplitude of the profile (generated by the Petviashvili method) with the corresponding amplitude of the numerical solution computed using Newton's method, \cite{DDMM}. The $L^2$ based, normalized shape error is also defined at each timestep for, lets say, the solitary wave $u$ as $SE^n=\inf_\tau \|U^n-u(\cdot,\tau)\|/\|u(\cdot,0)\|$. For the computation of the shape error we compute the time $\tau^\ast$ neat the timestep $t^n$ such that $\frac{d}{d\tau}\xi^2(\tau^\ast)=0$, where $\xi(\tau):=\|U^n-u(\cdot,\tau)\|/\|u(\cdot,0)\|$ using Newton's method and an initial approximation $\tau^0=t^n-\Delta t$. Then the shape error is the quantity $SE^n=\xi(\tau^\ast)$. Their computation is implemented as in e.g. \cite{DDMM}. Figure~\ref{Fig:errs1}(a) shows the temporal evolution of this amplitude error up to a final time $T = 100$, for both cPer and iPer. We observe that for the specific values of $\Delta x$ and $\Delta t$ the amplitude is conserved up to 10 decimal digits in both cases.

On the other hand, the shape error is computed by comparing the numerical solution with time translations of the initial profile with the prescribed speed and minimizing the differences (see \cite{DMII} for the details). The results displayed in Figure~\ref{Fig:errs1}(b) show a virtually constant evolution of this error, which in both cases is of order $\O(10^{-10})$. These results confirm the accuracy of the technique used to generate the solitary wave profiles and of the numerical code for the time evolution. The latter will be used for the experiments below.

\subsection{Numerical results}

In this section we compare the solitary waves of the following models:
\begin{itemize}
  \item The \acf{KdV} equation \eqref{eq:KdV}.
  \item The \acf{BBM} equation \eqref{eq:BBM}.
  \item The \acf{iBBM} equation \eqref{eq:iBBM}.
  \item The Peregrine \acf{cPer} system  \eqref{eq:p1}, \eqref{eq:p2}.
  \item The invariant Peregrine \acf{iPer} system \eqref{eq:ip1}, \eqref{eq:ip2}.
  \item The Serre equations \eqref{eq:ser1}, \eqref{eq:ser2},
\end{itemize}
from either the analytical formula (when possible) or the computations with the Petviashvili method. The comparison is established between them and with those of the Euler equations. The construction of the approximations to travelling wave solutions for the 2D Euler equations with free surface (to be considered as reference solutions for the approximate models) will be computed with two techniques. One by using the Tanaka's algorithm, \cite{Tanaka1986}, and the other by using asymptotic solutions given in, \cite{Fenton1972, Longuet-Higgins1974}. The study is focused on the amplitudes and shapes of the computed free surface elevation $\eta(\xi)$, provided by the models for the same prescribed value of the propagation speed parameter $c_{s}$.

\subsubsection{Solitary wave speed--amplitude relation}

\begin{figure}
\centering
\subfigure[Solitary wave amplitude--speed diagram]%
{\includegraphics[width=0.48\textwidth]{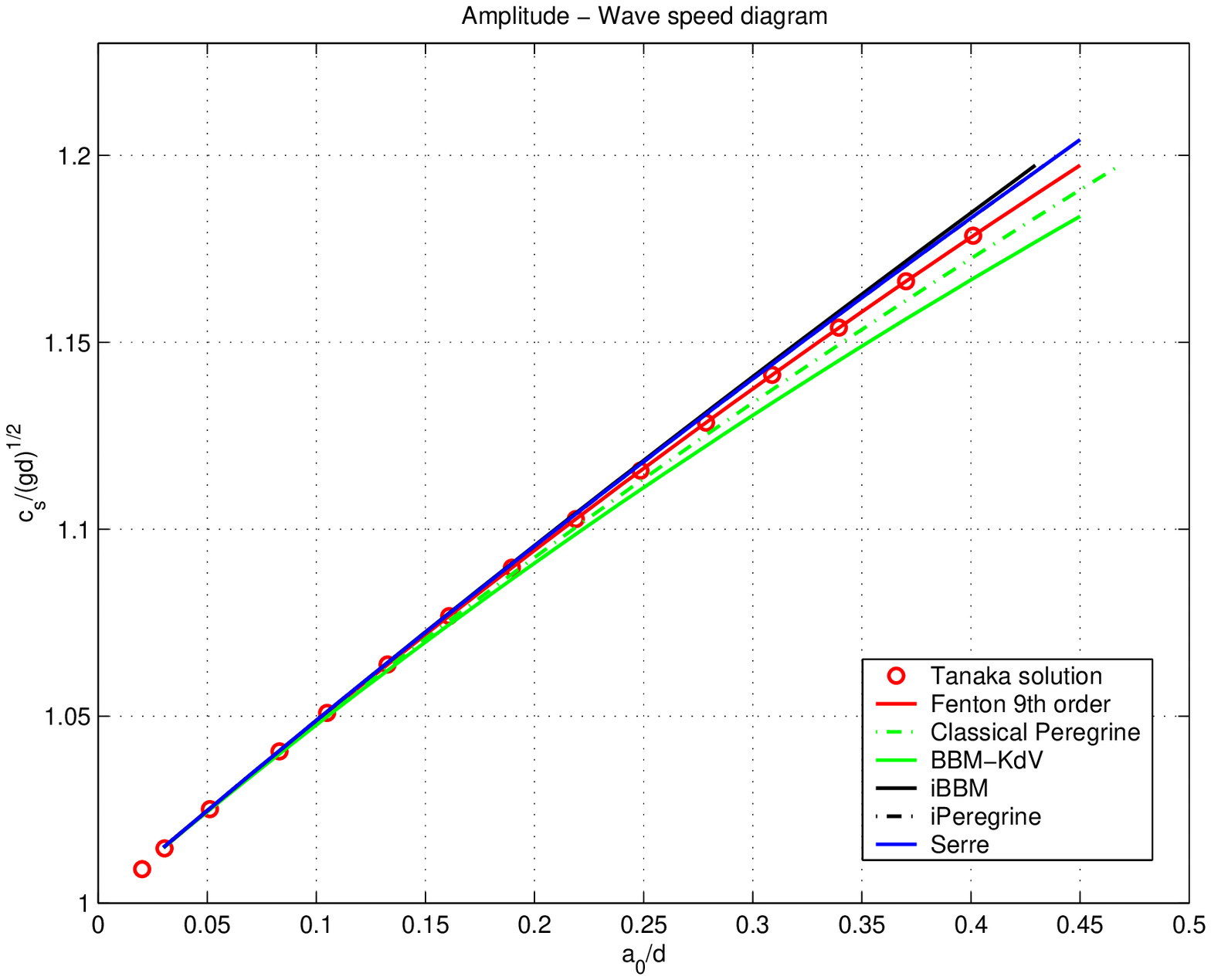}}
\subfigure[Magnification of (a)]%
{\includegraphics[width=0.48\textwidth]{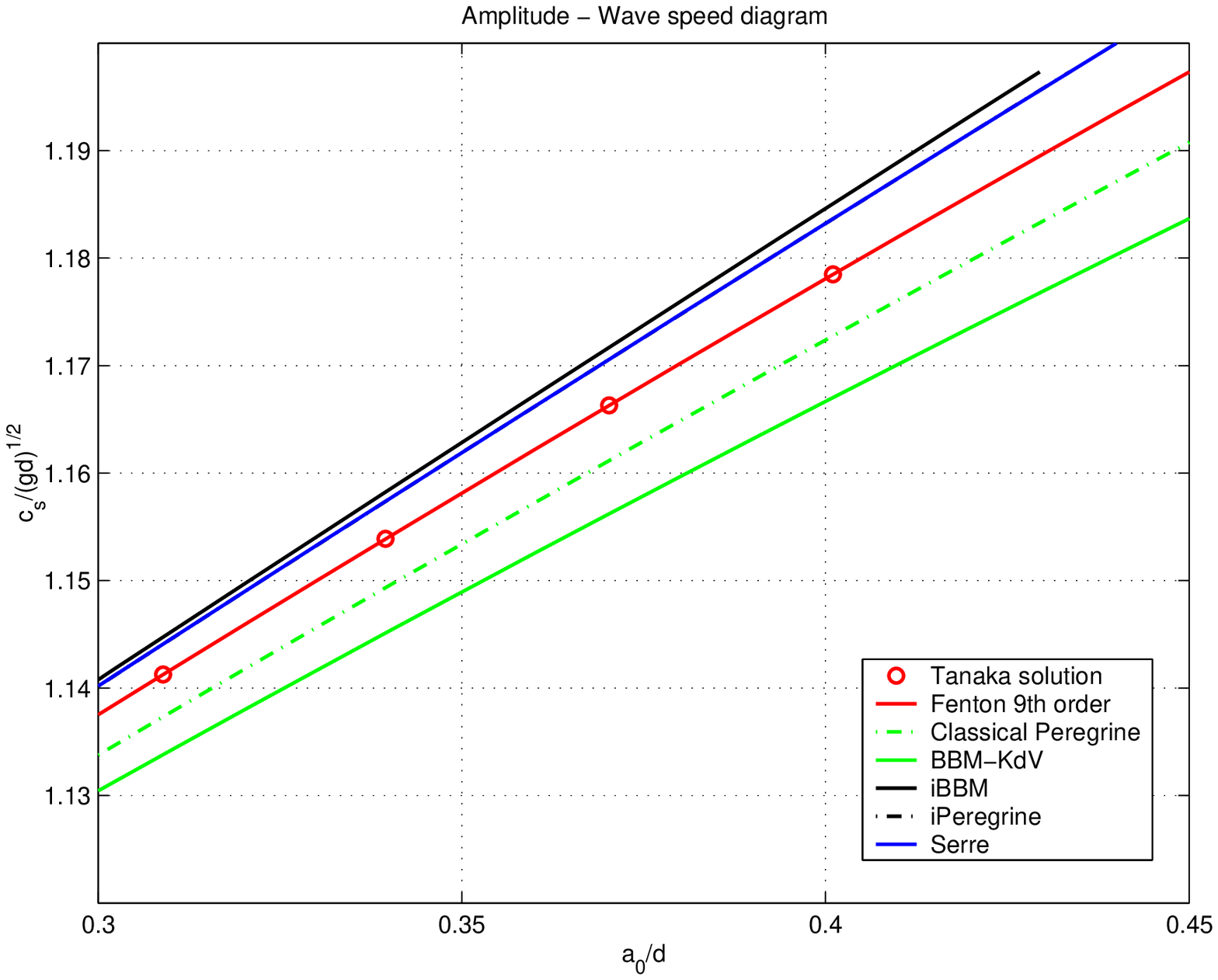}}
\caption{\em The curves corresponding to the iPeregrine system and the Serre system are superposed up to the graphical resolution. The Tanaka solution is represented with red circles, while the Fenton solution is depicted with the red solid line.}
\label{fig:swsa}
\end{figure}

Figure~\ref{fig:swsa}(a) shows an amplitude-wave speed diagram for the models considered in this paper. First, an approximate relation between the amplitude and speed for the solitary waves of the full Euler system is computed, by using the Tanaka's algorithm, \cite{Tanaka1986}, and Fenton's 9th order, \cite{Fenton1972, Longuet-Higgins1974}. They virtually give the same results and these are compared with the relation obtained by each of the models. We observe that the Serre equations (and, consequently, the other systems, that have weaker nonlinearities) are known to provide a relatively good approximation to the solitary wave solutions of the full Euler equations in a range of amplitudes not greater than $0.5$, cf. e.g. \cite{Li2004, Carter2011}. Therefore, our attention is focused on solitary waves with these amplitudes, as it is observed in Figure~\ref{fig:swsa}(a). Figure~\ref{fig:swsa}(b) shows a magnification for the largest amplitudes.

Corresponding to the comparison between the solitary waves of the models at hand, we note that the non Galilean invariant models, the \acf{BBM} equation and the \acf{cPer} system, tend to underestimate the solution speed for a given amplitude. On the other hand, the curves corresponding to their invariant counterparts along with the fully-nonlinear Serre equations lie above the reference solution. In particular, the results for the \acf{iBBM} equation are very close to those of Serre equations. A surprising fact is that the amplitude-speed relation given by the \acf{iPer} system is superposed with that of the Serre equations up to the graphical resolution.

\subsubsection{Solitary wave shape}

\begin{figure}%
\centering
\subfigure[$\frac{c_s}{\sqrt{gd}} = 1.05084689$  ($a/d \approx 0.1$)]%
{\includegraphics[width=0.49\textwidth]{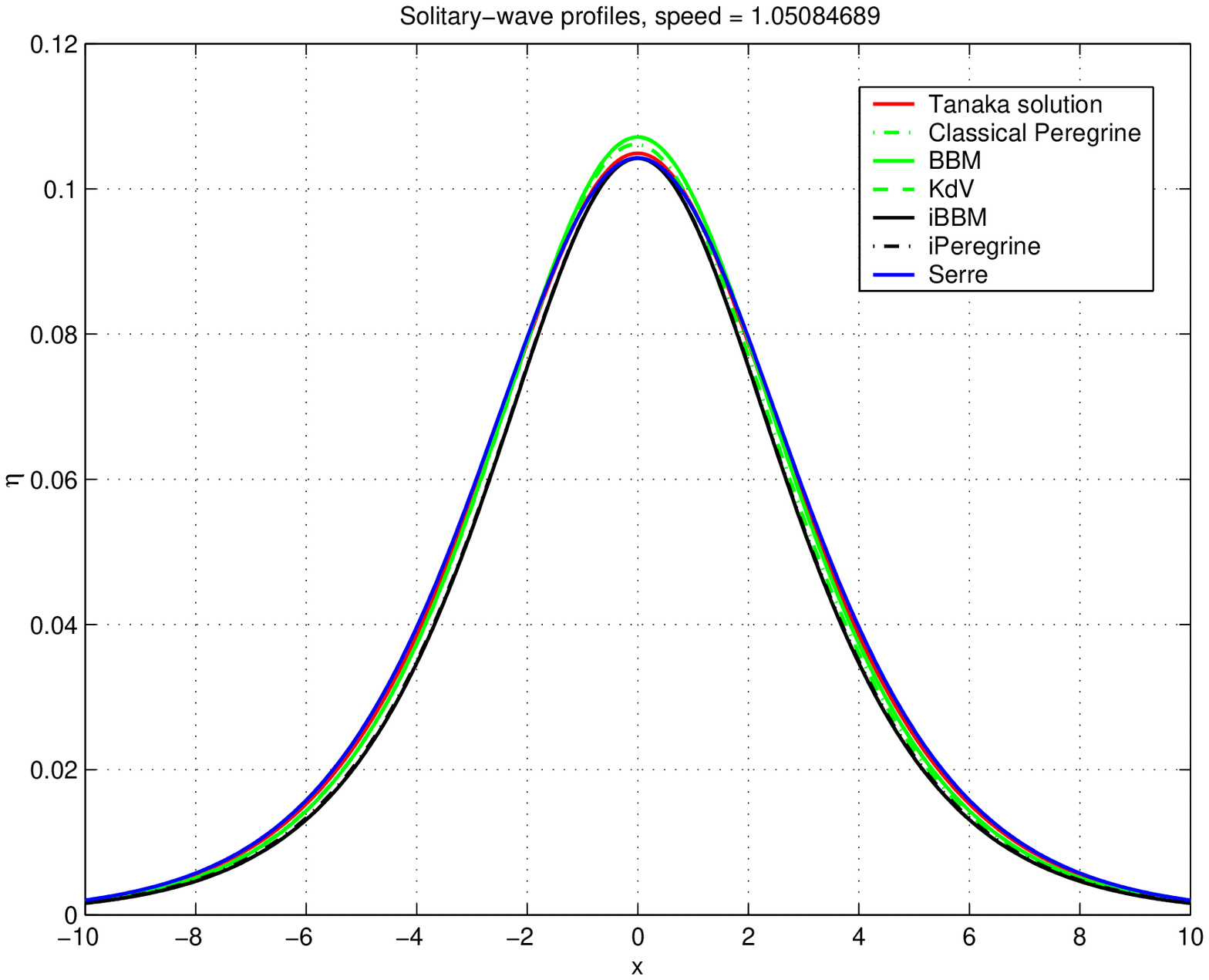}}
\subfigure[Magnification of (a)]%
{\includegraphics[width=0.49\textwidth]{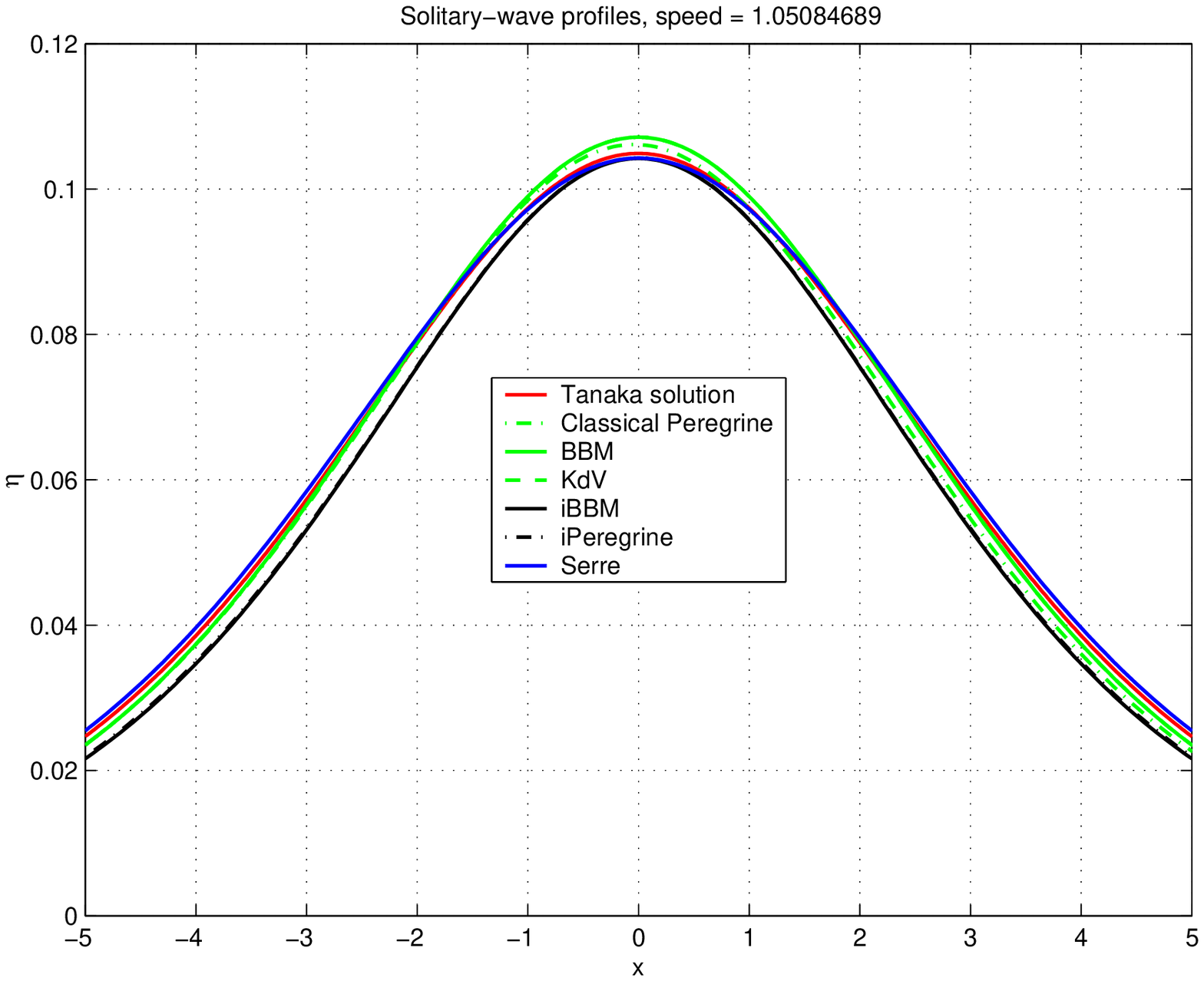}}
\subfigure[$\frac{c_s}{\sqrt{gd}} = 1.10269248$  ($a/d \approx 0.22$)]%
{\includegraphics[width=0.49\textwidth]{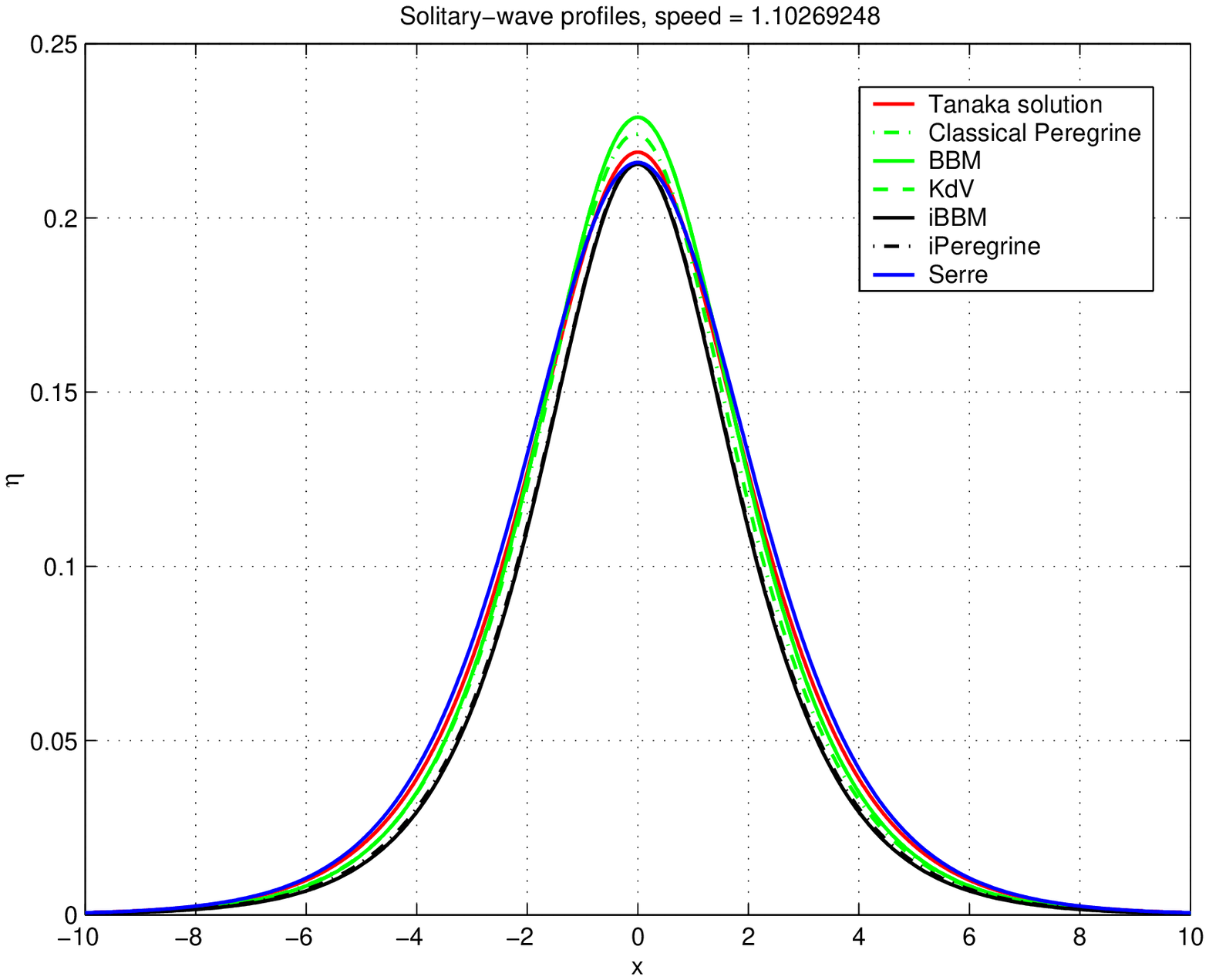}}
\subfigure[Magnification of (c)]%
{\includegraphics[width=0.49\textwidth]{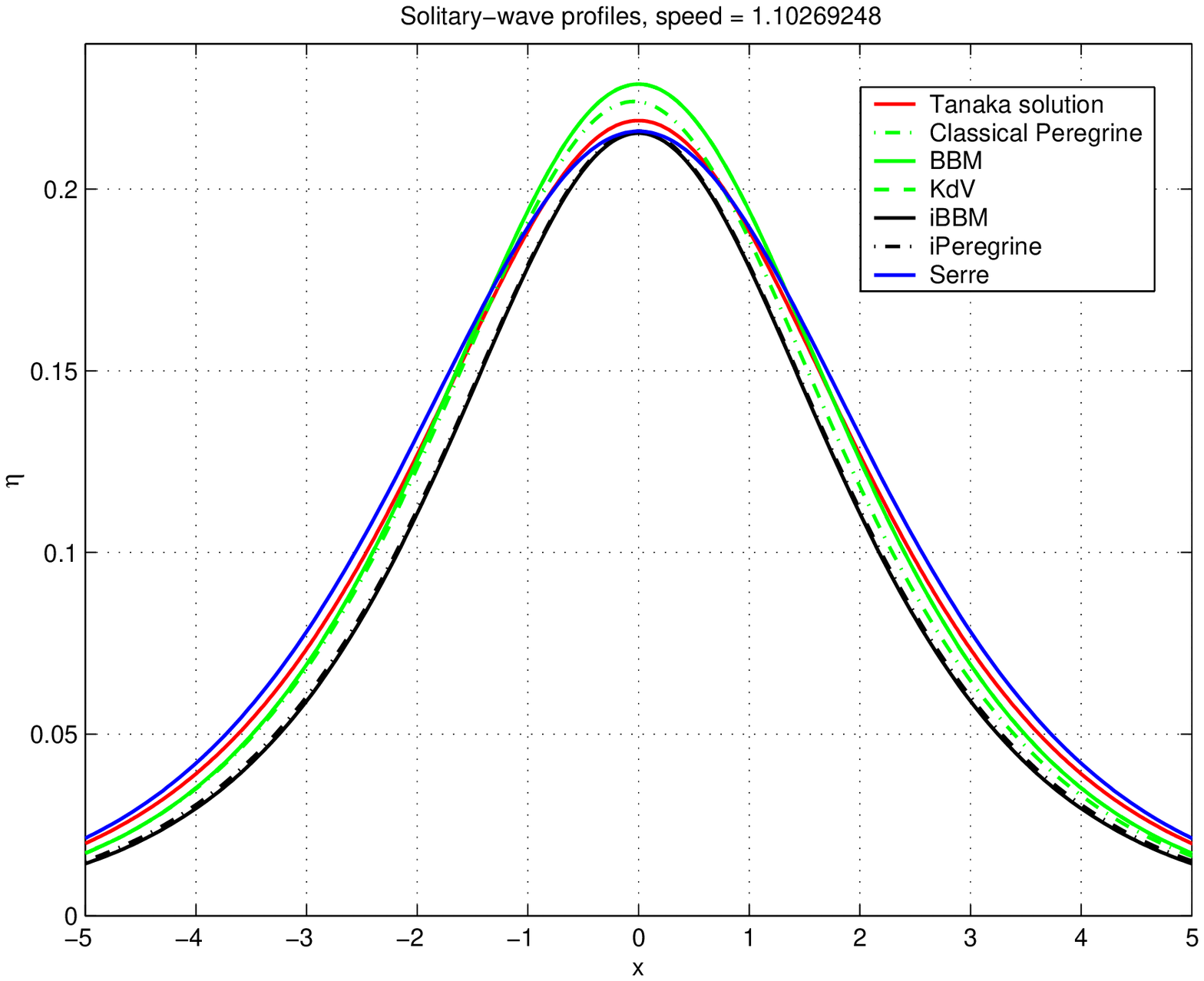}}
\subfigure[$\frac{c_s}{\sqrt{gd}} = 1.1784972$  ($a/d \approx 0.65$)]%
{\includegraphics[width=0.49\textwidth]{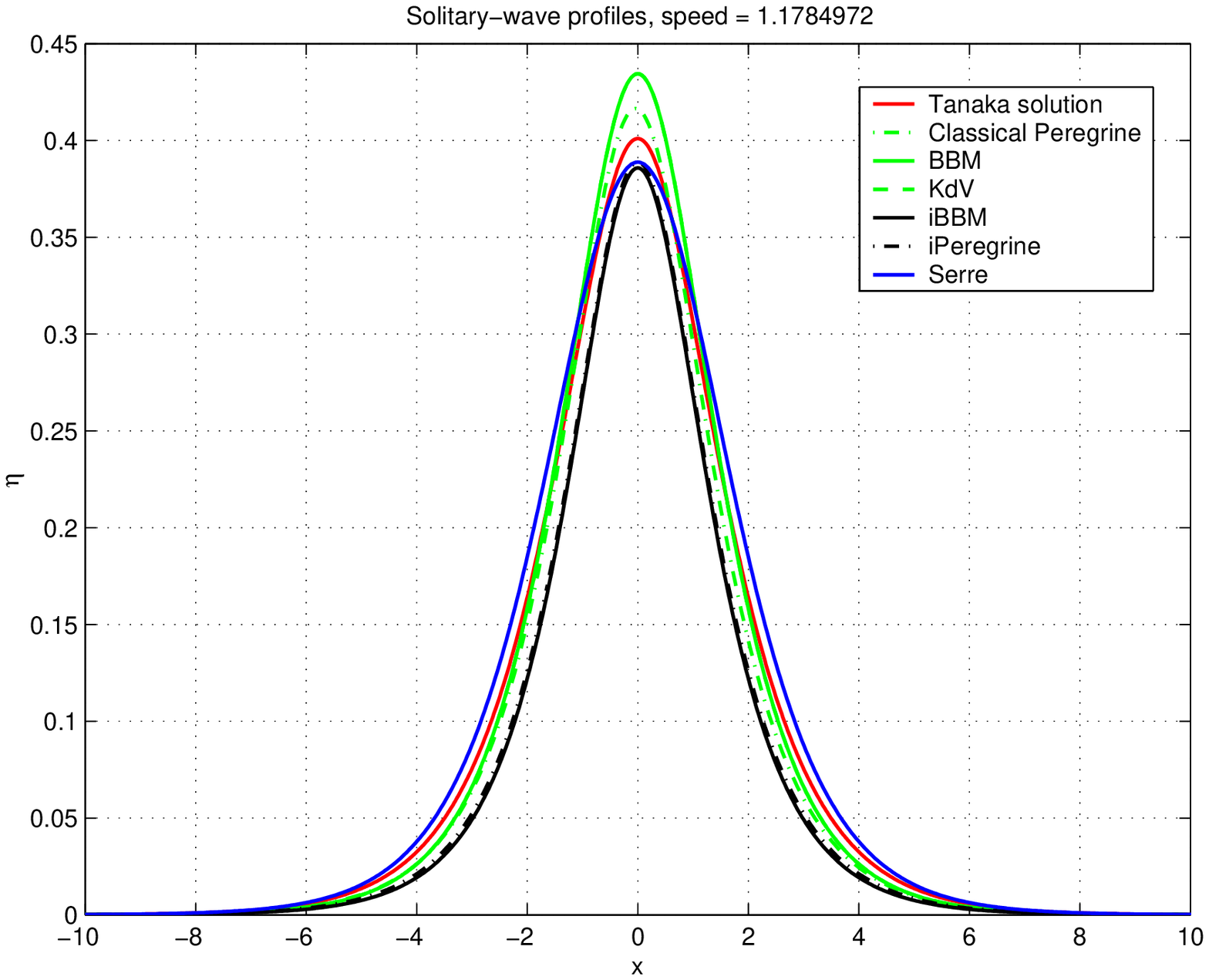}}
\subfigure[Magnification of (e)]%
{\includegraphics[width=0.49\textwidth]{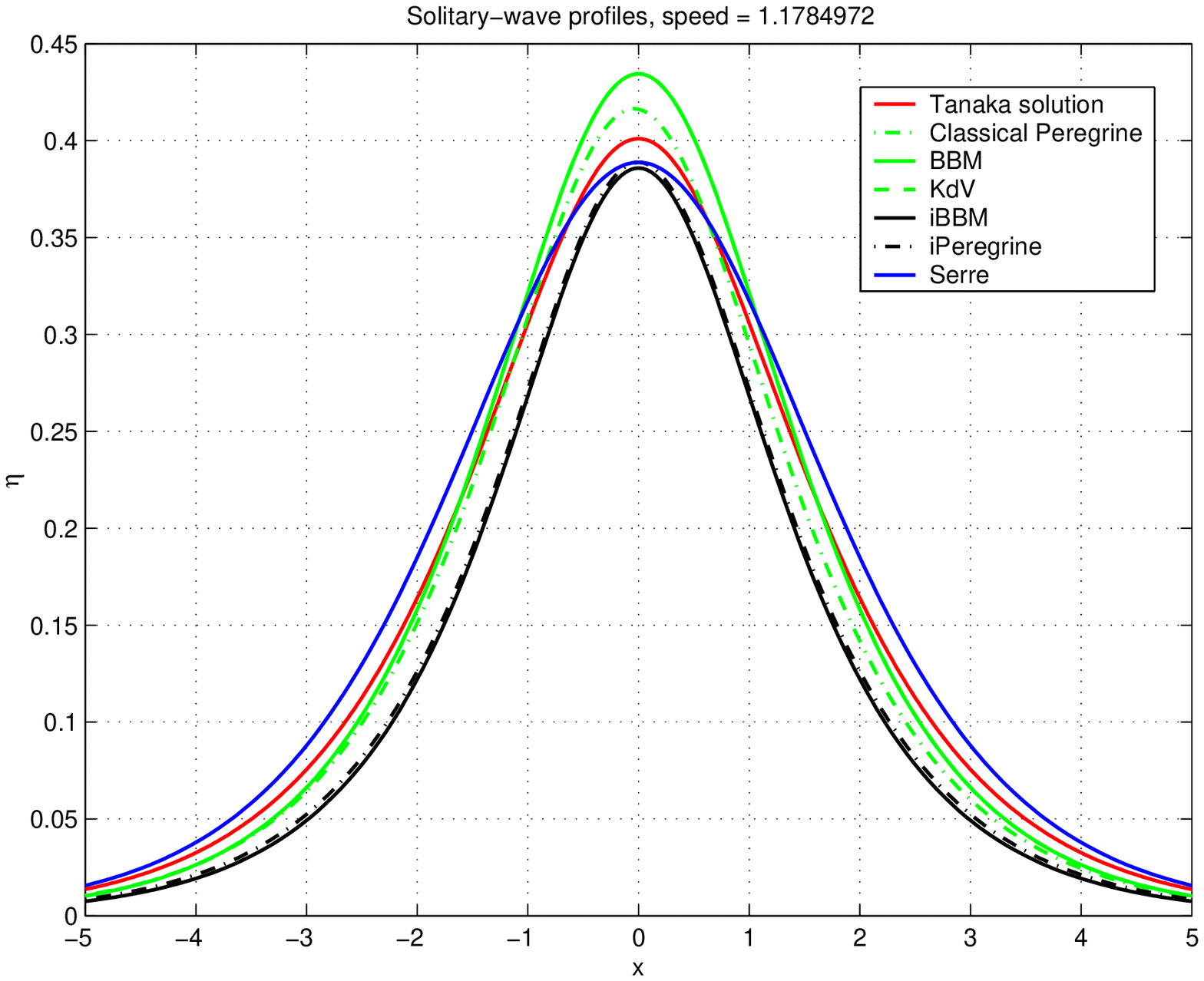}}
\caption{\em Solitary waves of different speeds $c_s$.}
\label{fig:sw}
\end{figure}

A second comparison between the shape of the computed solitary waves, is presented in Figures~\ref{fig:sw}(a)-\ref{fig:sw}(f). They illustrate the cases of small ($a/d \approx 0.1$, see Figure~\ref{fig:sw}(a) and a magnification on the wave crest in Figure \ref{fig:sw}(b)), moderate ($a/d \approx 0.22$, see Figures \ref{fig:sw}(c) and a magnification in Figure \ref{fig:sw}(d)) and large ($a/d \approx 0.4$, see Figures~\ref{fig:sw}(e) and \ref{fig:sw}(f)) solitary wave amplitudes of the models (within the range mentioned above). According to these results, it is observed that the \acf{iBBM} equation and the \acf{iPer} system approximate much better the amplitude of the reference solution (represented, in this case, by the Tanaka's solution) than the non-Galilean invariant counterparts, and they stay very close to the results of the Serre system near the crest.

As a measure of the level of approximation to solitary wave solutions of the Euler system, these and the previous results show the benefits of taking into account the invariantization process in the approximate models.

\section{Solitary waves dynamics}\label{sec:interact}

We complete the numerical experiments by studying the effects of the Galilean invariance property in the evolution of solitary waves. Specifically, we first compare, by numerical means, head-on collisions of two solitary waves of the invariant BBM and Peregrine equations with those of their corresponding not invariant models. Then, we will study the numerical solution of the four systems when a solitary wave profile of the Euler equations is used as initial condition.

It is known that the tails produced by the interaction of two solitary waves are sensitive to both the linear terms (characterizing the linear dispersion relation) and the nonlinearities. For example, two solitary waves of the KdV equation interact in an elastic way without producing dispersive tails at all, \cite{Zabusky1965}, while the collision of solitary waves of the BBM equation will produce dispersive tails and probably small-amplitude  nonlinear pulses as the main indication of an inelastic interaction, \cite{Bona1980, Bona1981}. In the new Galilean invariant models, the new nonlinear terms are of order $\eps\mu^2$ and their effects on the interaction will be studied here. The same code introduced in Section~\ref{sec:sols} is used for the numerical computations, as well as the Petviashvili method \eqref{eq:fpetv1}--\eqref{eq:fpetv2} to generate solitary wave profiles when necessary.

\subsection{Head-on collisions of solitary waves}

A first group of experiments concerns head-on collisions. The \acf{cPer} and the \acf{iPer} systems have been considered, by constructing, in both cases, two solitary waves on the interval $[-256, 256]$ with speeds  $c_{s,1} = 1.15$ and $c_{s,2} = 1.05$ (translated appropriately such that their maximum values are achieved on $x=-50$ and $x=50$ respectively) and travelling in opposite directions. These solitary waves are of small amplitude and their shapes are almost the same for both models. The code uses $N = 4096$ nodes, a spatial meshlength of $\Delta x=1.25\times 10^{-1}$ and $\Delta t=5\times 10^{-3}$ as the time step.

\begin{figure}%
\centering
\subfigure[Solution before the interaction ($t=25$)]{\includegraphics[width=0.49\textwidth]{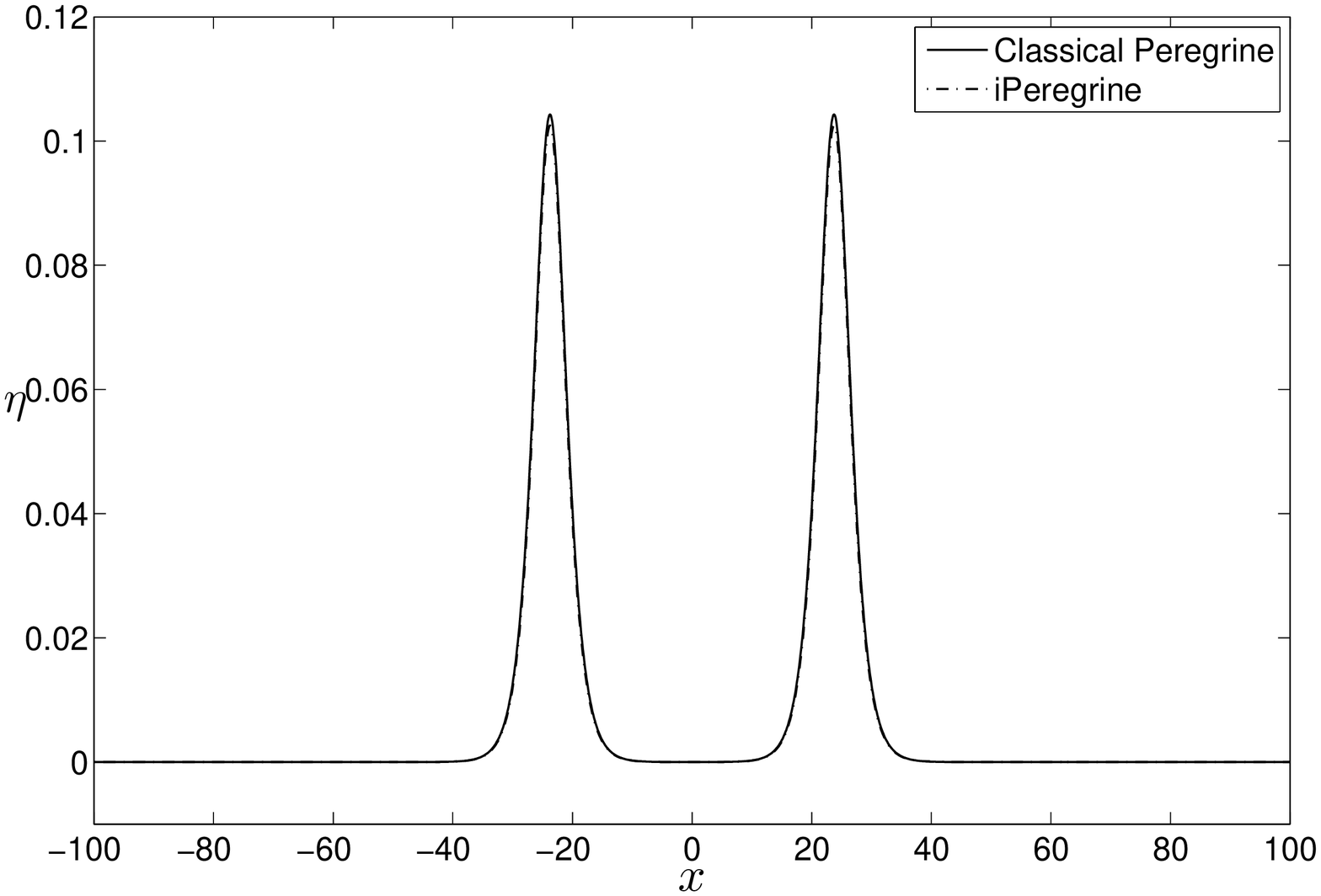}}
\subfigure[Solution during the interaction ($t=47.5$)]{\includegraphics[width=0.49\textwidth]{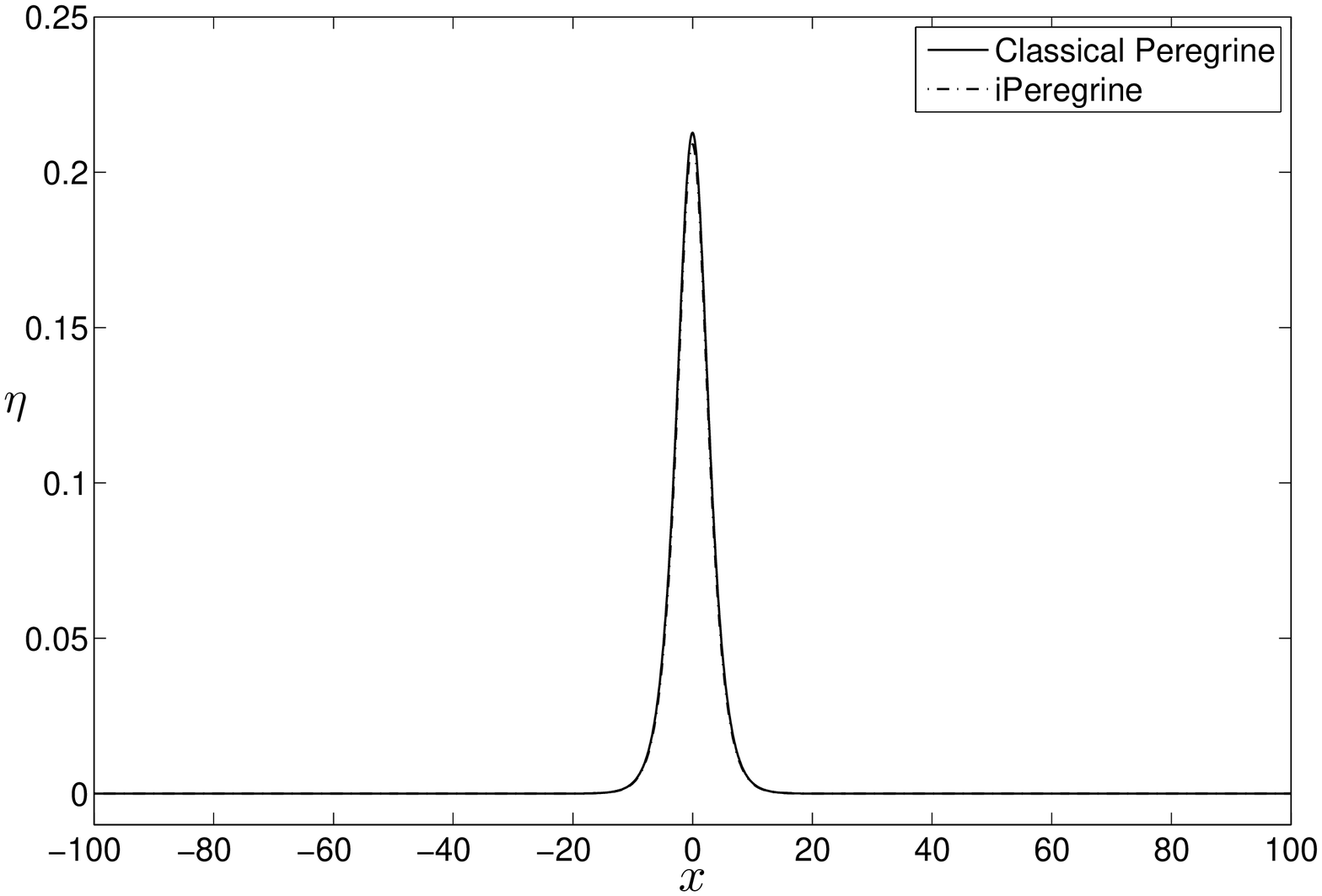}}
\subfigure[Solution after the interaction ($t=100$)]{\includegraphics[width=0.49\textwidth]{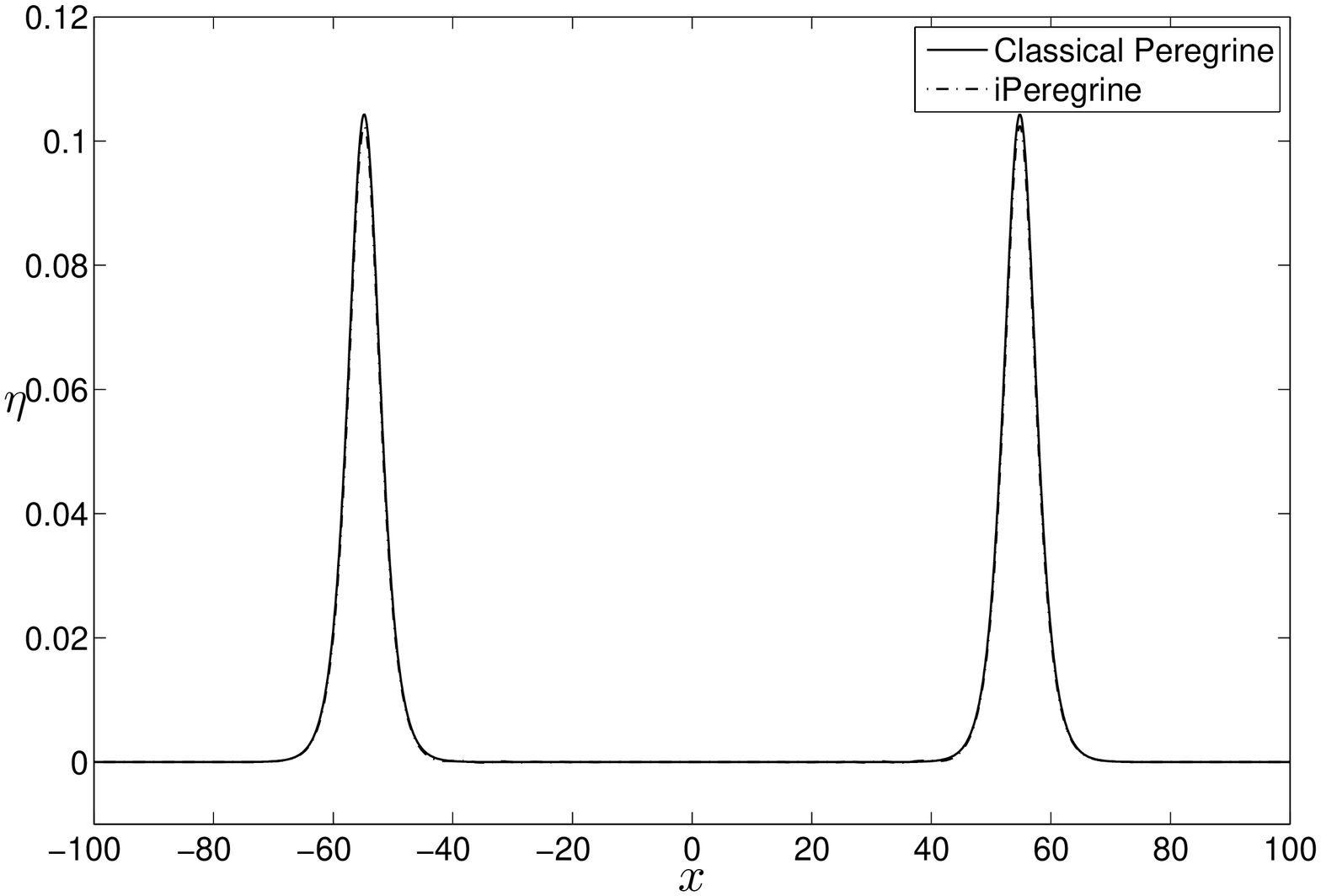}}
\subfigure[Magnification of the dispersive tail ($t=100$)]{\includegraphics[width=0.49\textwidth]{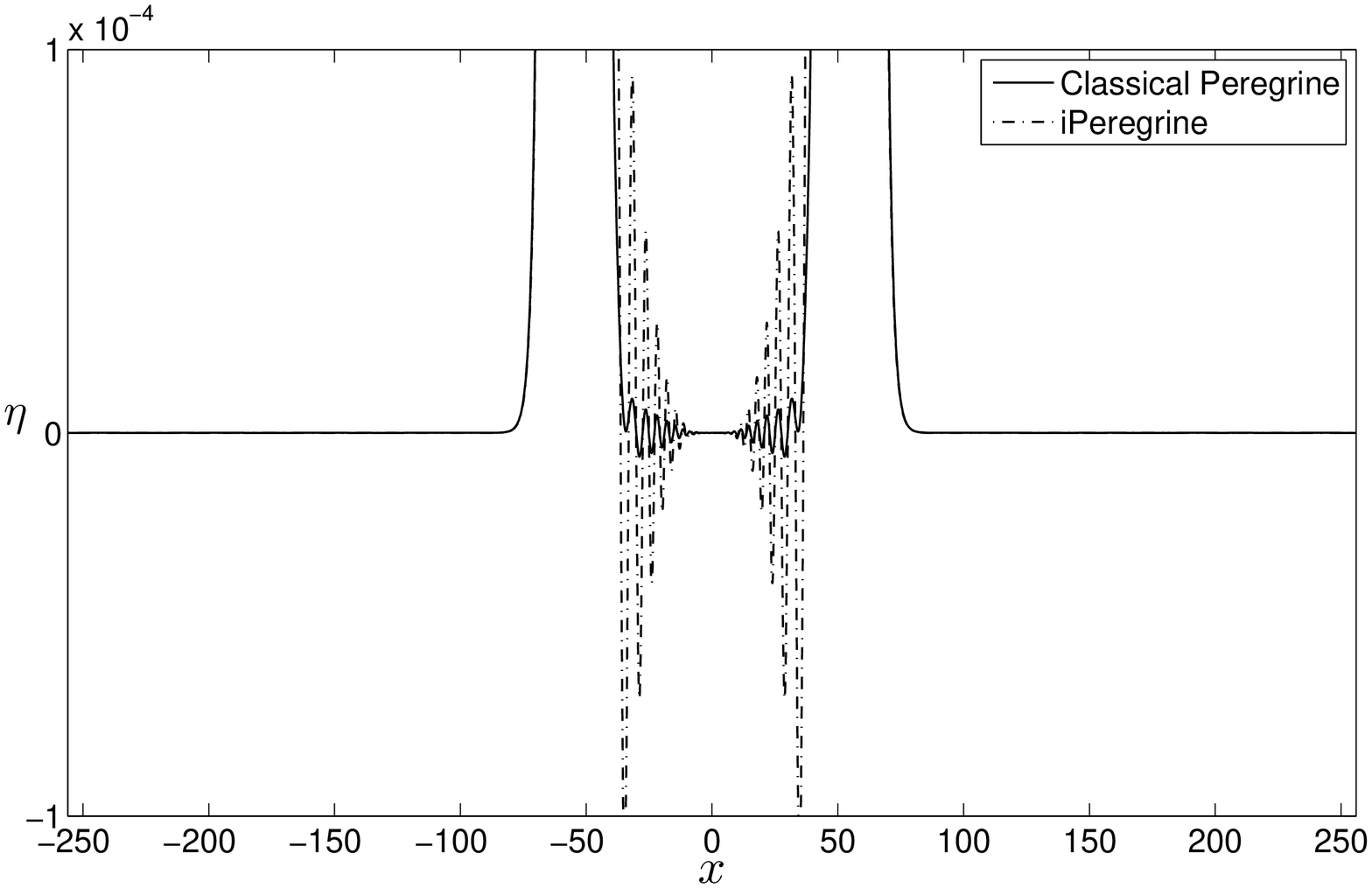}}
\subfigure[Solution after the interaction ($t=150$)]{\includegraphics[width=0.49\textwidth]{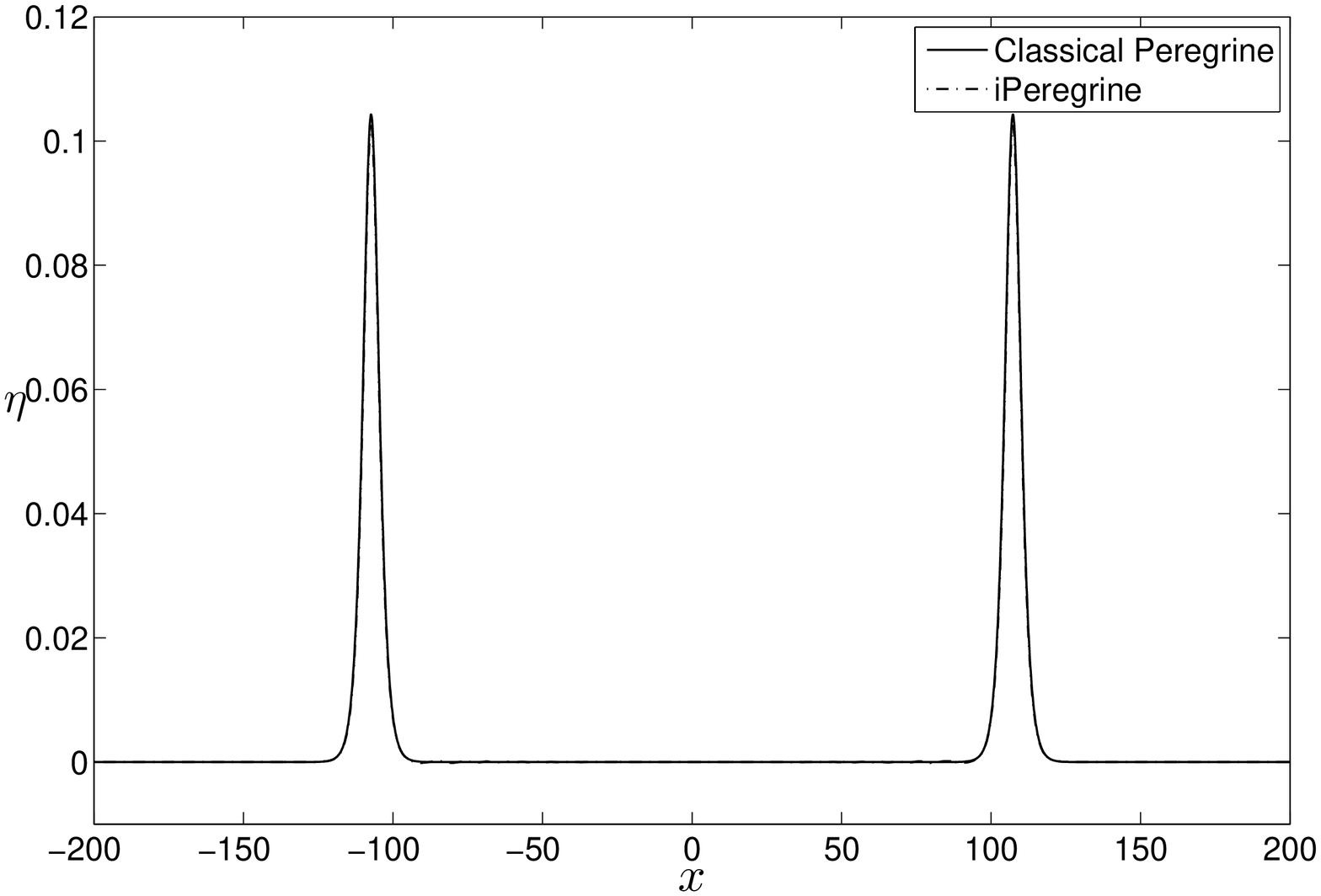}}
\subfigure[Magnification of the dispersive tail ($t=150$)]{\includegraphics[width=0.49\textwidth]{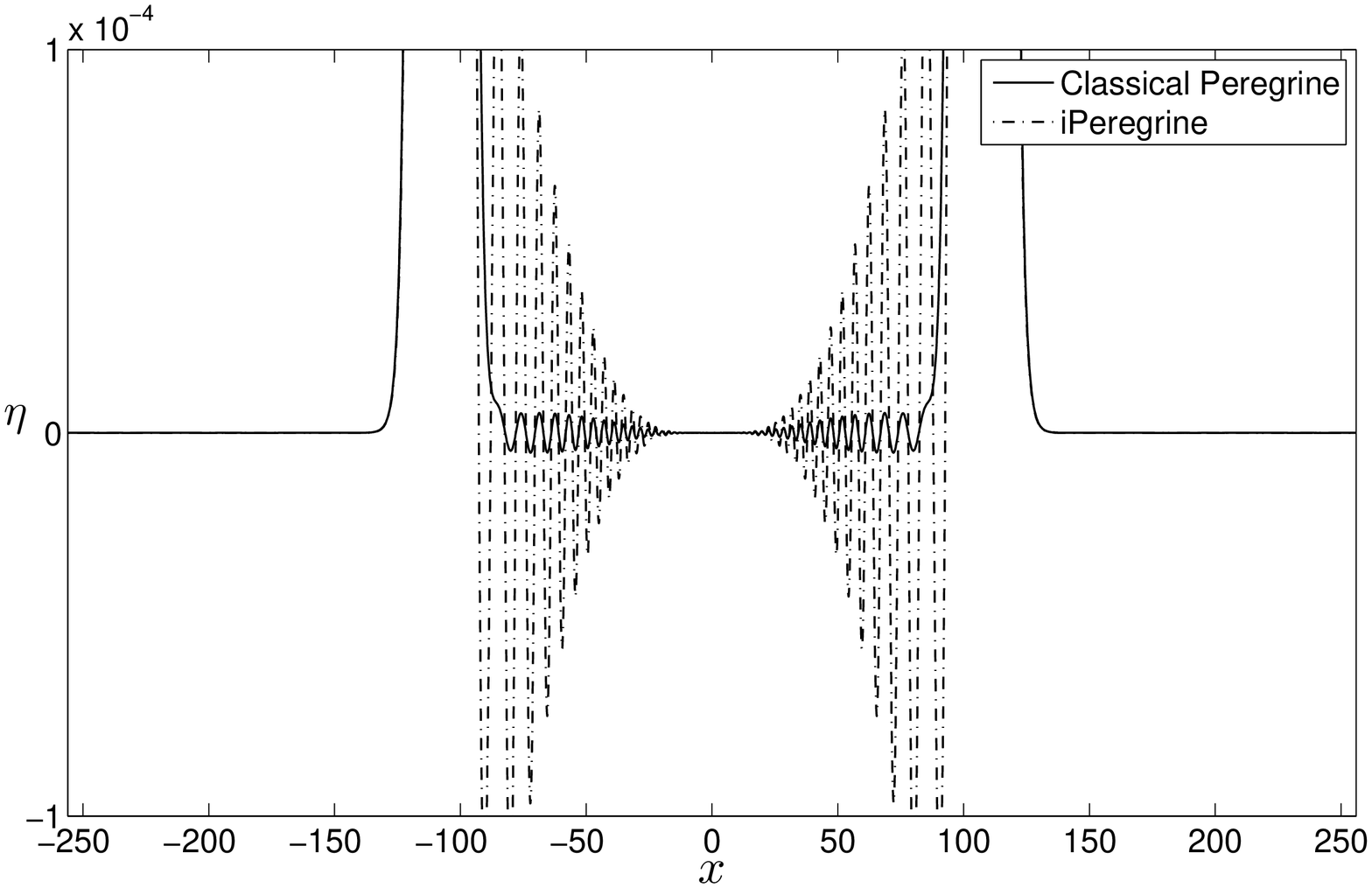}}
\caption{\em Head on collision of two solitary waves for the Peregrine and invariant Peregrine systems.}%
\label{Fig:HeadOn}%
\end{figure}

\begin{figure}%
\centering
\subfigure[]{\includegraphics[width=0.49\textwidth]{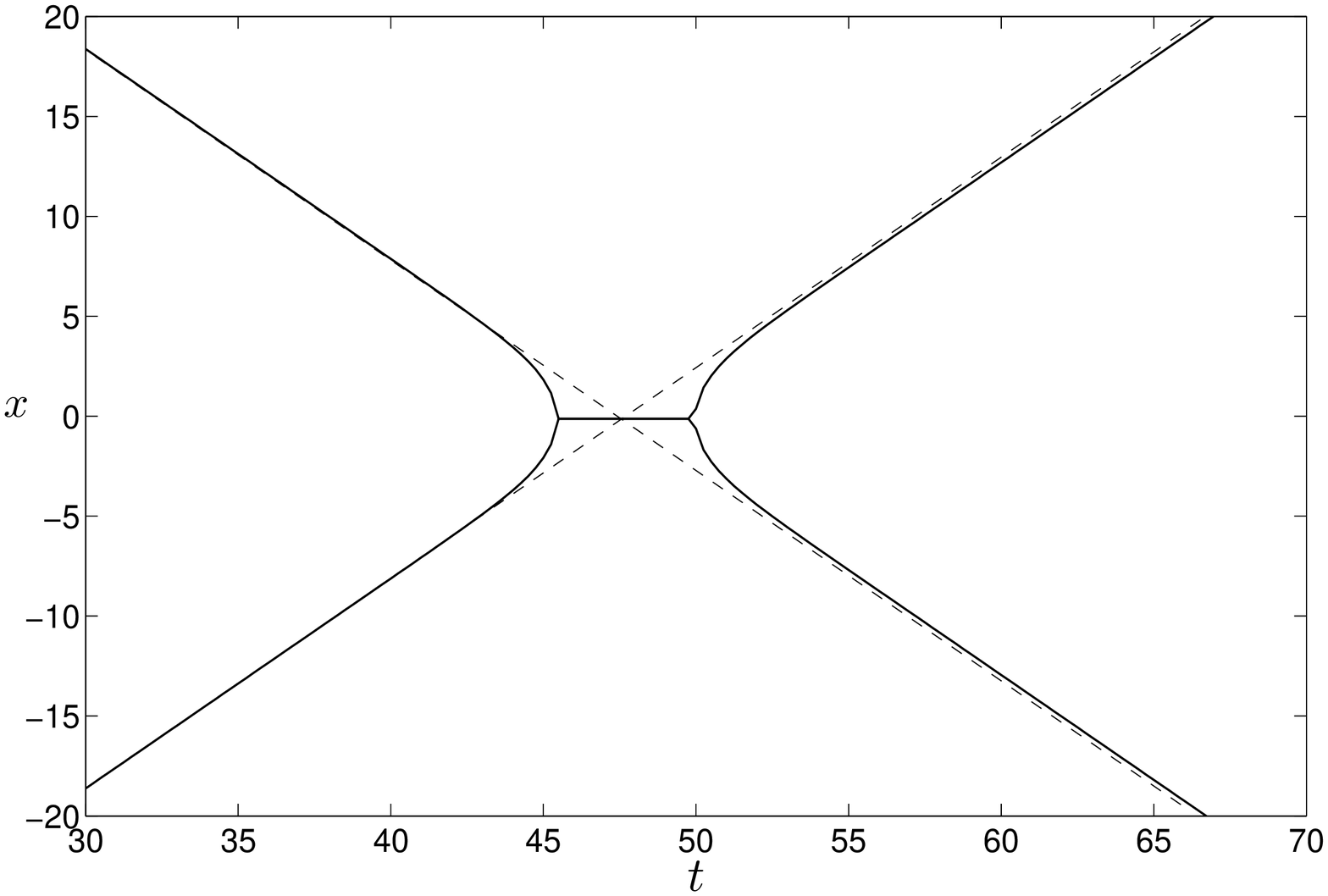}}
\subfigure[]{\includegraphics[width=0.49\textwidth]{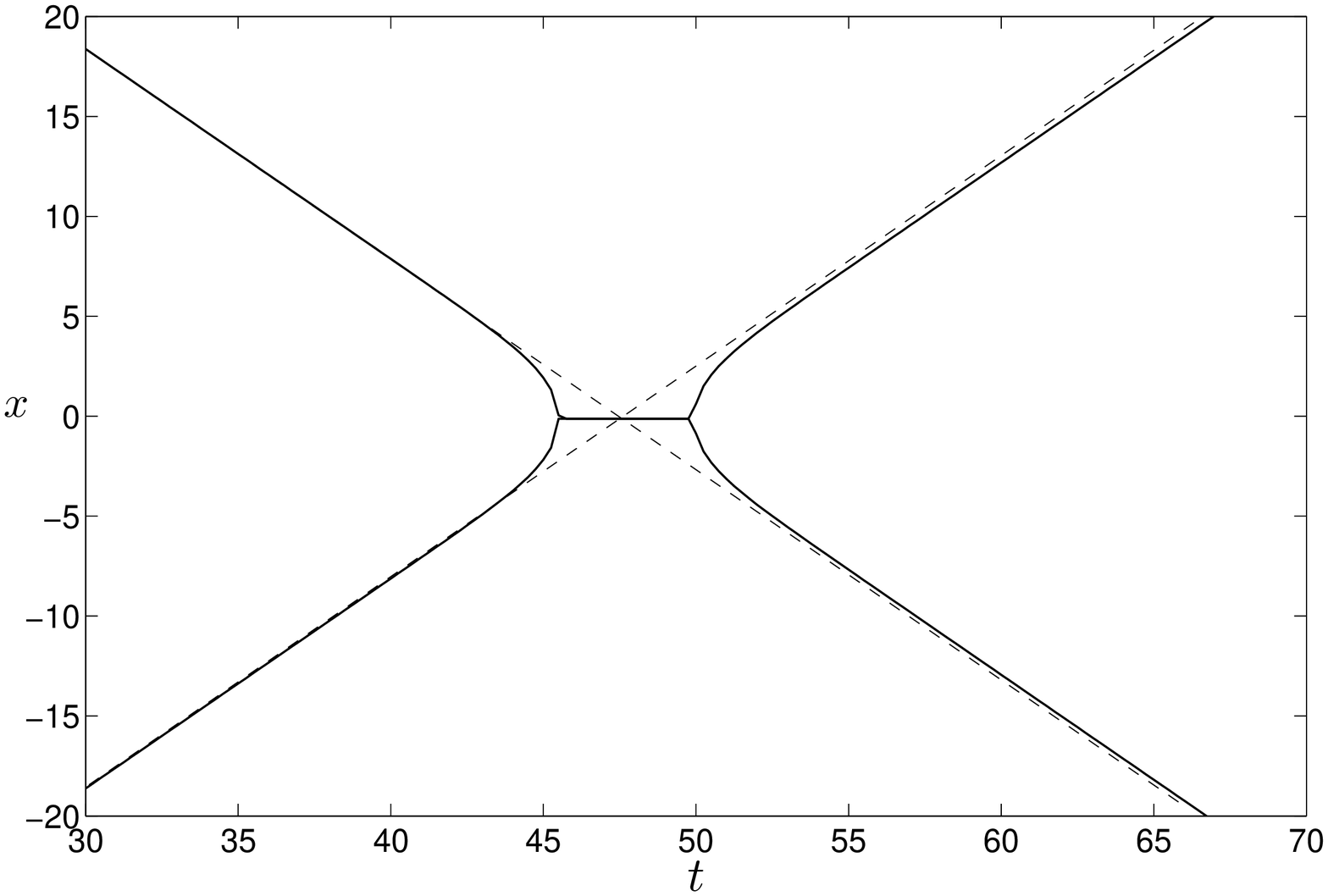}}
\caption{\em Phase diagram of the overtaking collision: (a) Peregrine system. (b) Invariant Peregrine system.}%
\label{Fig:dphase}%
\end{figure}

Figure~\ref{Fig:HeadOn} shows the $\eta-$ profile of the head-on collision for both models and at several times. A tail behind each solitary pulse after the collision is observed. This is larger in the case of the invariant Peregrine system (of the order of $10^{-4}$) than in the case of the classical Peregrine system (approx. $10^{-4}$), see Figure~\ref{Fig:HeadOn}(d) and (f). During the collisions, a similar phase shift also takes place, see Figure~\ref{Fig:dphase}.

\begin{figure}
  \centering
  \includegraphics[width=0.75\textwidth]{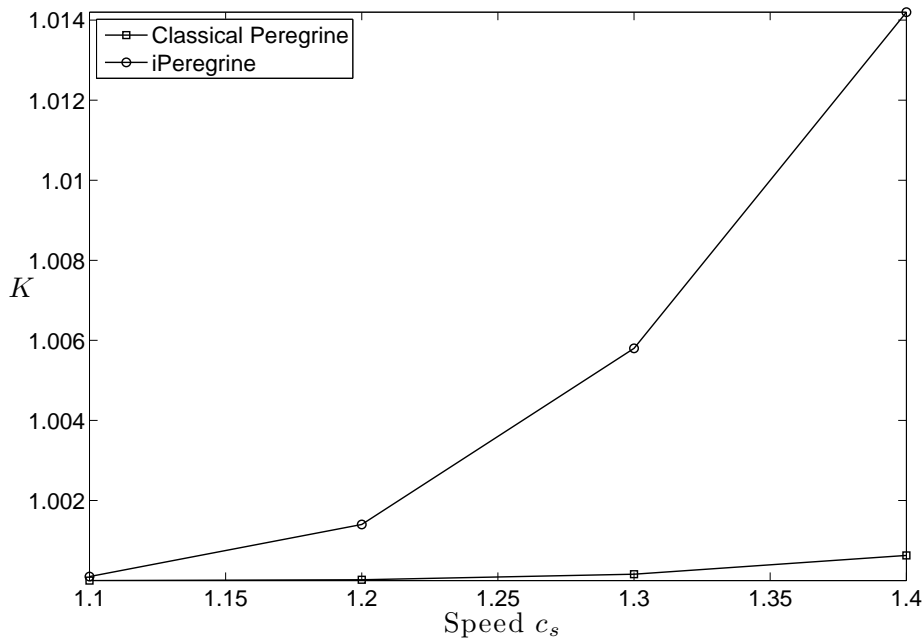}
  \caption{\em Symmetric head-on collision. Ratio $K$ of amplitudes before and after the collision, for Peregrine and invariant Peregrine systems.}
  \label{fig:inela}
\end{figure}

\begin{table}
\centering \subtable[Peregrine
system]{\begin{tabular}{|c|c|c|}\hline
$c_{s}$&$A_{init}$&$A_{after}$\\\hline
$1.1$&$0.2177418$&$0.2177417$\\
$1.2$&$0.4757297$&$0.4757202$\\
$1.3$&$0.7822906$&$0.7821674$\\
$1.4$&$1.1476304$&$1.1469096$
\\\hline
\end{tabular} }
\subtable[iPeregrine system]{
\begin{tabular}{|c|c|c|}\hline
$c_{s}$&$A_{init}$&$A_{after}$\\\hline
$1.1$&$0.21$&$0.209978$\\
$1.2$&$0.44$&$0.439365$\\
$1.3$&$0.69$&$0.686027$\\
$1.4$&$0.95$&$0.946593$
\\\hline
\end{tabular}
}
\caption{\em Symmetric head-on collision.}
\label{tab:tavela1}
\end{table}

A final comparison is established in terms of the degree of inelasticity of the interaction. This can be measured by using several parameters, \cite{Abdulloev1976, CourtenayLewis1979}. In each case, a symmetric head-on collision has been implemented; that is, two solitary waves with the same speed $c_{s}$ travelling in opposite directions. After the interaction, both solitary waves emerge with similar amplitudes $A_{after}$, but below the initial one $A_{init}$, as can be observed in Table~\ref{tab:tavela1}. Then a ratio $K$ of the amplitude of the waves after the collision to their amplitude before the collision has been computed. Figure~\ref{fig:inela} shows the behavior of this value, as a function of the speed parameter $c_{s}$ and for both systems. The results reveal a higher inelastic collision in the case of the invariant system, in accordance to what is observed in the full water wave model \cite{CGHHS}.

\subsection{Overtaking interactions of solitary waves}

\begin{figure}%
\centering
\subfigure[Solution before the interaction ($t=250$)]%
{\includegraphics[width=0.49\textwidth]{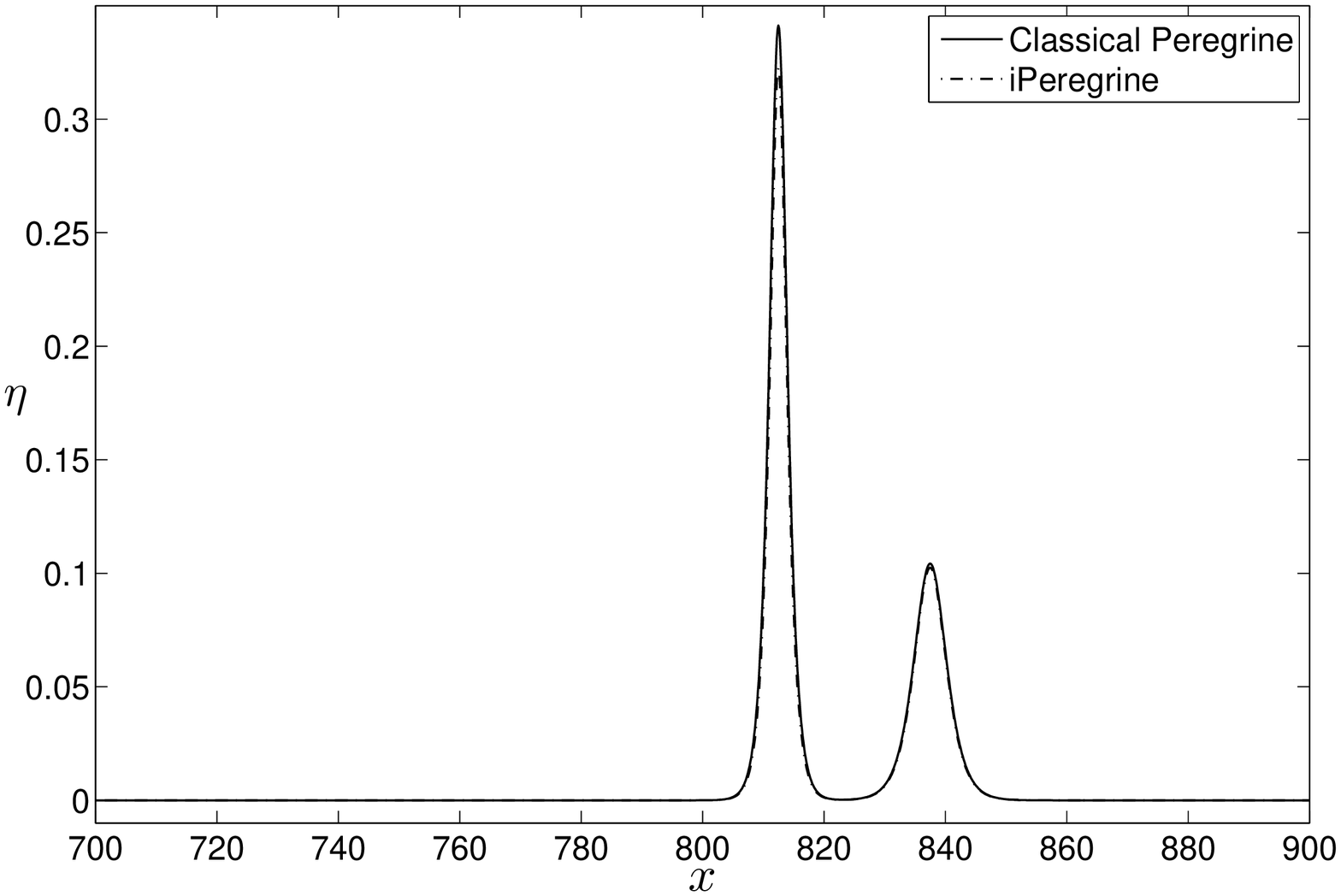}}
\subfigure[Solution during the interaction ($t=750$)]%
{\includegraphics[width=0.49\textwidth]{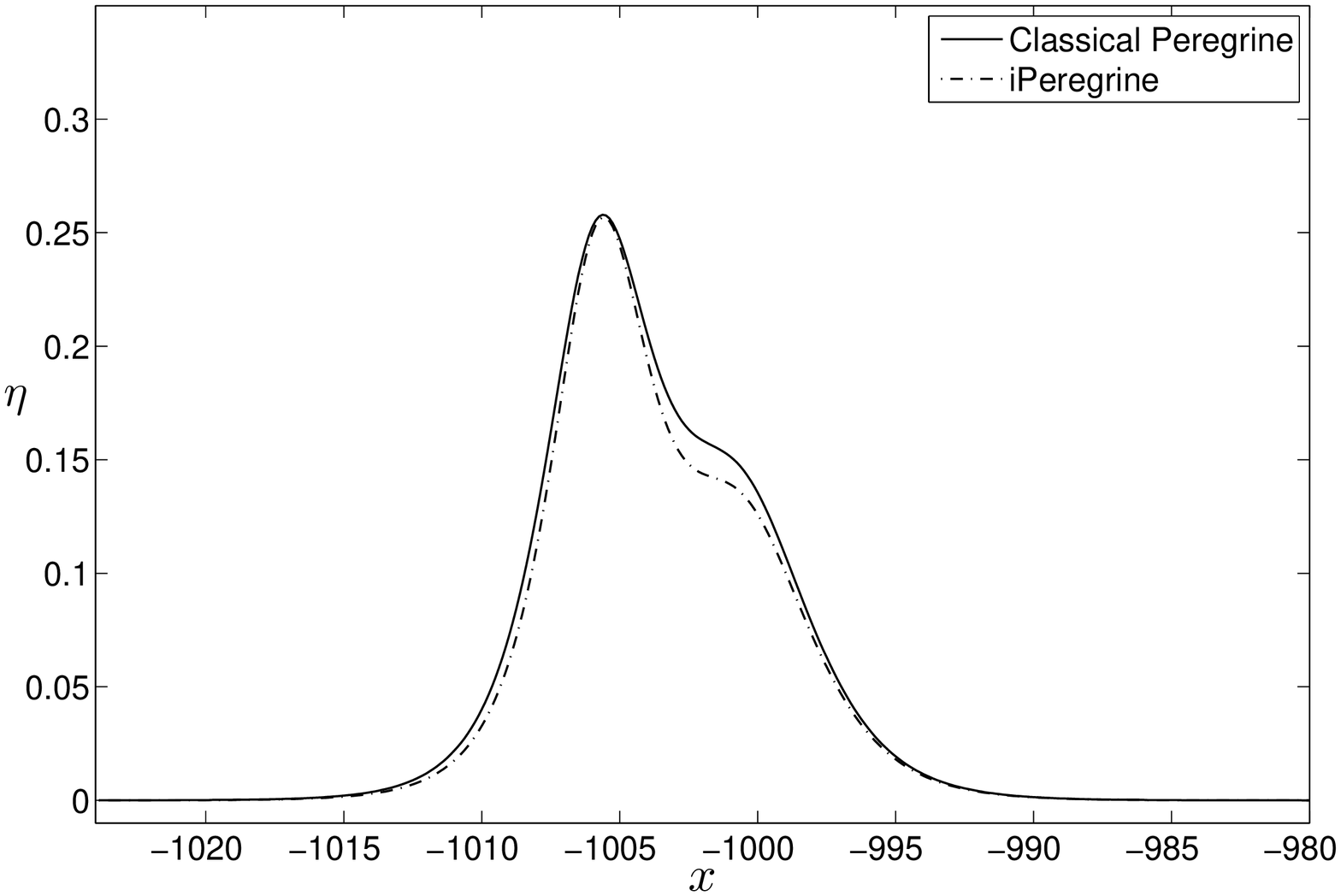}}
\subfigure[Solution after the interaction ($t=950$)]%
{\includegraphics[width=0.49\textwidth]{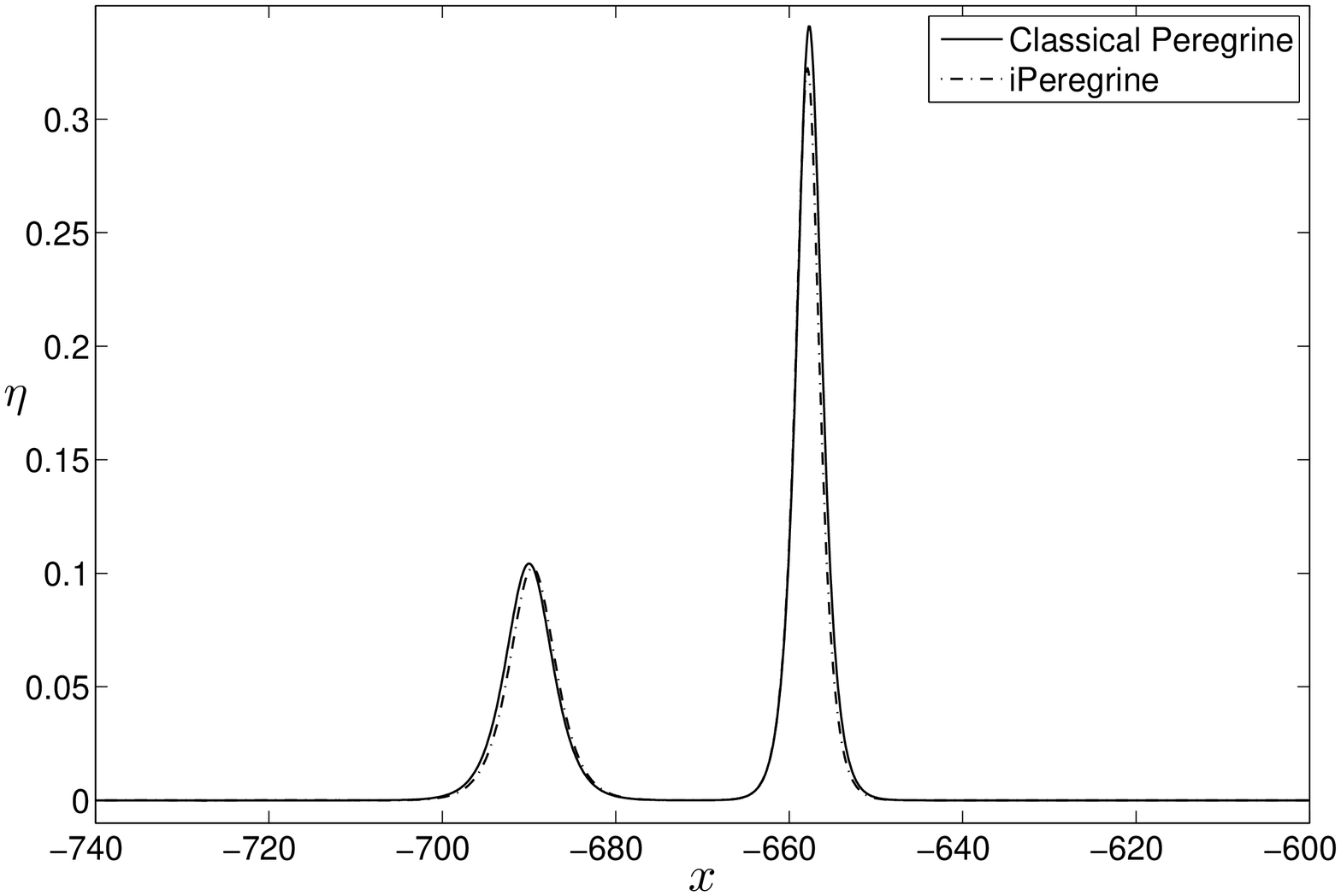}}
\subfigure[Solution after the interaction ($t=1750$)]%
{\includegraphics[width=0.49\textwidth]{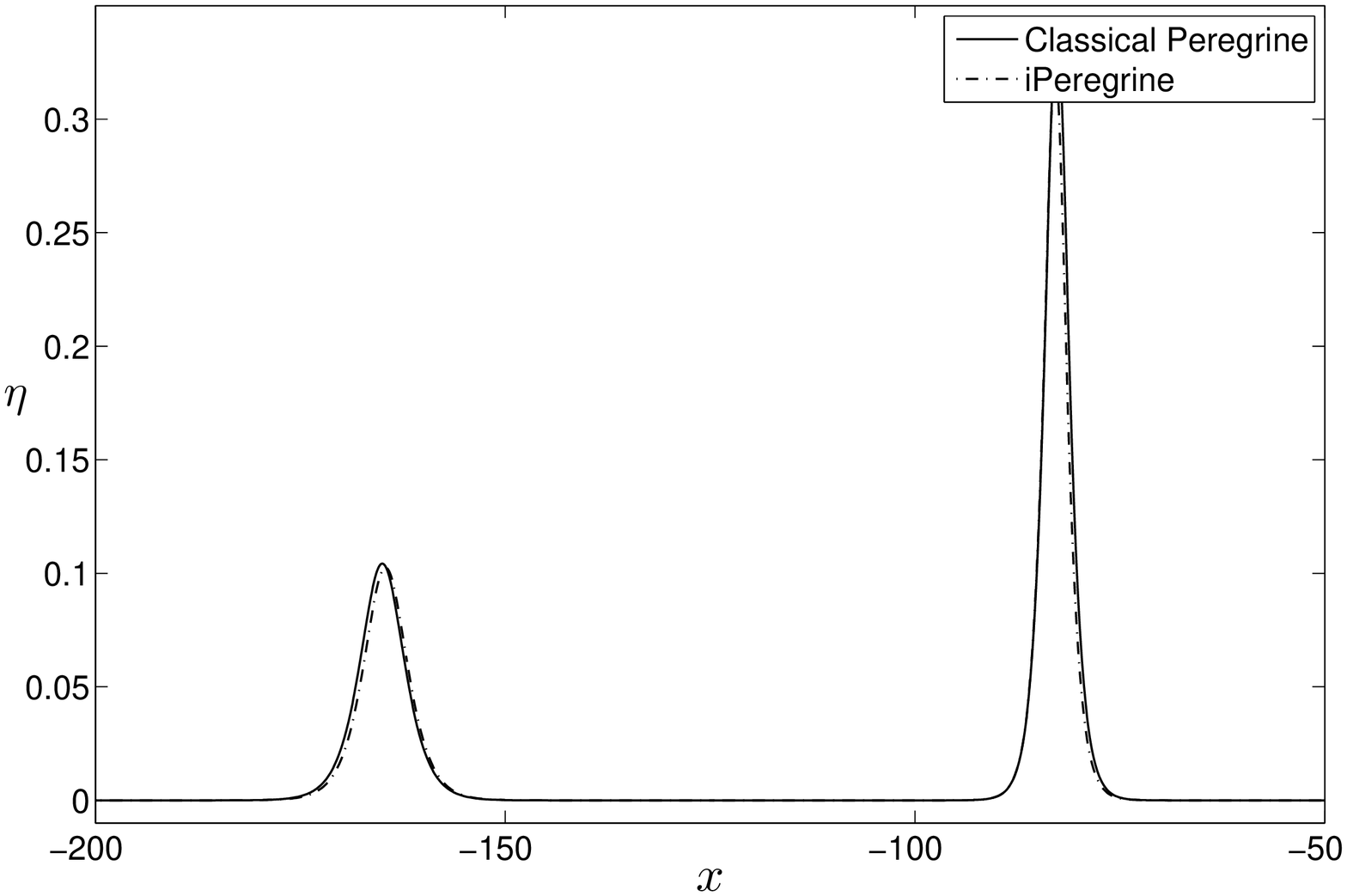}}
\subfigure[Magnification of the dispersive tail ($t=1750$)]%
{\includegraphics[width=0.49\textwidth]{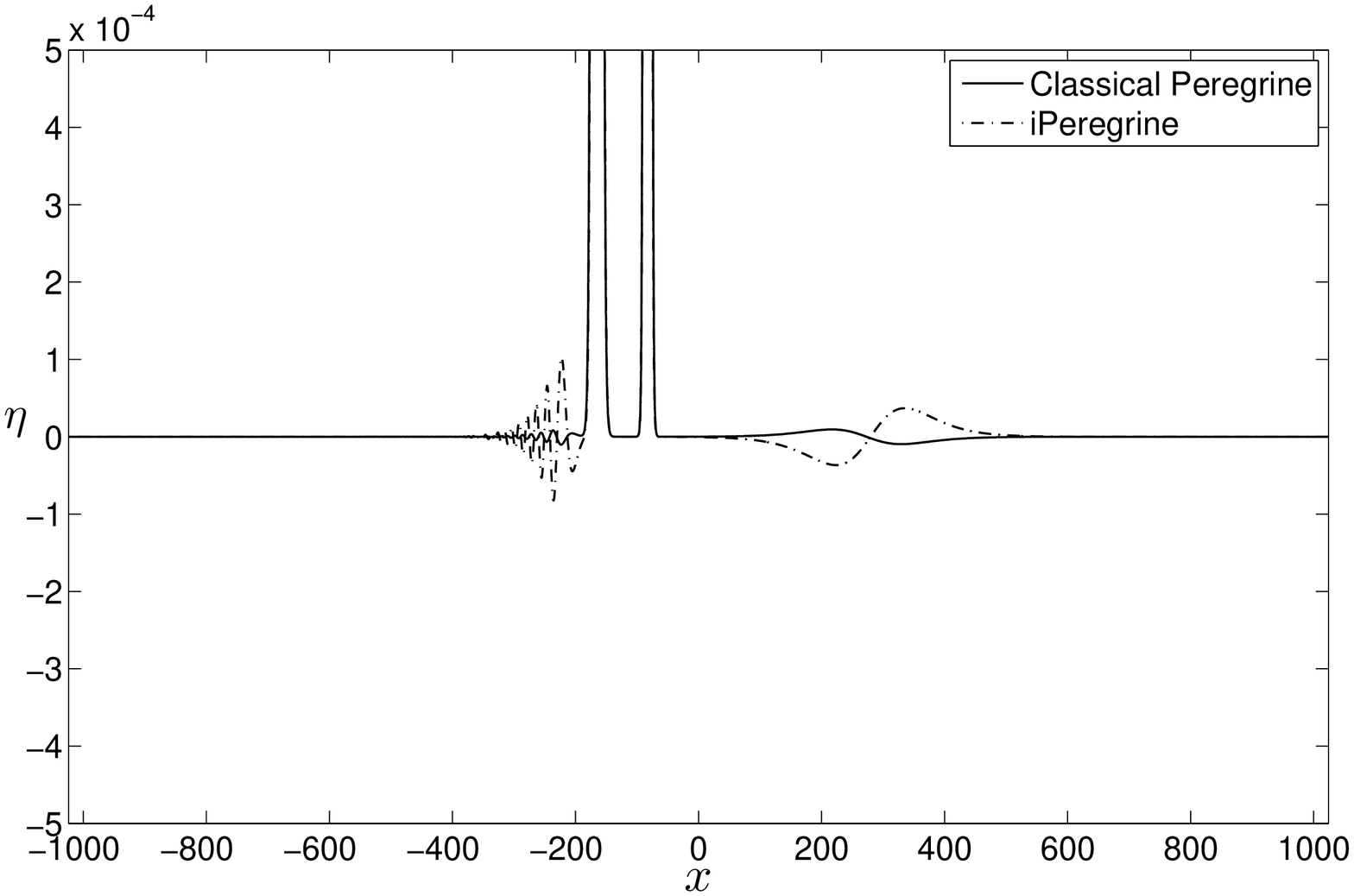}}
\caption{\em Overtaking collision of two solitary waves traveling to the right for \acf{cPer} and \acf{iPer} systems.}%
\label{Fig:over}%
\end{figure}

In order to study the overtaking collision of solitary waves, we first consider the classical and the invariant Peregrine systems and compute two solitary waves on the interval $[-1024, 1024]$ with speeds $c_{s,1} = 1.15$ and $c_{s,2} = 1.05$ (translated appropriately such as their maximum values are achieved at $x=-50$ and $x=50$ respectively) travelling in the same direction, with $N = 16384$ and timestep of $\Delta t= 5\times 10^{-3}$. In Figure~\ref{Fig:over} we observe that the basic characteristics of the interaction are similar for both systems, (cf. also \cite{ADM2}). Specifically, the interaction is again inelastic; after the collision a tail, of apparent dispersive nature, behind the smallest wave (moving to the right) and a small N-shape wavelet (moving to the left) are generated. Moreover, a small phase shift and a change in shape can be observed in the solitary pulses (Figure~\ref{Fig:dphase}). The wavelet generated in the case of the classical Peregrine system has an inverse N-shape while the invariant version has an N-shape.

\begin{figure}%
\centering
\subfigure[Solution before the interaction ($t=250$)]{\includegraphics[width=0.49\textwidth]{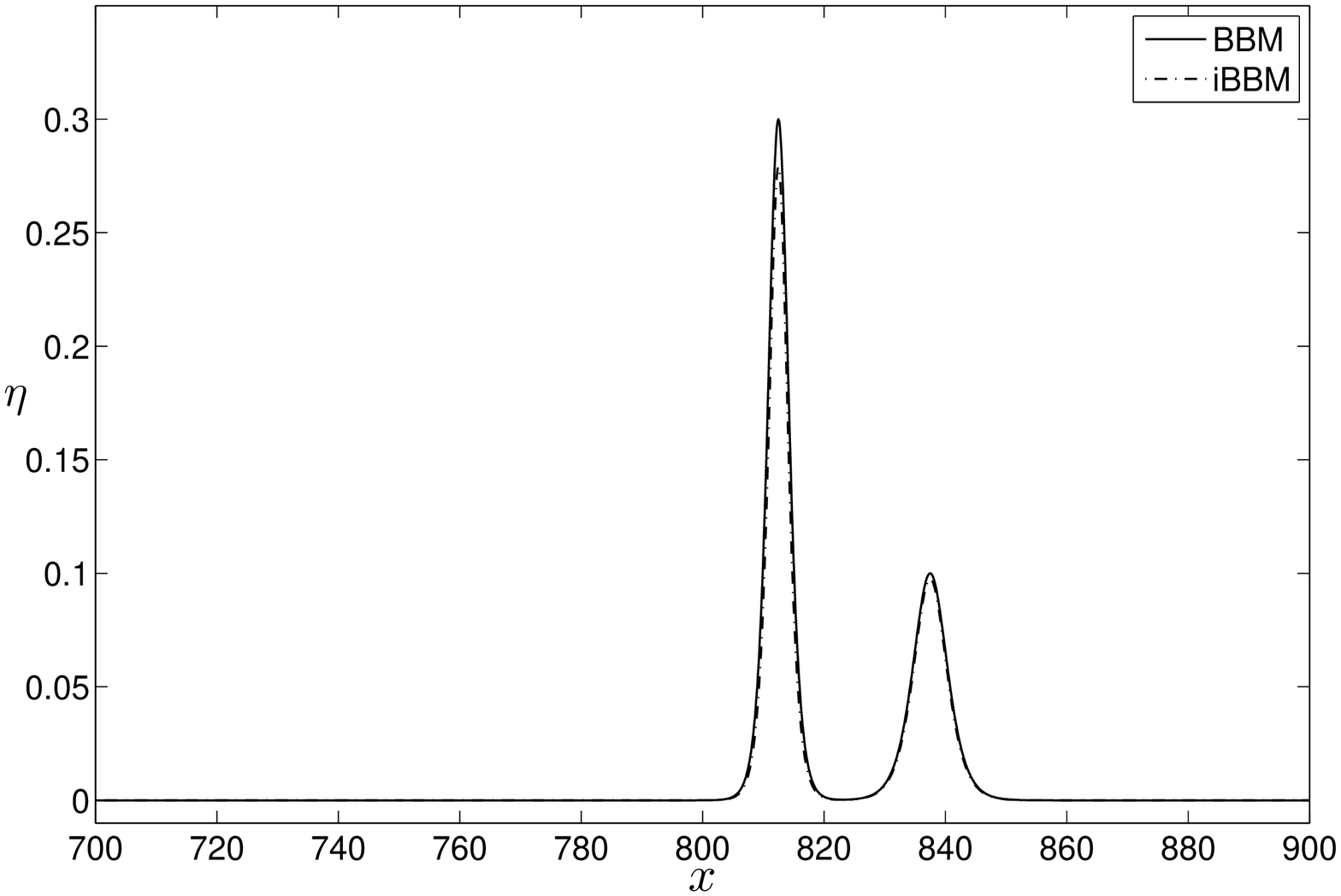}}
\subfigure[Solution during the interaction ($t=750$)]{\includegraphics[width=0.49\textwidth]{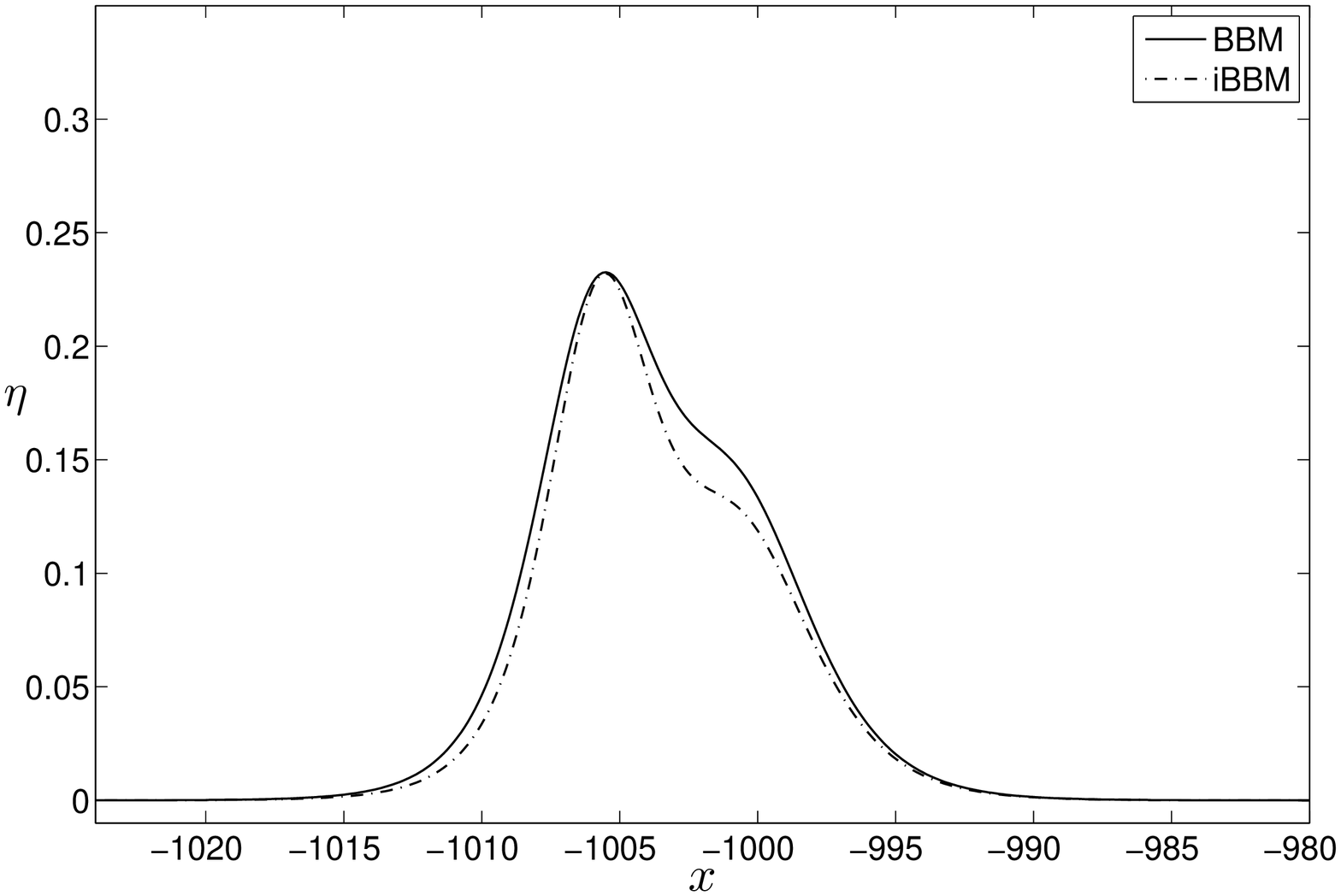}}
\subfigure[Solution after the interaction ($t=950$)]{\includegraphics[width=0.49\textwidth]{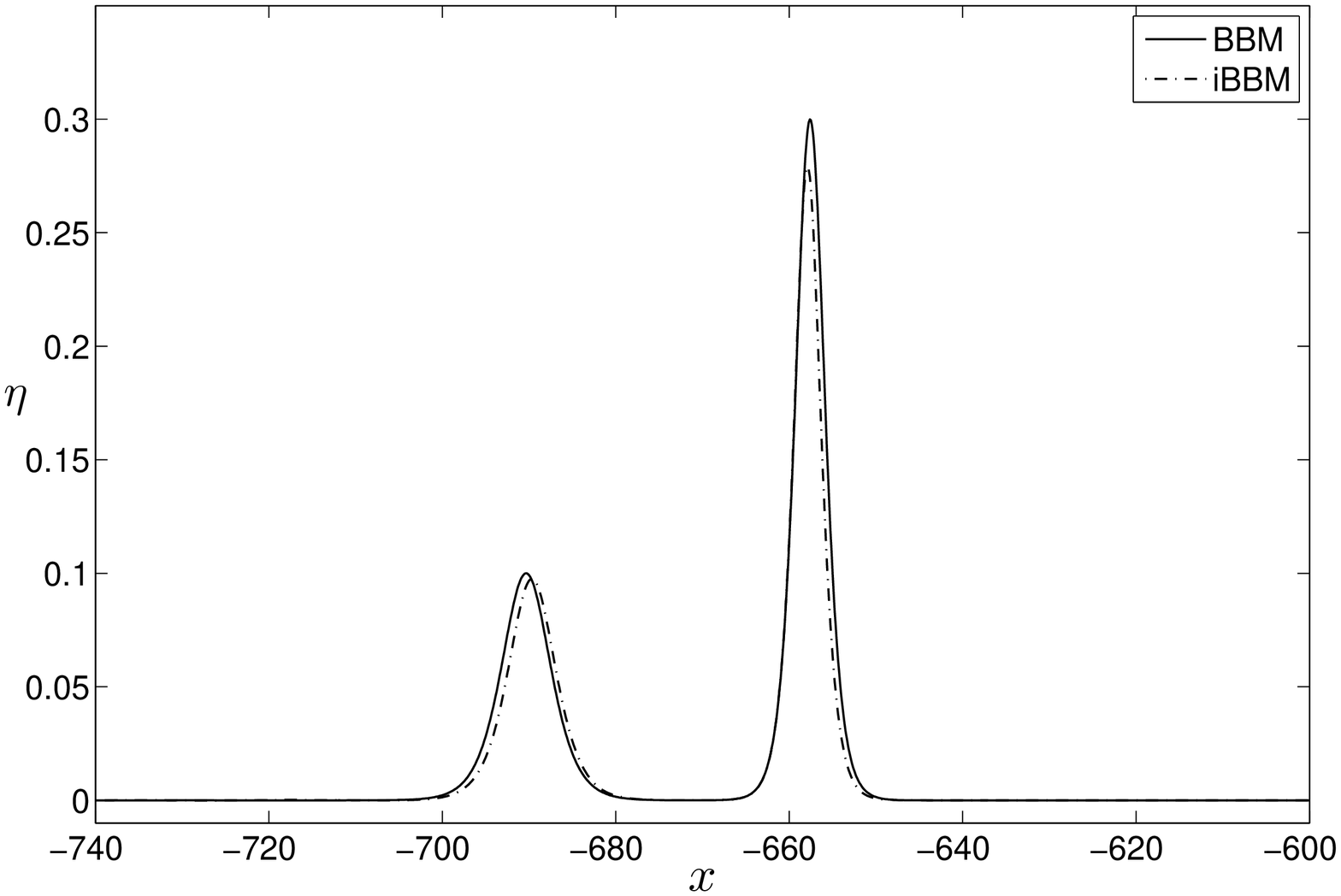}}
\subfigure[Solution after the interaction ($t=1750$)]{\includegraphics[width=0.49\textwidth]{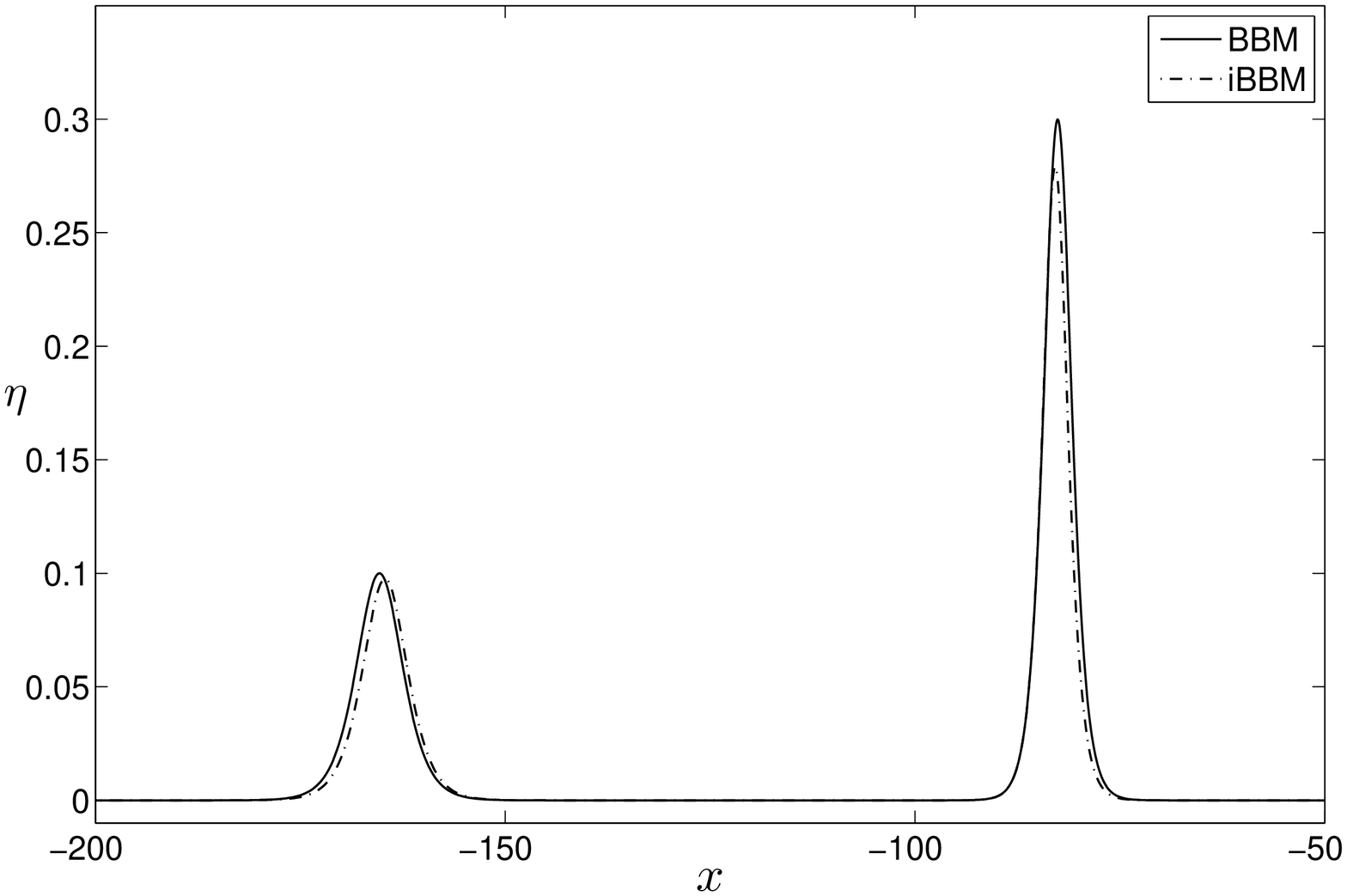}}
\subfigure[Magnification of the dispersive tail ($t=1750$)]{\includegraphics[width=0.49\textwidth]{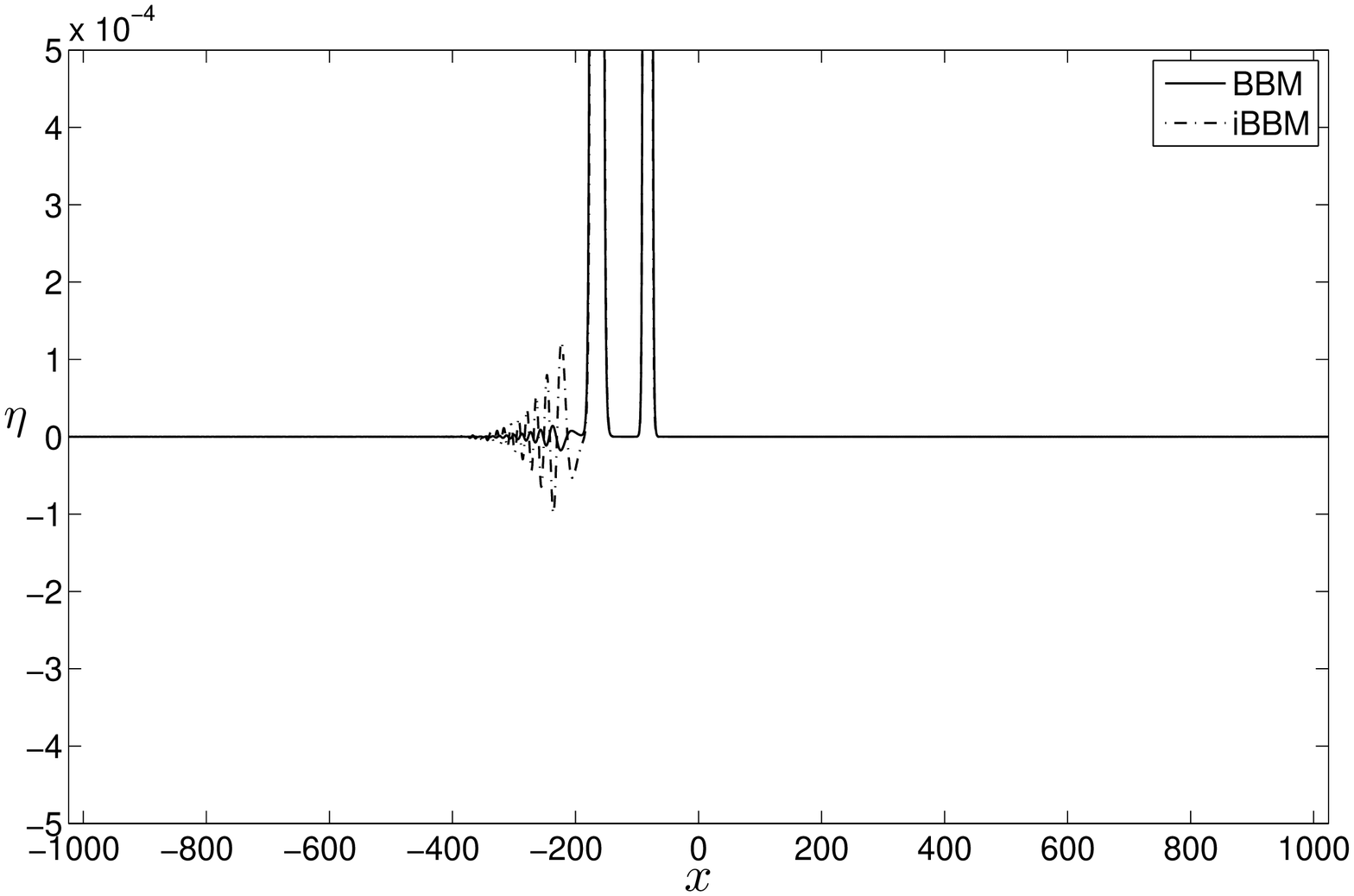}}
\caption{\em Overtaking collision of two solitary waves for \acf{BBM} and \acf{iBBM} models.}%
\label{Fig:overb}%
\end{figure}

With the same input data, an overtaking collision for the case of the \acs{BBM} and \acs{iBBM} equations is shown in Figure~\ref{Fig:overb} at several times. In this case, on the contrary, no N-shape wavelet is observed and only a tail behind the waves appears.
The tail following the solitary waves of the iBBM appears to be larger again as an effect of the high-order nonlinear terms. Also, the phase shift observed in the case of the iBBM is larger, indicating again a larger amount of nonlinearity and inelasticity in the specific interaction.

We can conclude that these two groups of results show that, as expected, the $\O(\eps\mu^{2})$ nonlinear terms of the invariant models do not change significantly, in a qualitative sense, the behaviour of the interactions of solitary waves provided by the corresponding not Galilean invariant system.

\subsection{Comparison with laboratory data}

\begin{figure}%
\centering
\subfigure[Solution before the interaction ($t=18.3$s)]{\includegraphics[width=0.49\textwidth]{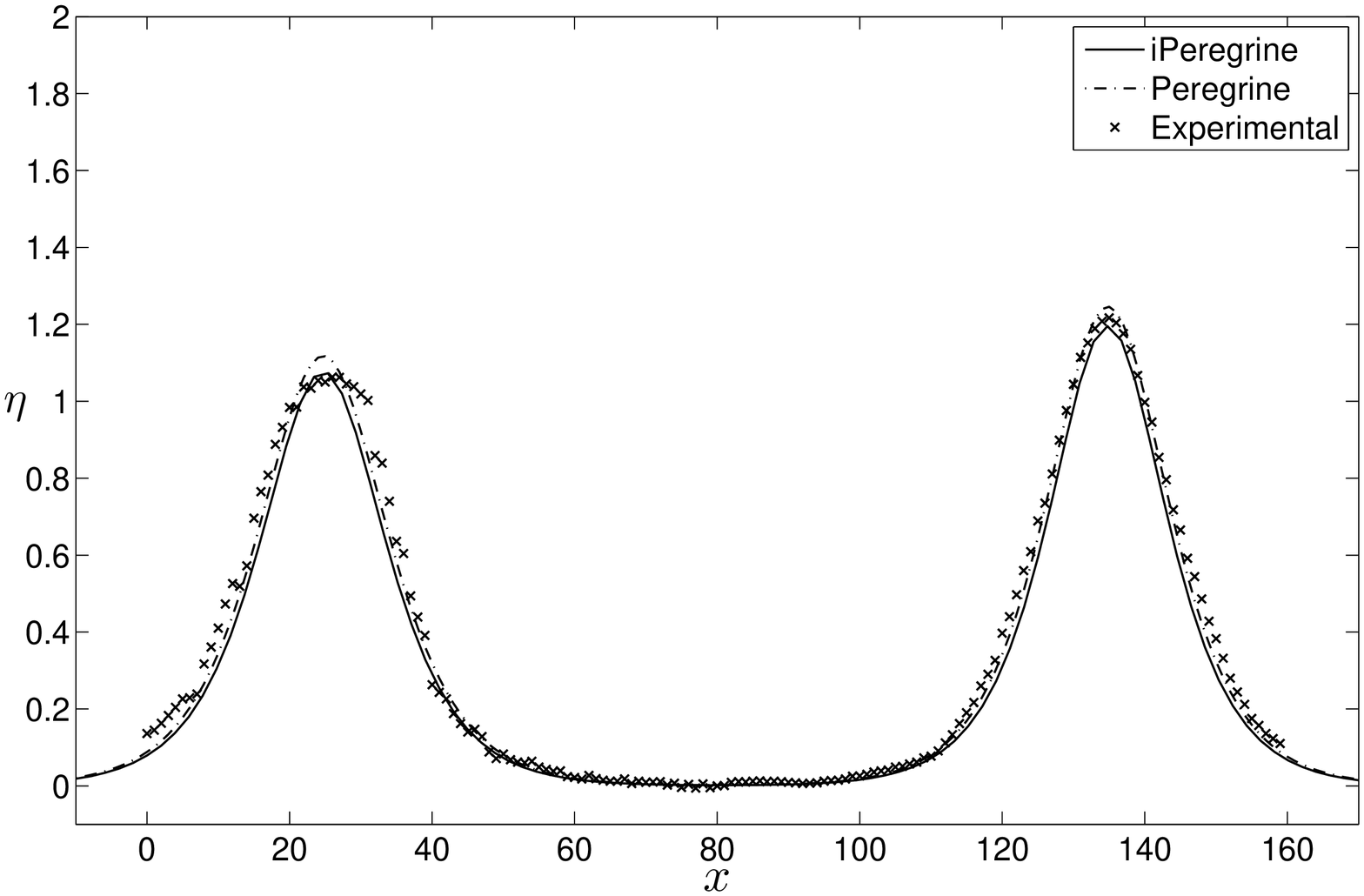}}
\subfigure[Solution during the interaction ($t=18.8$s)]{\includegraphics[width=0.49\textwidth]{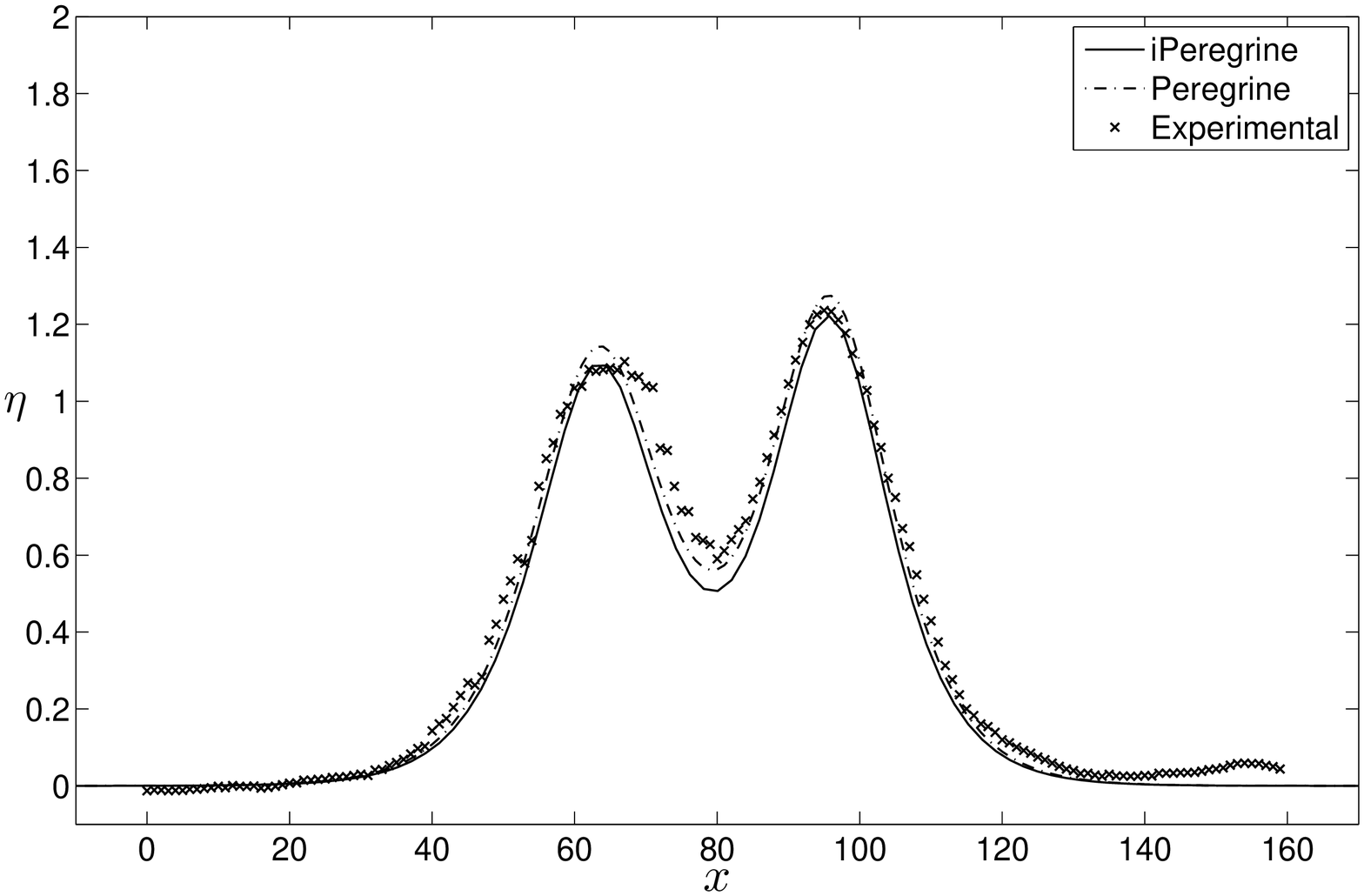}}
\subfigure[Solution after the interaction ($t=19.0$s)]{\includegraphics[width=0.49\textwidth]{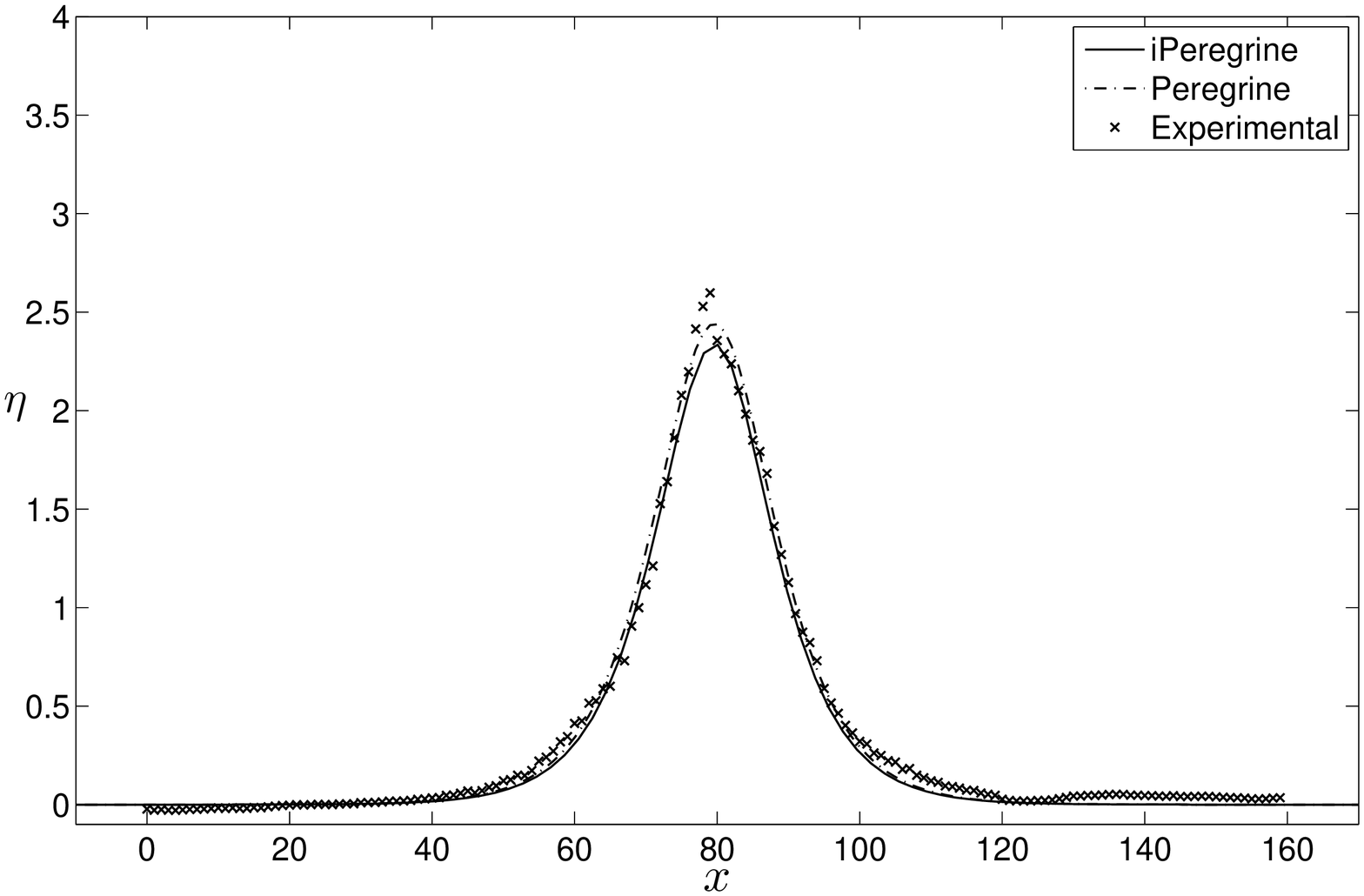}}
\subfigure[Solution after the interaction ($t=19.2$s)]{\includegraphics[width=0.49\textwidth]{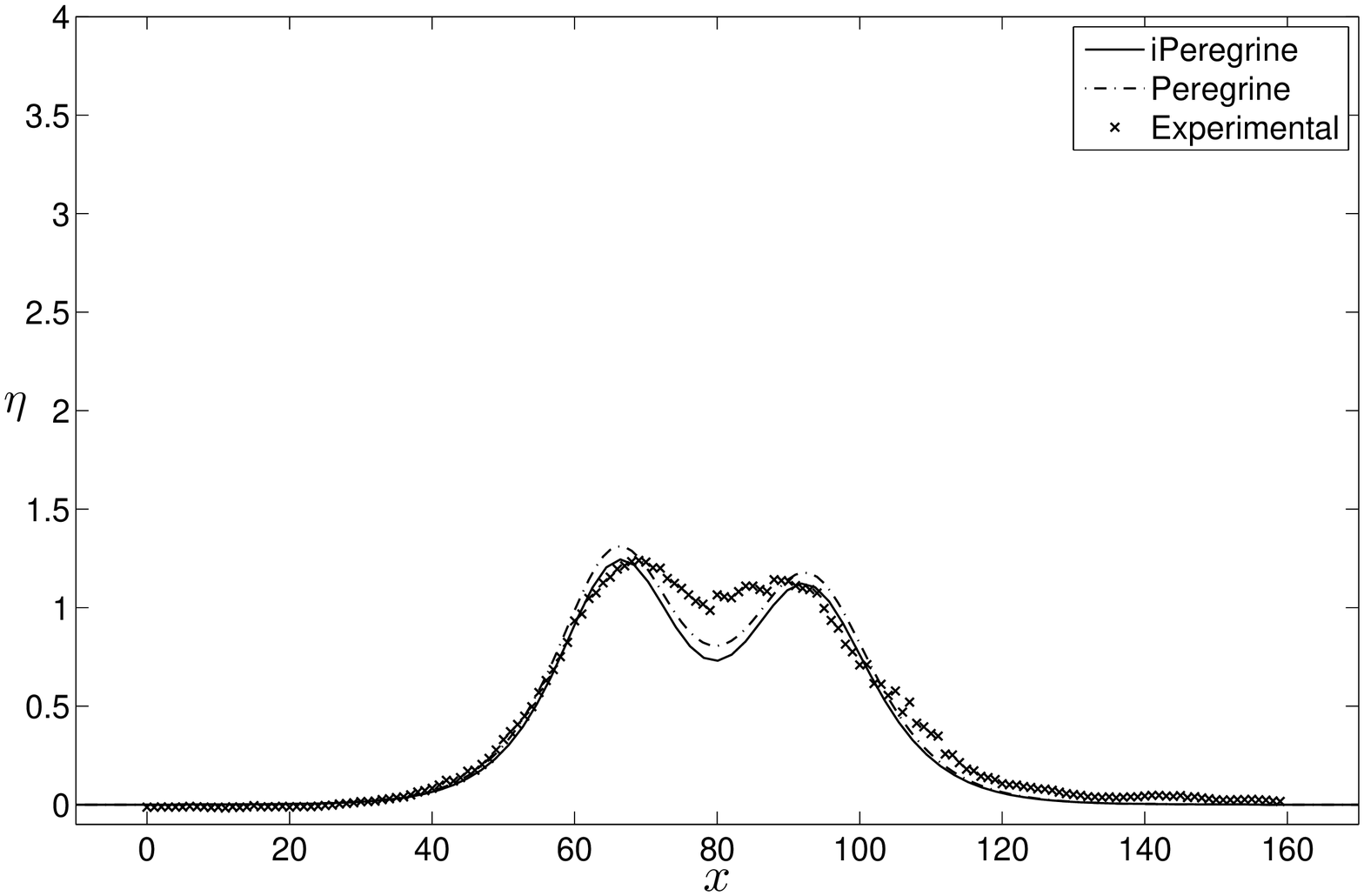}}
\subfigure[Solution after the interaction ($t=19.3$s)]{\includegraphics[width=0.49\textwidth]{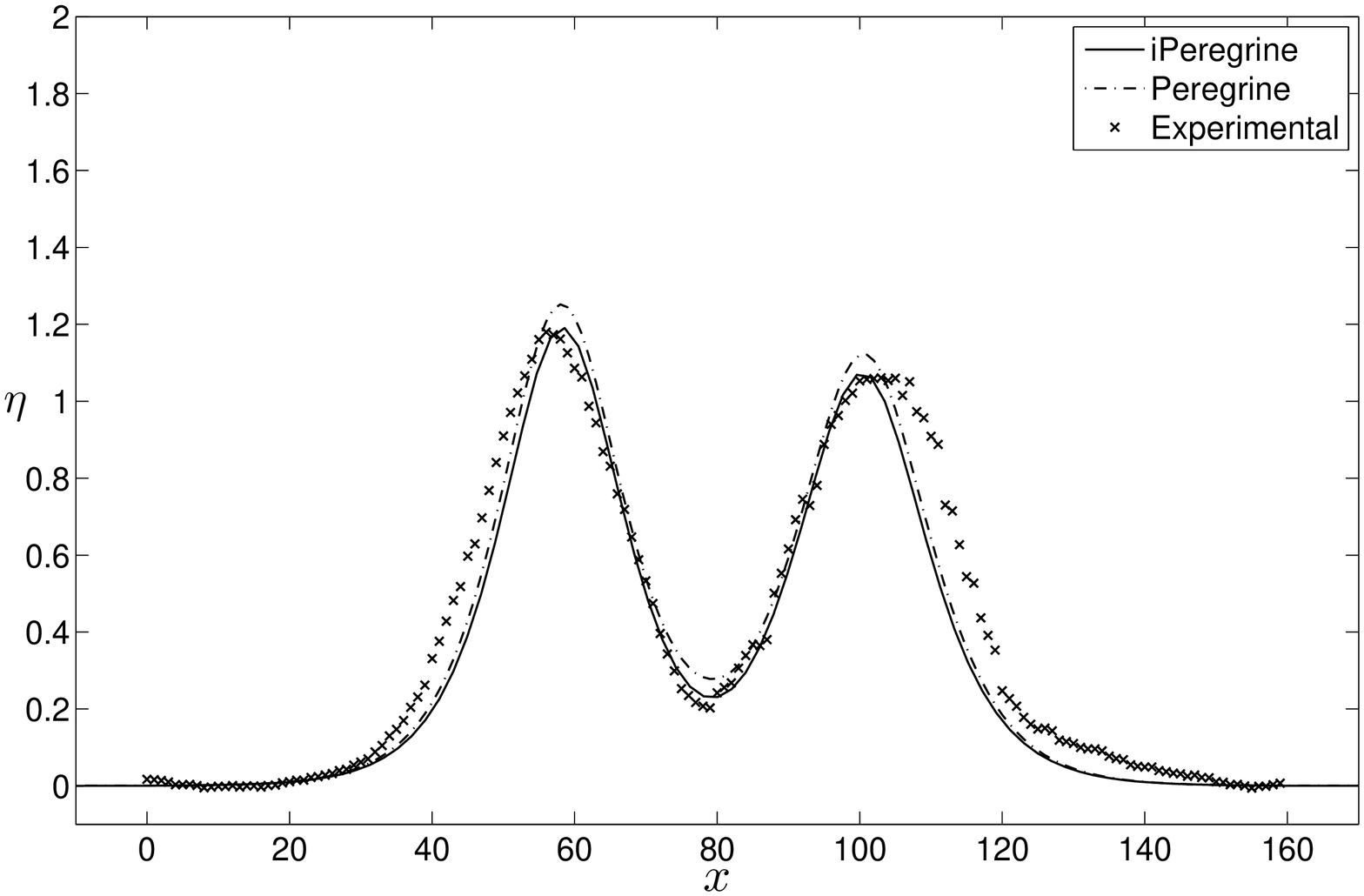}}
\subfigure[Magnification of the dispersive tail ($t=20.5$s)]{\includegraphics[width=0.49\textwidth]{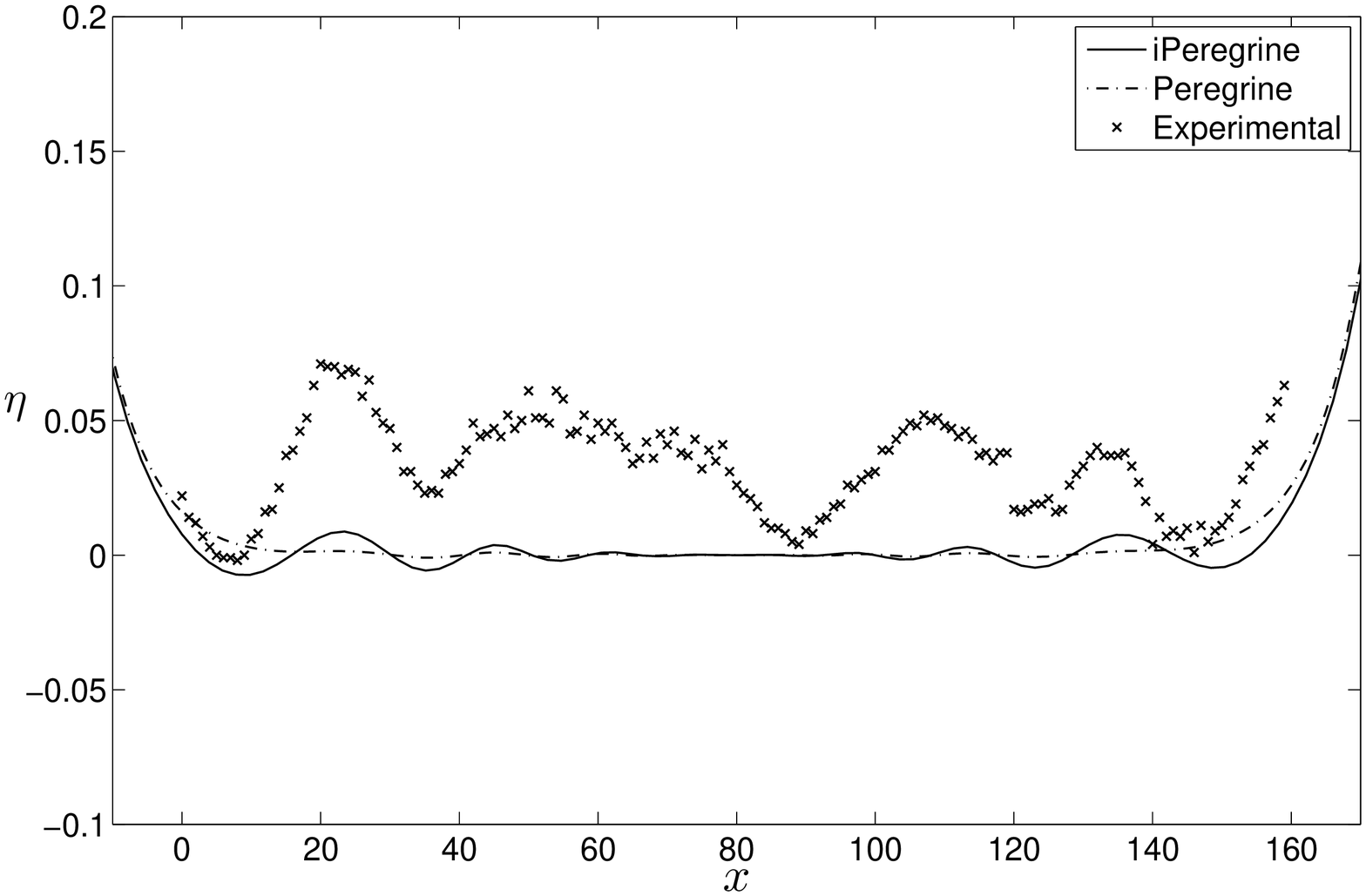}}
\caption{\em Head-on collision of two solitary waves and comparison with laboratory data.}%
\label{Fig:experp}%
\end{figure}

After observing the inelastic collisions in the previous section it rises the question of which model is more realistic. To answer this question we compare the numerical solution of the cPer and the iPer systems during a head-on collision with some experimental data from \cite{CGHHS}. Specifically, we consider two solitary waves of speeds $c_{s,1}=0.7721$ m/s and $c_{s,2}=0.7796$ m/s, respectively and translated such as fit with the laboratory data at $t=18.3$. These solitary waves are almost identical for both models and the head-on collision appeared to be be very similar compared to the laboratory data. A closer look of the dispersive tails shows that, again, after their interaction a larder dispersive tail is generated in the case of the iPer system. Comparing with the experimental data (see Figure \ref{Fig:experp}) we observe that the collision described by the iPer system is very close to the one described by the cPer and to the laboratory data as well. Moreover, the iPer system demonstrates a better performance in describing the dispersive tail showing that the iPer system could be realistic too. 

\subsection{Comparison with Euler equations}

\begin{figure}
\centering
\subfigure[Solution of the BBM equation at $T = 100$]%
{\includegraphics[width=0.49\textwidth]{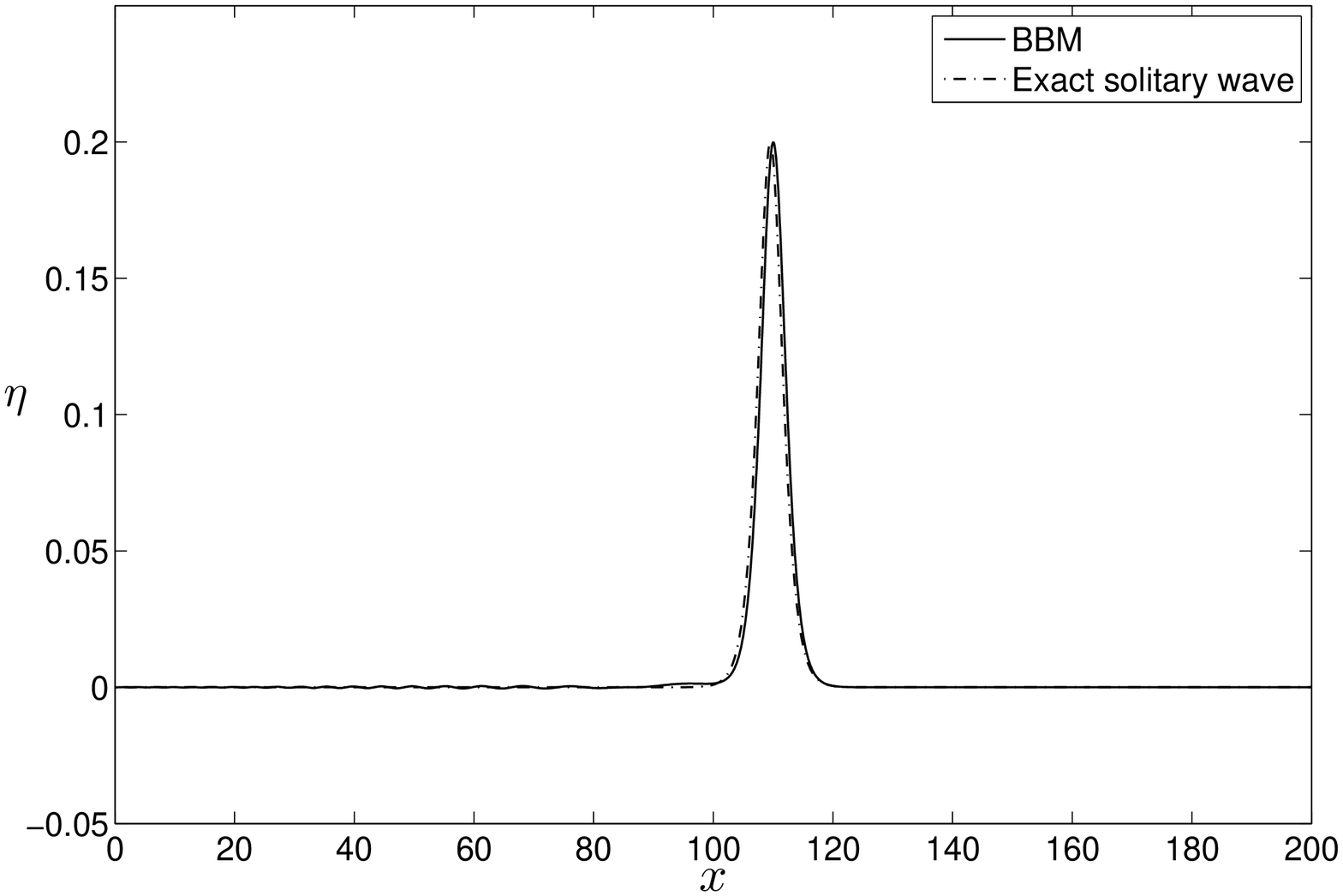}}
\subfigure[Solution of the iBBM equation at $T = 100$]%
{\includegraphics[width=0.49\textwidth]{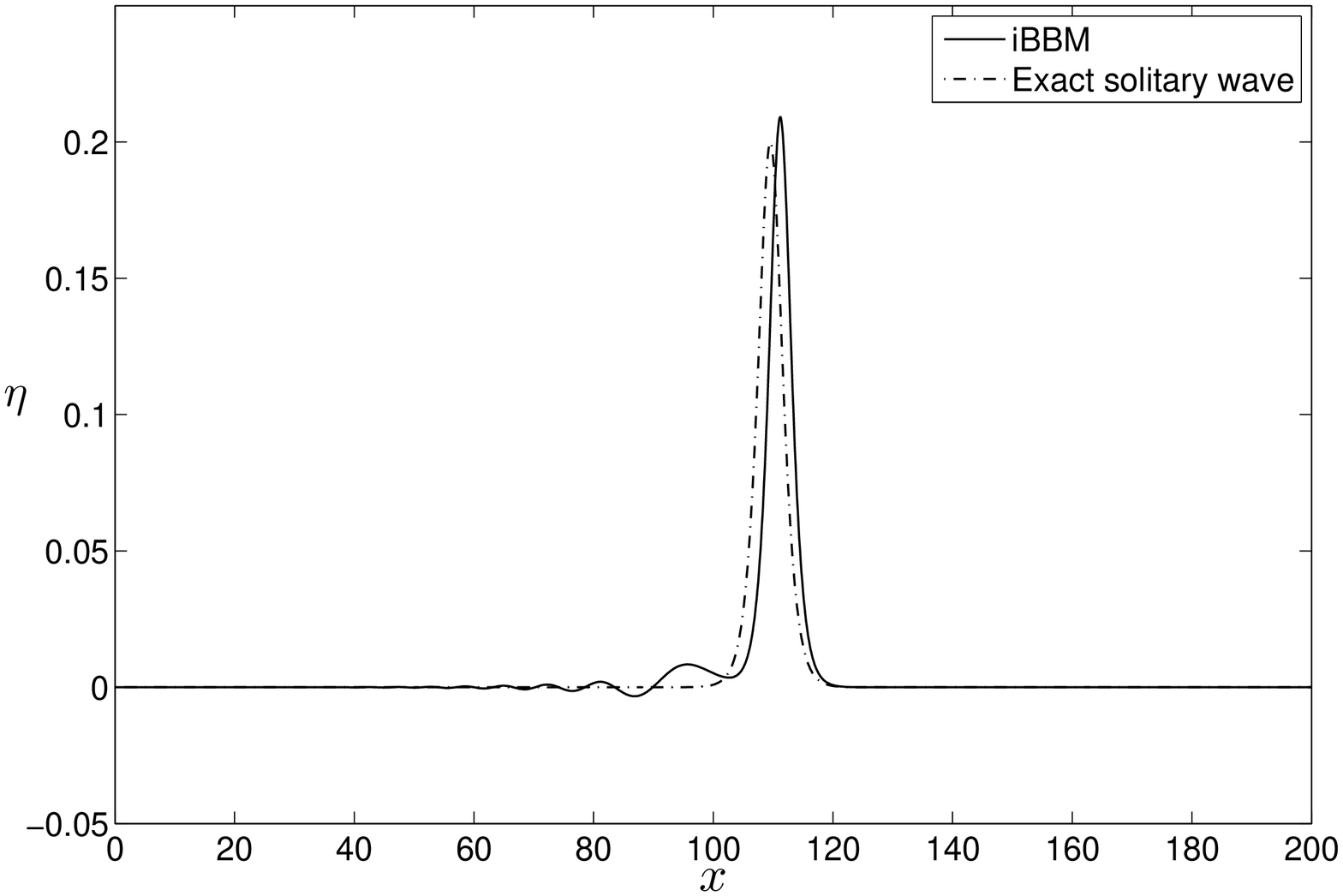}}
\subfigure[Solution of the Peregrine system at $T = 100$]%
{\includegraphics[width=0.49\textwidth]{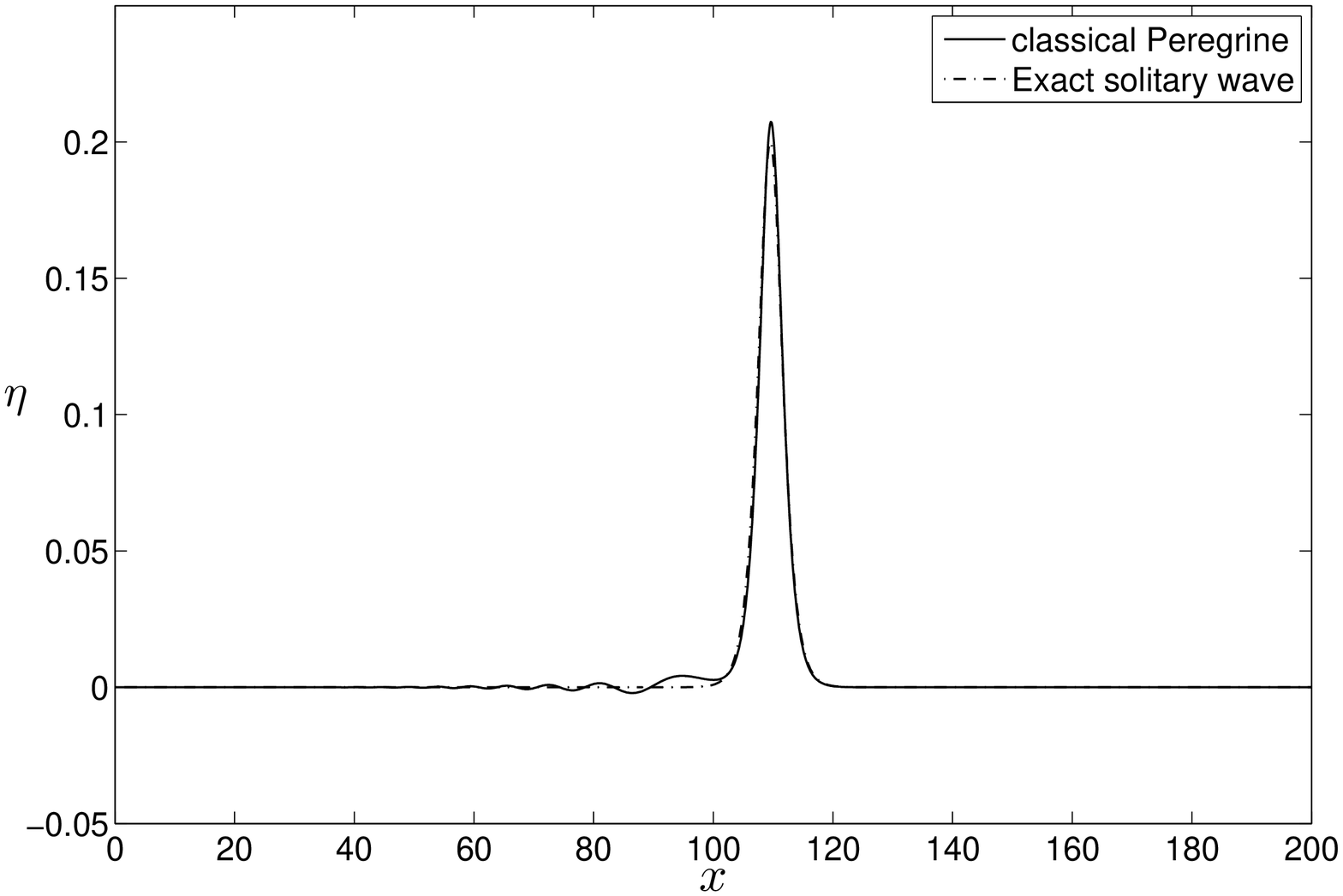}}
\subfigure[Solution of the iPeregrine system at $T = 100$]%
{\includegraphics[width=0.49\textwidth]{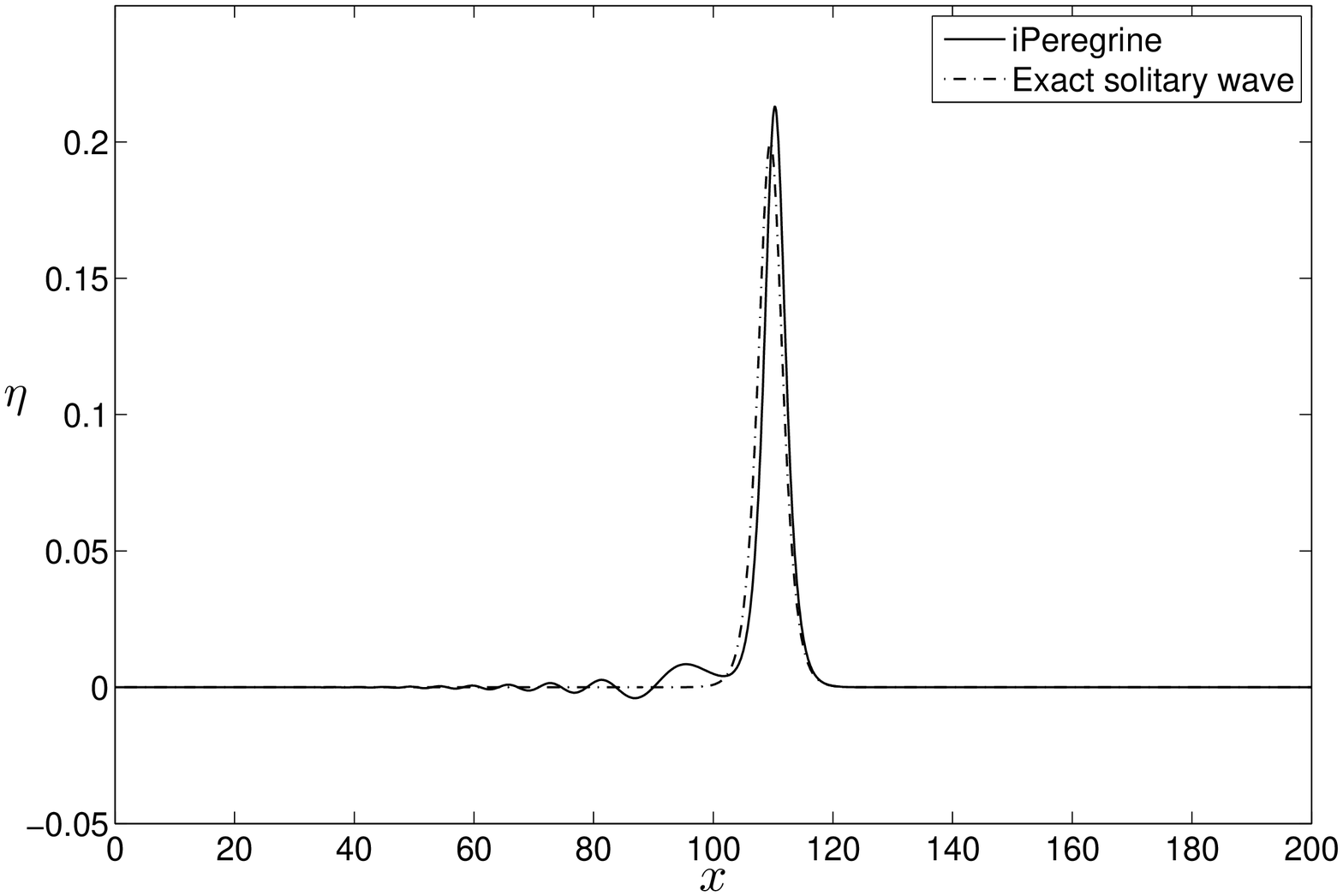}}
\caption{\em Evolution of a solitary wave of the Euler equations, with BBM, iBBM, cPer and iPer models. ($A = 0.2$)}
\label{Fig:compeu}
\end{figure}

Finally, we study the evolution of a solitary wave of the Euler equations when we use it as initial condition to the approximate models. Specifically, we consider an approximate solitary wave $\Phi_h(x)$  of amplitude $A = 0.2$ (and speed $c_s \approx 1.095490471188718$) obtained by using Fenton's ninth order asymptotic solution. In the case of the Peregrine and the iPeregrine systems we use for initial velocity $u_0(x) = \nicefrac{c_s\eta_0(x)}{d+\eta_0(x)}$. We remind that due to the mass conservation property this formula is exact (see equation \eqref{eq:etau} for the velocity given the surface elevation $\eta$ for both Peregrine and iPeregrine systems). In Figure~\ref{Fig:compeu} we present the solution at $T = 100$. We observe that the initial condition is resolved into a new solitary wave followed by a dispersive tail. In the case of the classical models, the dispersive tails appear to be smaller.

\begin{figure}
\centering
\subfigure[Shape error]%
{\includegraphics[width=0.49\textwidth]{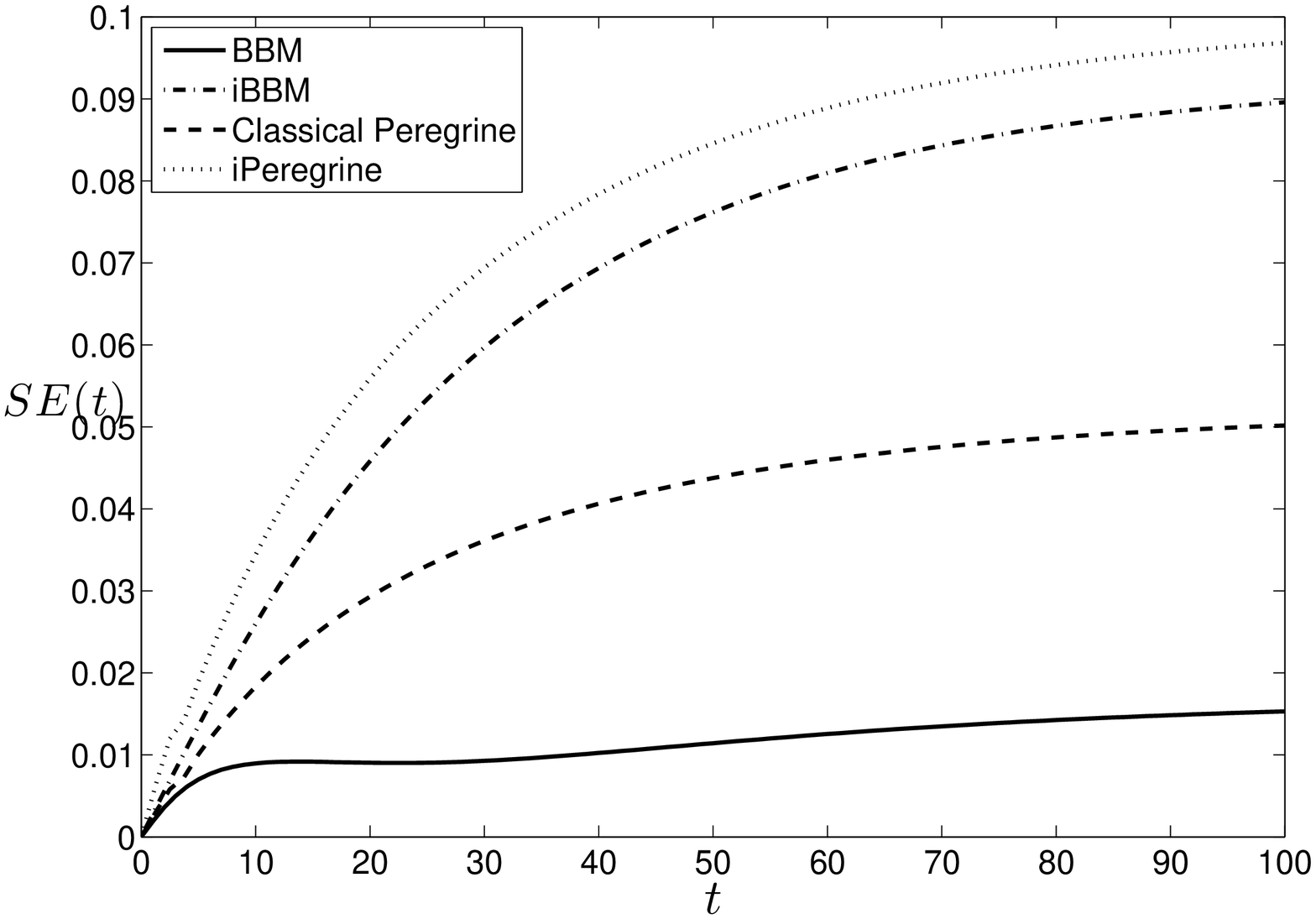}}
\subfigure[Amplitude]%
{\includegraphics[width=0.49\textwidth]{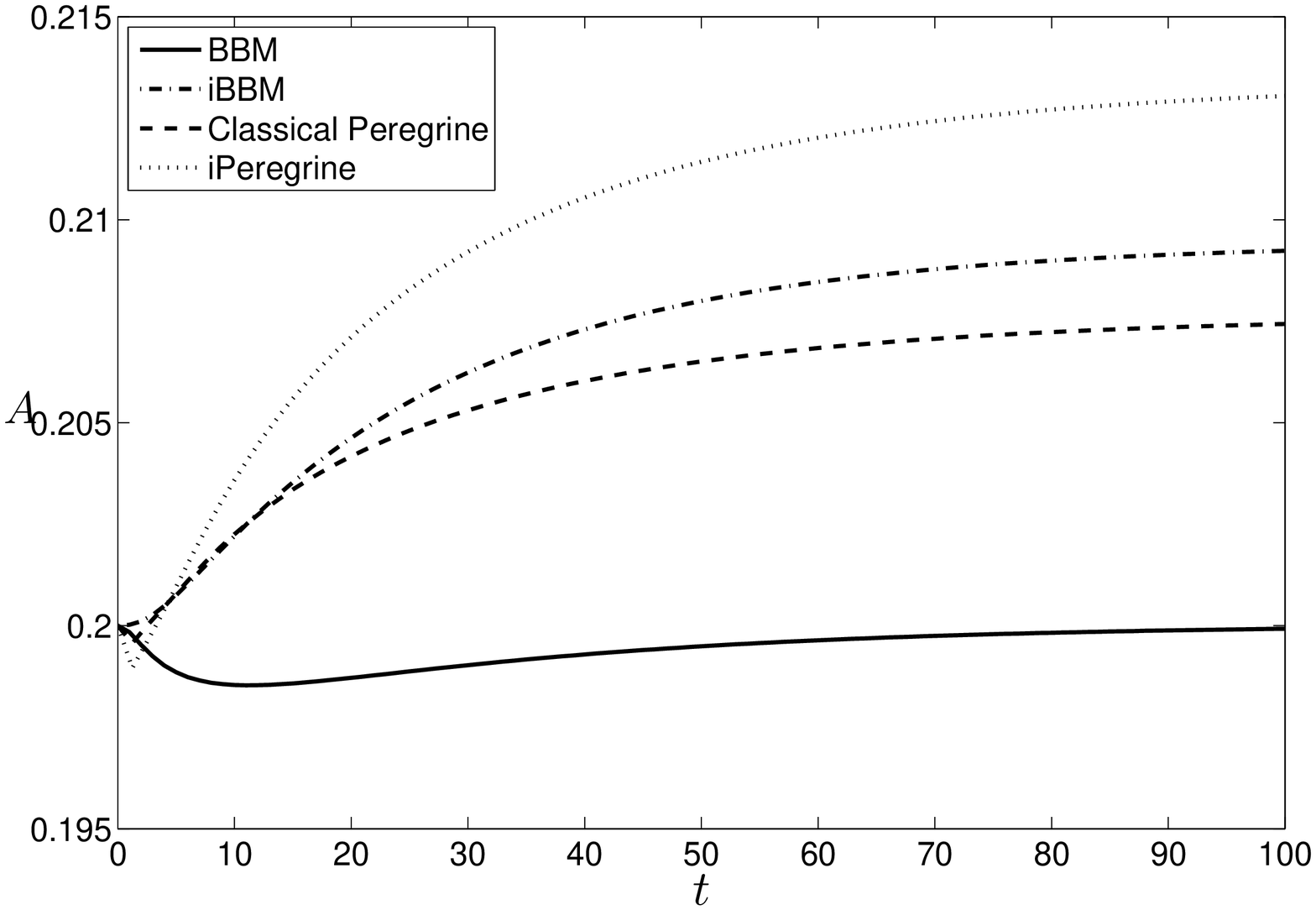}}
\caption{\em Shape error and amplitude of the solution as a function of time}%
\label{Fig:compeu2}
\end{figure}

Figure~\ref{Fig:compeu2} (a) shows the shape error of the solution (i.e. how much different is the solution from being the exact solitary wave of the Euler equations) while Figure~\ref{Fig:compeu2} (b) presents the amplitude of the solution as a function of time $t$. From these two figures we observe that in the case of classical models the emerging solitary waves are closer in shape and amplitude to the original solitary wave solution of the Euler equations than the respective solitary waves of the invariant models.

\section{Conclusions}\label{sec:concl}

In the present work the influence of Galilean invariance in several equations arising in water wave modelling is studied. We modify not Galilean invariant models in order to include this fundamental property. The technique introduced here consists of adding higher-order terms from the approximation of the governing equations. These terms are asymptotically negligible and consequently, the modified models are still valid in the appropriate regime. As a case study, corresponding modifications of two not Galilean invariant models, the \acf{BBM} equation and the classical Peregrine system, are presented. The comparison with reference solutions to the full Euler equations shows that this extra-term is beneficial for the description of the travelling wave solutions in several ways. First, this modification improves the solitary waves amplitude-speed relation which lies closer to the Tanaka's and Fenton's solutions. In this regard, we obtain a surprising performance of the iPeregrine system \eqref{eq:ip1}, \eqref{eq:ip2} with the amplitude-speed relation undistinguishable from the fully-nonlinear Serre equations \eqref{eq:ser1}, \eqref{eq:ser2}. Moreover, the amplitudes of solitary wave solutions to the invariant models are closer to the corresponding full Euler solutions than the classical counterparts. The comparison is finished off with a numerical study of head-on and overtaking collisions. Compared to the behavior observed in the not Galilean invariant equations, the higher order nonlinear terms incorporated in the new models do not affect qualitatively the inelastic character of the interactions. However, a relevant difference in the degree of inelasticity is observed, being higher in the case of the Galilean invariant versions. This behavior is closer to what has been observed in the case of the full Euler equations and in laboratory experimentals.

\subsection*{Acknowledgments}
\addcontentsline{toc}{section}{Acknowledgments}

D.~\textsc{Dutykh} acknowledges the support from French ``Agence Nationale de la Recherche'', project ``MathOc\'ean'' (Grant ANR-08-BLAN-0301-01) along with the support from ERC under the research project ERC-2011-AdG 290562-MULTIWAVE. The support from the University of Valladolid during the stay of D.~\textsc{Dutykh} in July, 2011 is equally acknowledged. A.~Duran has been supported by MICINN project MTM2010-19510/MTM.

The authors would like to thank Professors Didier \textsc{Clamond}, Vassilios \textsc{Dougalis} and Cesar \textsc{Palencia} for very helpful discussions.

\addcontentsline{toc}{section}{References}
\bibliographystyle{abbrv}
\bibliography{biblio}

\end{document}